\documentclass[12pt,onepage,oneside]{article}
\usepackage{fullpage}
\usepackage{amsmath,amsthm}
\usepackage{color,soul}
\usepackage{natbib}
\usepackage{hyperref}
\usepackage{booktabs}
\usepackage{amsfonts}
\usepackage{graphicx}
\usepackage[export]{adjustbox}
\usepackage{cleveref}
\usepackage{pdflscape}
\usepackage{multirow}
\usepackage[footnotesize,bf]{caption}
\usepackage{subcaption}
\usepackage[stable]{footmisc}
\usepackage{bm}
\usepackage{booktabs}
\usepackage{lineno}
\usepackage{makecell}
\usepackage{subcaption}
\usepackage{pdflscape}
\usepackage{xcolor}
\usepackage{longtable}
\usepackage{float}
\usepackage{threeparttable}
\usepackage{booktabs}
\usepackage{afterpage}
\usepackage{tabularx}

\usepackage{array}
\usepackage{courier}
\usepackage[normalem]{ulem}

\usepackage{setspace,caption}
\usepackage[margin=1in]{geometry}
\usepackage{footmisc}

\usepackage{cancel}

\usepackage{calc}
\newlength{\depthofsumsign}
\let\oldFootnote\footnote
\newcommand\nextToken\relax

\renewcommand\footnote[1]{%
    \oldFootnote{#1}\futurelet\nextToken\isFootnote}

\newcommand\isFootnote{%
    \ifx\footnote\nextToken\textsuperscript{,}\fi}

\renewcommand{\emph}[1]{#1}

\newtheorem{proposition}{Proposition}

\hypersetup{
    colorlinks,
    linkcolor={blue!50!black},
    citecolor={blue!50!black},
    urlcolor={blue!50!black}
}

\title{\textbf{Common Idiosyncratic Quantile Factors and Asset Prices}\thanks{We are grateful to the editor, Stephan Siegel, and anonymous reviewers for their useful comments and suggestions, which greatly improved the paper. We also thank Mykola Babiak, Mattia Bevilacqua, Michael Ellington, Deniz Erdemlioglu, Alex Kostakis, Jeroen Rombouts, participants at STAT of ML (Prague), Financial Econometrics Conference (Lancaster), 16th Annual Society for Financial Econometrics Conference (Rio de Janeiro) and seminars at Charles University (Prague), University of Sussex (Brighton), Lancaster University and University of Liverpool and various other seminars and workshops for valuable discussions and comments. Support from the Czech Science Foundation under the 24-11555S project is gratefully acknowledged. The Common Idiosyncratic Quantile (CIQ) factors we use are available for download and use at \url{https://github.com/matejnevrla/Common-Idiosyncratic-Quantile-Risk}}}
\author{%
Jozef {\sc Barun\'{i}k}$^{\rm a,b}$, and
Mat\v ej {\sc Nevrla}$^{\rm c,b}$\thanks{Corresponding author. Email address: matej.nevrla@liverpool.ac.uk}
\vspace{5mm} \\
\small $^{\rm a}$ Institute of Economic Studies, Charles University \vspace{0mm}\\
\small $^{\rm b}$ IITA, The Czech Academy of Sciences \vspace{0mm}\\
\small $^{\rm c}$ University of Liverpool Management School \vspace{0mm}\\
}
\begin{document}
%%\linenumbers
\maketitle
%\newpage
\begin{abstract}

\noindent  We investigate whether the tails of firm-level idiosyncratic return distributions are driven by common shocks. We use quantile factor analysis to extract such common idiosyncratic quantile factors with asymmetric pricing effects and we find a significant premium for innovations to the lower-tail factor: high-beta stocks outperform low-beta stocks by around 7-8\% per year. This premium remains significant even when controlling for standard factors, idiosyncratic volatility and tail-risk measures. The downside factor strengthens when intermediary capital is weak and market liquidity is low, and it predicts aggregate market excess returns.
\noindent 
\\

\end{abstract}

\newpage

\section{Introduction}
\label{sec:introduction}

Conventional wisdom holds that firm-level idiosyncratic risk, particularly rare, firm-specific tail shocks, can be diversified away. However, periods of market stress tend to exhibit a striking regularity: many firms experience unusually negative returns at the same time, and these fluctuations cannot be explained by volatility alone. This means that tail events, which appear diversifiable in isolation, become systemic in aggregate. In this paper, we examine this phenomenon and ask whether a common factor exists in the tails of firm-level idiosyncratic returns that is relevant for investors. We also consider what economic state this tail comovement reflects.

To address these issues, we introduce a non-parametric measure of common idiosyncratic tail risk. After removing standard linear factors from individual stock returns, we estimate a quantile factor model of \cite{chen2021} on rolling windows to extract the dominant factor of the cross-sectional conditional quantiles of the idiosyncratic return distributions. We refer to this latent factor as the common idiosyncratic quantile (CIQ) factor. Our focus is on innovations, and we orient the factors so that they positively correlate with the corresponding realized cross-sectional quantile of idiosyncratic returns. This means that a decline in the lower-tail factor corresponds to a general deterioration in downside idiosyncratic conditions. The key empirical object is each stock's exposure to innovations in downside CIQ, which is estimated using time-series regressions and is then used to form out-of-sample portfolios.

Our first set of results shows that downside idiosyncratic tail comovement is priced effectively. Sorting stocks according to their exposure to lower-tail CIQ innovations reveals a consistent pattern of average returns and a high-minus-low spread of approximately 7-8\% per year. In contrast, exposure to median and upper-tail CIQ innovations does not earn a premium and has small, statistically insignificant risk price. This asymmetry is difficult to reconcile with a purely volatility-driven narrative. The premium remains substantial even when controlling for standard factor models, common idiosyncratic volatility \citep{HERSKOVIC2016} and significant downside and tail-risk measures, such as the tail-risk factor of \cite{kelly2014}. The premium is also robust across quantile thresholds: it is concentrated in the lower tail and diminishes as the quantile threshold approaches the centre of the distribution.

We explain these results with an economic mechanism based on intermediary asset pricing and systemic fragility. In intermediary models such as \citet{HeKrishnamurthy2013}, \citet{BrunnermeierSannikov2014}, and \citet{HeKellyManela2017}, shocks to intermediary net worth tighten risk-bearing constraints, raising the price of risk. We demonstrate that the cross-sectional distribution of idiosyncratic stock returns encompasses a related state variable: a shared element in firms' left-tail idiosyncratic quantiles. In a simple intermediary fragility setting, a single constraint multiplier shifts the lower-tail CIQ factor's quantiles across assets; the implied quantile representation links innovations in this fragility state to innovations in the lower-tail CIQ factor. Consistent with this interpretation, we find that lower-tail CIQ deteriorates when intermediary capital is weak and market liquidity is low, and that firms with fragile trading conditions and limited financial slack have the strongest downside-CIQ exposures.

Finally, the same state variable is important for aggregate risk premia. Innovations in the lower-tail CIQ factor forecast future market excess returns; when downside idiosyncratic tail risk worsens, subsequent aggregate returns increase. This predictive relationship is economically significant and remains consistent in both in-sample and out-of-sample tests. In contrast, the upper-tail CIQ factor exhibits weaker and less stable predictability. Overall, the evidence supports the idea that markets price common downside idiosyncratic tail risk, which becomes systematic when tied to intermediary constraints and liquidity conditions.

Our results contribute to several areas of literature. Firstly, we present a novel, distribution-sensitive measure of common idiosyncratic risk which does not depend on parametric tail assumptions or options data and which distinguishes between downside and upside tail comovement. Secondly, we demonstrate that exposure to innovations in downside CIQ carries a significant premium that differs from that of standard factors, volatility-based idiosyncratic risk and existing tail-risk measures. Thirdly, we provide economic content for the factor by linking it to intermediary risk-bearing capacity and liquidity, and by demonstrating that firm-level exposures correspond with balance sheet and trading fragility characteristics. Fourthly, we demonstrate that innovations in downside CIQ forecast the aggregate equity premium, thereby connecting cross-sectional pricing and time-series predictability via a single stress-related state variable.

\section{Common Idiosyncratic Quantile (CIQ) Factors: Asymmetric Idiosyncratic Risk}
\label{sec:ciq_factors}

Standard linear factor models summarise risk through variation, implicitly treating downside and upside outcomes as symmetric mirror images. Our approach, however, focuses on common shifts in the tails of the cross-sectional distribution of idiosyncratic returns. This approach retains directional information in adverse and favourable states, providing a more detailed characterisation of systematic risk than volatility-based measures.

For each month $m$ and every stock $i$, we use a rolling window of 60 months and retain stocks for which there are complete observations within the window. Within each window, we estimate a baseline factor model and save the residuals
\begin{align}
	\label{eq:ret_ts}
	r_{i,t} = \alpha_{i, m} + \beta^{MKT}_{i, m} MKT_t + \beta^{SMB}_{i, m} SMB_t + \beta^{HML}_{i, m} HML_t + e_{i, t, m}, \quad t=m-59,\ldots,m.
\end{align}
We use the three-factor model of \citet{FAMA19933} (FF3) as our baseline model as it aligns our approach with the existing literature on commonality in idiosyncratic risks and tail exposures \citep{ang2006vol,kelly2014}. Moreover, augmenting the baseline factor set has only a modest effect on the time variation in realised idiosyncratic quantiles, and FF3 provides a transparent benchmark for evaluating the additional information provided by tail-based factors. In Section~\ref{subsec:stability} we demonstrate that our pricing results remain robust when using richer linear specifications, including the five- and six-factor models of \citet{FAMA20151,FAMA2018234}.\footnote{Appendix~\ref{sec:cs_quantiles} reports variance ratios for realised idiosyncratic quantiles across alternative factor models, showing that residual quantiles exhibit significant time variation.}

We study common tail variation using the residuals $e_{i,t}$ from Equation \eqref{eq:ret_ts}. As in principal component analysis, we standardize these residuals by their within-window standard deviation.\footnote{This standardization makes the cross-sectional quantiles comparable across stocks and stabilizes estimation in the rolling windows.} For a probability level $\tau\in(0,1)$ in the current month $m$, we estimate a common idiosyncratic quantile factor $CIQ_{t,m}(\tau)$ via
\begin{align}
	\label{eq:res_ciq}
	%\forall \tau: e_{i,t} = \gamma_i (\tau) CIQ_t (\tau) + u_{i, t} (\tau)
	e_{i,t,m} = \gamma_{i,m} (\tau) CIQ_{t, m} (\tau) + u_{i, t, m} (\tau), \quad t=m-59,\ldots,m
\end{align}
where $u_{i,t,m}(\tau)$ satisfies the quantile restriction $P\{u_{i,t,m}(\tau)<0\mid CIQ_{t, m}(\tau)\}=\tau$. We estimate $CIQ_{t, m}(\tau)$ and $\gamma_{i, m}(\tau)$ using the quantile factor analysis procedure of \citet{chen2021}.\footnote{Details of the algorithm and implementation are provided in Appendix~\ref{sec:qfm}.}
At each $\tau$, we use only the first (most informative) factor. In our applications, the factor-number selection procedures in \citet{chen2021} overwhelmingly select a single factor for the residual-quantile panels.

In the baseline analysis, we focus on three CIQ factors. We capture common movements in adverse idiosyncratic outcomes using the \textit{lower-tail} factor at $\tau = 0.2$, $CIQ^{LT} \equiv CIQ(0.2)$. This choice is motivated by (i) evidence that disappointment-type preferences are linked to approximately the worst 20\% of outcomes in several asset-pricing settings\footnote{See, for example, \citet{GiglioKellyPruitt2016}, \citet{Delikouras_Kostakis_2019}, \citet{massacci2025factor}, and \citet{FARAGO201869}.}, (ii) its close relationship to other downside quantiles below the median (e.g., $\tau = 0.1$ and $\tau = 0.3$), and (iii) the practical trade-off between focusing on tail events and retaining sufficient observations for precise estimation. As a counterpart capturing common upside movements, we define the \textit{upper-tail} factor at $\tau = 0.8$, $CIQ^{UT} \equiv CIQ(0.8)$, and we define the \textit{central} factor at $\tau = 0.5$, $CIQ^{C} \equiv CIQ(0.5)$, to summarize movements in the middle of the idiosyncratic return distribution.

Quantile factors can only be identified up to sign. We impose a normalization that gives their movements an economically meaningful direction. Specifically, each CIQ factor is oriented so that it is positively correlated with the corresponding cross-sectional realized quantile of idiosyncratic returns. For the lower-tail factor, this means that a decline in $CIQ_t^{LT}$ indicates a deterioration in downside idiosyncratic outcomes across firms. For the upper-tail factor, an increase in $CIQ_t^{UT}$ corresponds to improved residual upside potential. This directional information is precisely what cannot be preserved by symmetric dispersion measures.

In line with the volatility-factor literature, we focus on innovations in conditional idiosyncratic quantile factors,
\begin{equation}
\label{eq:ciq_diff}
\Delta CIQ_{t, m} (\tau) \equiv CIQ_{t, m} (\tau) - CIQ_{t-1, m} (\tau), \quad t=m-58,\ldots,m.
\end{equation}
and conduct our baseline analysis using $\Delta CIQ_{t, m}(\tau)$ rather than levels.\footnote{We report pricing results using levels or AR(1) innovations in Section~\ref{subsec:stability} and obtain qualitatively similar conclusions in both cases.} This choice reflects the standard idea that investors are compensated for exposure to innovations in systematic risk, as set out in the ICAPM of \citet{Merton1973} and commonly used in volatility-based settings \citep{ang2006vol,HERSKOVIC2016}. In our setting, factor levels can combine persistent distributional features with higher-frequency stress fluctuations. First differences emphasize the latter and yield a sharper proxy for shifts in tail conditions.\footnote{Differences are computed within each 60-month rolling window using information up to time $m$. The resulting 59 differences are used to estimate stock-level exposures, which are then related to out-of-sample returns in month $m+1$. For time-series predictability tests, the final difference from each window, $\Delta CIQ_{m, m}(\tau)$, is used to predict the market return in month $m+1$.} We repeat the steps in Equations (\ref{eq:ret_ts}) to (\ref{eq:ciq_diff}) until the whole dataset is exhausted.

We estimate quantile-specific factors using idiosyncratic returns rather than raw returns due to several reasons. Firstly, this approach is consistent with the existing literature on common movements in idiosyncratic volatility and tail risk.\footnote{For example, \citet{ang2006vol} construct idiosyncratic volatility relative to FF3, and \citet{kelly2014} study tail-risk exposures in a similar residual-return framework.} Secondly, quantiles depend on both location and scale, so removing well-documented linear risk exposures isolates distributional variation beyond the mean. Thirdly, linear factor models provide a familiar benchmark against which the additional content of tail-based factors can be evaluated. Finally, since exposures to common linear factors are often easy to hedge, working with residuals enables us to study tail variation that is not mechanically spanned by standard factor structures.

Our framework is nonparametric: it relies on conditional quantiles of observed returns and does not impose a parametric structure for nonlinear dependence.\footnote{For instance, \citet{gorodnichenko2017level} study joint level and volatility factors, and other approaches use copulas to model tail dependence \citep{amengual2020,oh2017modeling}.}
Quantiles are informative about a range of distributional features, including but not limited to volatility. Under a location-scale model, common movements in quantiles are tightly linked to common movements in volatility, consistent with the factor structure in idiosyncratic volatility (e.g., \citealp{ang2006vol}). In that special case, principal components applied to squared residuals (PCA-SQ) can recover a volatility factor.

When higher-order features matter, however, volatility-based methods may miss economically relevant tail risks. Quantile factor models remain informative by directly targeting distributional shifts beyond the first two moments. Appendix~\ref{sec:beyond_vol} illustrates this distinction in a simple theoretical example.

\subsection{Data and Descriptive Statistics}

We estimate CIQ factors using monthly CRSP stock returns from January 1963 to December 2024. We include common stocks (share codes 10 and 11), adjust for delistings following \citet{bali2016empirical}, and exclude penny stocks with prices below \$1 to mitigate microstructure-related biases.\footnote{See, e.g., \citet{AMIHUD200231}.} With a 60-month rolling window, the first set of returns that we are able to predict comes in January 1968.

\begin{figure}[ht!]
\caption{CIQ Factors} 
\centering
\scriptsize
\begin{minipage}{\textwidth} 
The figure captures the lower-tail ($\tau = 0.2$), central ($\tau = 0.5$), and upper-tail ($\tau = 0.8$) $CIQ(\tau)$ factors, alongside the PCA-SQ factor. The factors are estimated on monthly idiosyncratic returns with respect to the three-factor model developed by \cite{FAMA19933} on the basis of either the quantile factor analysis suggested by \cite{chen2021} (CIQ) or a principal component analysis using squared idiosyncratic returns (PCA-SQ). Factors are estimated using a rolling window of 60 months. In each window, the last estimated value is plotted. CIQ-C a PCA-SQ are standardized to have zero mean and same volatility  as the CIQ-LT factor. The data come from the CRSP and cover the period from January 1968 to December 2024. We exclude penny stocks with prices below \$1. The shaded areas represent NBER recessions.
\end{minipage}
\vspace{1em}
\includegraphics[scale=0.45]{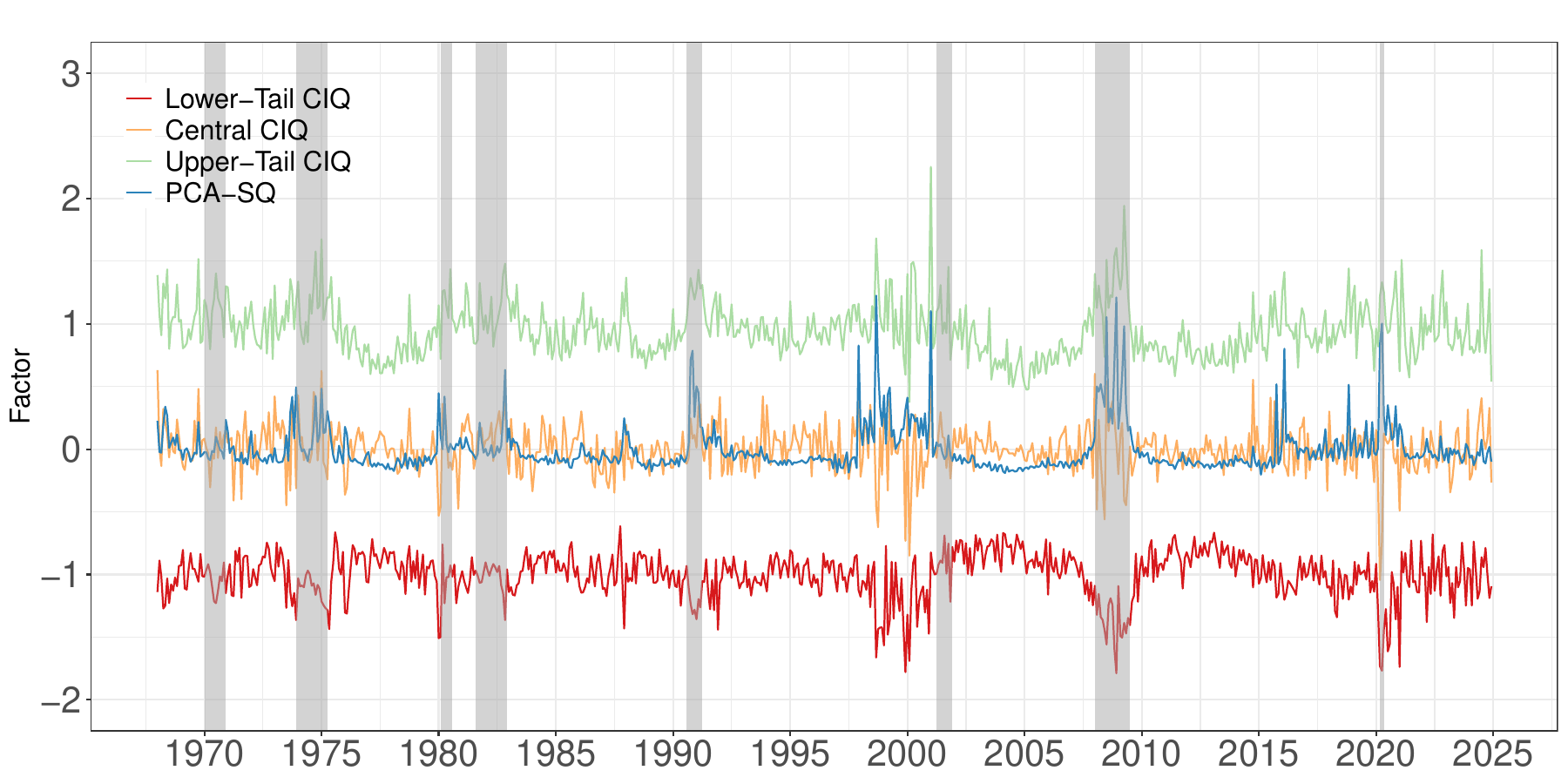}
\label{fig:ciq_factors}
\end{figure}

Figure~\ref{fig:ciq_factors} plots the lower-tail, central, and upper-tail CIQ factors alongside a volatility-based benchmark (PCA-SQ).\footnote{The PCA-SQ factor is estimated in a similar way to the CIQ factors, using principal component analysis on a balanced panel of squared FF3 residuals over a rolling window of 60 months. Only the last observation from each window is plotted.} Two features stand out. First, CIQ factors exhibit pronounced variation and spikes around major intermediary-based stress episodes, such as that between 2007 and 2009. This is consistent with state-like shifts in cross-sectional tail conditions. Second, the lower- and upper-tail factors are not mirror images, indicating that downside and upside tail states are distinct.

Panel A of Table~\ref{tab:ciq_short_summary} summarizes the distributional properties of $\Delta CIQ$ factors.\footnote{Table~\ref{tab:ciq_summary} in Appendix~\ref{sec:ciq_summary} reports summary statistics and the full correlation structure across $\Delta CIQ(\tau)$.} Standard deviations are large relative to means, kurtosis exceeds three, and first-order autocorrelations are negative, indicating that tail innovations are short-lived and mean reverting. Panel B shows that $\Delta CIQ^{LT}$ and $\Delta CIQ^{UT}$ are only weakly correlated (0.13), rejecting the view that the two series simply rescale a common dispersion factor. Correlations with $\Delta$PCA-SQ are economically meaningful and vary by tail, as expected if quantiles partly reflect variance but also contain tail-specific information. Correlations with prominent idiosyncratic volatility and tail-risk factors are moderate to small, highlighting that CIQ innovations are related to, but not subsumed by, existing measures. Panel C links $\Delta CIQ$ to forward-looking uncertainty measures. Higher $\mathrm{VIX}^2$ at the end of month $t$ predicts a decline in the lower-tail factor in $t+1$, i.e., a worsening of downside idiosyncratic tail conditions, with similar patterns for its components. In contrast, the upper-tail factor exhibits weak or no predictive relationship with these measures. A similar asymmetry appears for narrative-based volatility \citep{MANELA2017137}: topics related to \textit{Securities Markets} and \textit{Intermediation} predict deteriorations in lower-tail idiosyncratic conditions, while the upper tail remains largely unrelated.

\begin{table}[t!]
\caption{Summary of the $\Delta$CIQ Factors} 
\centering
\scriptsize
\begin{minipage}{\textwidth} 
The table provides summary statistics of the estimated lower-tail ($\tau = 0.2$), central ($\tau = 0.5$), and upper-tail ($\tau = 0.8$) $\Delta CIQ (\tau)$ factors. Factors are estimated using idiosyncratic returns with respect to the three-factor model of \cite{FAMA19933} and using the quantile factor analysis of \cite{chen2021}. Factors are estimated using a 60-month moving window, with differences being computed during that period and the last value being used. In Panel A, we report descriptive statistics of the $\Delta$CIQ$(\tau)$ factors, including their means, standard deviations, skewness, kurtosis and autocorrelation coefficients of order between one and three. In Panel B, we report contemporaneous correlations between the $\Delta$CIQ$(\tau)$ factors themselves, as well as with other related factors: differences of the PCA-SQ factor ($\Delta$PCA-SQ), differences of the common idiosyncratic variance factor ($\Delta$CIV) of \cite{HERSKOVIC2016}, differences of the cross-sectional bivariate idiosyncratic volatility ($\Delta$CBIV) factor of \cite{Han_Li_2025}, differences of the tail risk factor ($\Delta$TR) of \cite{kelly2014}, and differences of the sentiment factor of \cite{baker2006,baker2007}. In Panel C, we report correlations between the $\Delta$CIQ$(\tau)$ factors at time $t+1$ and forward-looking variance measures at time $t$. First, we report correlations with the squared CBOE volatility index ($\textrm{VIX}^2$) and its two components: conditional variance (CV) and variance premium (VP) of \cite{BEKAERT2014181}. Second, we employ the squared news-implied volatility index ($\textrm{NVIX}^2$) of \cite{MANELA2017137} and its topic decomposition. The data generally cover the period from January 1968 to December 2024 with the following exceptions: January 1968 to December 2022 for the CBIV, January 1968 to December 2023 for the sentiment, February 1990 to December 2024 for the VIX, February 1990 to February 2022 for the CV and VP, and January 1968 to April 2016 for the NVIX. * indicates $p < 0.1$, ** indicates $p < 0.05$ and *** indicates $p < 0.01$.
\end{minipage}
\vspace{1em}
%\resizebox{\textwidth}{!}{
\begin{tabular}{lccc}
  \toprule
  \midrule
 & Lower-Tail & Central & Upper-Tail \\ 
 \midrule
 %\multicolumn{4}{l}{\textit{\textbf{Panel A:}} Descriptive statistics} \\ % align: l,c,r
 \textbf{Panel A:}  \textit{Descriptive Statistics} &&& \\
 \cmidrule(lr){2-4}
 Mean $\times 10^3$ & -3.21 & -41.48 & -8.34 \\ 
  St. Dev. & 0.19 & 1.46 & 0.23 \\ 
  Skewness & 0.06 & -0.35 & -0.05 \\ 
  Kurtosis & 5.09 & 6.41 & 6.86 \\ 
  AR(1) & -0.44 & -0.27 & -0.42 \\ 
  AR(2) & -0.02 & 0.07 & -0.07 \\ 
  AR(3) & 0.03 & -0.02 & 0.07 \\ 
  \midrule
  %\multicolumn{4}{l}{\textit{\textbf{Panel B:}} Correlations} \\ % align: l,c,r
  \textbf{Panel B:} \textit{Contemporaneous Correlations} &&& \\
   \cmidrule(lr){2-4}
   Lower-Tail & 1.00 & 0.22*** & 0.13*** \\ 
   Central & & 1.00 & 0.41*** \\ 
   Upper-Tail & &  & 1.00 \\
    \cmidrule(lr){2-4}
    $\Delta$PCA-SQ & -0.42*** & 0.14*** & 0.50*** \\ 
    $\Delta$CIV & -0.26*** & 0.10*** & 0.29*** \\ 
    $\Delta$CBIV & -0.00 & 0.07* & -0.02 \\ 
    $\Delta$TR & 0.09** & -0.03 & -0.24*** \\ 
    $\Delta$Sentiment & 0.10*** & 0.02 & 0.02 \\
   \midrule
     \textbf{Panel C:} \textit{Forward Correlations} &&& \\
   \cmidrule(lr){2-4}
   $\textrm{VIX}^2$ & -0.17*** & -0.19*** & 0.07 \\ 
    - CV & -0.19*** & -0.21*** & 0.07 \\ 
    - VP & -0.12** & -0.18*** & 0.07 \\
    \cmidrule(lr){2-4}
  $\textrm{NVIX}^2$ & -0.13*** & -0.07* & 0.04 \\ 
  - Securities Markets & -0.14*** & -0.06 & 0.06 \\ 
  - Intermediation & -0.10** & -0.03 & 0.03 \\ 
  - Unclassified & -0.14*** & -0.08* & 0.05 \\ 
  - Government & 0.02 & 0.01 & 0.03 \\ 
  - Natural Disaster & -0.02 & 0.02 & 0.04 \\ 
  - War & 0.04 & 0.07* & -0.01 \\
   \midrule
   \bottomrule
\end{tabular}
%}
\label{tab:ciq_short_summary}
\end{table}

Finally, CIQ factors capture realised cross-sectional tail quantiles more precisely than volatility does: the lower-tail CIQ factor explains a greater proportion of the variation in realised lower-tail quantiles than cross-sectional volatility. Furthermore, monthly Anderson–Darling tests consistently reject the normality of the cross-sectional idiosyncratic return distribution, emphasising the need to move beyond variance-based summaries. The incremental value of CIQ is particularly evident in the lower tail, whereas upper-tail quantiles are closer to those that can be captured by volatility alone.\footnote{Panel B of Table~\ref{tab:quantiles_summary} in Appendix~\ref{sec:cs_quantiles} reports the corresponding regression results.}

\section{Economic Mechanism: Intermediary Risk-Bearing Capacity and Common Idiosyncratic Tail Risk}

We interpret CIQ risk as operating through an intermediary-based mechanism, whereby financial intermediaries provide liquidity by absorbing inventory risk, subject to balance-sheet constraints. When intermediary capital deteriorates or aggregate uncertainty rises, these constraints tighten, reducing the willingness of intermediaries to absorb sell-side order flow. Consequently, negative idiosyncratic shocks are more likely to propagate into common downside tail realizations across assets, resulting in elevated price impact and transaction costs.

In this environment, the lower-tail CIQ factor captures situations in which the risk-bearing capacity of intermediaries is impaired and sell-side liquidity dries up across the board. The significant impact of downside illiquidity and its dispersion reflects the uneven distribution of liquidation pressure among firms: when intermediaries are under pressure, heterogeneous sell pressure results in correlated downside tail outcomes. In contrast, positive idiosyncratic shocks do not face comparable balance sheet frictions, meaning that upside CIQ innovations are weakly related to liquidity conditions and do not command a risk premium.

From an asset-pricing perspective, exposure to lower-tail CIQ innovations means exposure to states in which liquidity provision is costly and intermediary capital is scarce. In such states, the stochastic discount factor is high, meaning that assets which perform poorly when lower-tail CIQ deteriorates require a higher expected return to compensate for this. The core mechanism is state-dependent intermediary price impact, which amplifies downside idiosyncratic realisations through forced selling, but which has limited relevance for upside quantiles.

Dynamic intermediary asset-pricing models formalise this intuition. In \citet{HeKrishnamurthy2013}, \citet{BrunnermeierSannikov2014}, and \citet{HeKellyManela2017}, intermediary net worth or leverage acts as a state variable that drives equilibrium risk premium. In the systemic-risk literature, measures such as CoVaR and SRISK similarly summarize a high-dimensional system into a low-dimensional fragility index \citep{AdrianBrunnermeier2016,AcharyaEtAl2017,GiglioKellyPruitt2016}. We interpret CIQ risk in this spirit: it is a reduced-form proxy for the shadow cost of intermediaries' risk-bearing capacity.

\subsection{A Reduced-Form Representation.}

Let $\varepsilon_{i,t}$ denote the idiosyncratic return from a linear factor model $r_{i,t}=\alpha_i+\beta_i'f_t+\varepsilon_{i,t}$, where $f_t$ collects benchmark factors (e.g., Fama--French factors). We assume a latent systemic state $s_t\in\mathbb{R}$ that summarizes the tightness of intermediaries' balance sheet constraints. In fragile states (high $s_t$), risk-bearing capacity is scarce: intermediaries absorb less inventory, funding becomes more expensive, and sell-side price impact rises. These conditions affect firm-level downside risk through tighter credit lines, higher margins and haircuts, stricter lending standards, and weaker secondary-market liquidity. Consequently, firm-specific negative shocks are more likely to result in extreme residual losses, particularly for firms with high leverage, low cash reserves, or illiquid equity.

We model the equilibrium response of idiosyncratic returns to the liquidity demand absorbed by competitive intermediaries as follows:
\begin{equation}
\varepsilon_{i,t} \;=\; \lambda_t (s_t) \, u_{i,t} + \xi_{i,t}, \label{eq:idio_micro}
\end{equation}
where $u_{i,t}$ is idiosyncratic liquidity demand, $\xi_{i,t}$ is idiosyncratic noise orthogonal to liquidity demand, and $\lambda_t(s_t)>0$ is a common price-impact coefficient (the shadow cost of balance-sheet capacity). We assume price impact increases when constraints bind, $\partial \lambda_t/\partial s_t>0$.

In order to capture the effects of \textit{asymmetric} forced selling, we introduce a one-sided stress component to liquidity demand,
\begin{equation}
u_{i,t} \;=\; \tilde u_{i,t} \;-\; \ell_i\,J_t(s_t)\,\eta_{i,t},
\label{eq:forced_selling}
\end{equation}
where $\tilde u_{i,t}$ is baseline demand, $J_t(s_t)\ge 0$ is a common stress intensity, $\eta_{i,t}\ge 0$ is an idiosyncratic nonnegative shock, and $\ell_i\ge 0$ captures liquidation sensitivity / liquidity fragility (e.g., low turnover depth, unstable turnover, unexplained volume, drawdown sensitivity). We assume $J_t(s_t)$ rises when constraints bind, $\partial J_t/\partial s_t>0$.

A key implication of this asymmetry assumption is that higher fragility $s_t$ generates a stronger coordinated downward shift in many firms' left-tail idiosyncratic outcomes in comparison to right-tail shifts. We capture this with a parsimonious quantile representation,
\begin{equation}
\label{eq:quantile_state}
\epsilon_{i,t} = b_i (\tau) s_t + \nu_{i,t}(\tau)
%Q_{\varepsilon_{i,t}}(\tau \mid \mathcal{F}_{t-1}) \;=\; \alpha_i(\tau) + b_i(\tau)\, s_t,
\end{equation}
where $b_i(\tau)$ is more negative for some sufficiently low $\tau = \tau_L$ when constraints disproportionately amplify downside outcomes, and $b_i(\tau)\approx 0$ for the median and upper tail, and $\nu_{i,t}(\tau)$ satisfies the appropriate quantile restriction. Under this structure, the quantile factor extracted from the data is a reduced-form proxy for the latent state, $CIQ_t(\tau)\approx G_\tau(s_t)$ for some monotone mapping $G_\tau(\cdot)$ that is steeper in the lower tail. Because $s_t$ is persistent, levels of $CIQ_t(\tau)$ contain slow-moving components. Asset prices, however, respond to news about future investment opportunities and risk-bearing capacity. Standard intertemporal asset-pricing logic therefore motivates focusing on innovations, $\Delta CIQ_t(\tau) \equiv CIQ_t(\tau)-CIQ_{t-1}(\tau) \;\approx\; G_\tau'(s_{t-1})\,\Delta s_t$, so that $\Delta CIQ_t(\tau_L)$ isolates ``news'' about aggregate downside idiosyncratic conditions rather than slowly evolving tail levels.

\subsection{Intermediary Constraints and the Pricing Kernel}
\label{sub:intermediary_constraints_and_the_pricing_kernel}

In a constrained-intermediary setting,\footnote{Appendix~\ref{app:model} formalizes a simple environment in which a cross-section of balance-sheet constraints aggregates into a single state $s_t$. Under standard conditions, the representative constrained intermediary's shadow cost of risk-bearing is proportional to $s_t$, and the lower-tail CIQ factor identifies this state.} the stochastic discount factor (SDF) admits the representation
\begin{equation}
  m_{t+1} = \overline{m}_{t+1} + \psi_t\,\Delta s_{t+1},
  \label{eq:sdf_es}
\end{equation}
where $\psi_t>0$ in constrained states. When constraints tighten, $\psi_t$ increases and the SDF assigns more weight to states with large portfolio losses; see, e.g., \citet{BrunnermeierPedersen2009} or \citet{HeKrishnamurthy2013}. A standard linearization implies expected excess returns satisfy
\begin{equation}
  \mathbb{E}_t[r_{i,t+1}] = \lambda_M \beta^M_i + \lambda_{s,t}\,\beta^{s}_i,
  \label{eq:pricing_ciqa}
\end{equation}
where $\beta^{s}_i = \mathrm{Cov}_t(r_{i,t+1},\Delta s_{t+1})/\mathrm{Var}_t(\Delta s_{t+1})$ is exposure to risk-bearing-capacity news and $\lambda_{s,t}$ is its (time-varying) price of risk.

Quantile factor analysis applied to idiosyncratic returns recovers the dominant common component in lower-tail quantiles. Accordingly, lower-tail CIQ innovations at $\tau_L$ (e.g., $\tau_L = 0.2$) estimate an affine transformation of $\Delta s_t$:
\begin{proposition}[Quantile factor representation]
\label{prop:quantile_factorinnov}
Under standard large-$N$ conditions for approximate factor models, the lower-tail CIQ innovation at $\tau_L$ consistently estimates an affine transformation of $\Delta s_t$,
\[
  \Delta CIQ_t^L \equiv \Delta CIQ_t(\tau_L) = \kappa\,\Delta s_t + \nu_t,
\]
with $\kappa\neq 0$ and an error term $\nu_t$.
\end{proposition}
For the proof, see Proposition \ref{prop:quantile_factor} and discussion in Appendix \ref{app:model}. Empirically, we find that a single CIQ factor accounts for a significant proportion of the variation in realized lower-tail idiosyncratic quantiles, and is distinct from volatility-based factors such as CIV or PCA-SQ. The estimated exposure to the $\Delta$CIQ$(\tau_L)$ factor,
\[
  \beta^{CIQ}_i =
  \frac{\mathrm{Cov}(r_{i,t+1},\Delta CIQ^L_{t+1})}{\mathrm{Var}(\Delta CIQ^L_{t+1})},
\]
therefore corresponds to exposure to systemic fragility innovations. Assets whose returns are more adversely affected by these innovations must offer higher average returns as compensation. In the empirical analysis that follows, we operationalize this mechanism by estimating each stock's exposure to lower-tail CIQ innovations, and testing whether a higher level of exposure results in a higher return.

\subsection{CIQ Risks in the Economy}
\label{subsec:ciq_economy}

The mechanism outlined in the previous section implies that lower-tail CIQ innovations should be linked to proxies for intermediary constraints, uncertainty, and market-wide liquidity. Lower-tail CIQ betas should be larger for firms that are more exposed to sell-side illiquidity and have less financial slack. Because constraints bind primarily on the downside, pricing implications should be significant for lower $\tau$ values, and insignificant for upper-tail values. Motivated by theories in which adverse states amplify downside outcomes, we test whether $\Delta CIQ$ innovations are predictable from intermediary balance-sheet conditions, aggregate uncertainty, and market-wide liquidity.

We estimate various versions of
\begin{equation}
\Delta CIQ_{t+1}^{j}
=
\beta_0 + \beta_1 \Delta CIQ_{t}^{j}
+ \beta_2 ICF_t + \beta_3 VIX_t^2
+ \beta_4 X_t + \beta_5 (ICF_t \times X_t)
+ \varepsilon_{t+1},
\label{eq:pred_reg}
\end{equation}
for $j\in\{LT,UT\}$, where $ICF_t$ is the intermediary capital factor of \cite{HeKellyManela2017}, $VIX_t^2$ controls for aggregate uncertainty, and $X_t$ denotes alternative illiquidity measures based on \cite{AMIHUD200231}: changes in average illiquidity $\Delta Avg(ILLQ)_t$, downside and upside illiquidity $\Delta Avg(ILLQ)_t^{-}$ and $\Delta Avg(ILLQ)_t^{+}$ (computed within negative- and positive-return stocks, respectively), and the change in illiquidity dispersion $\Delta Var(ILLQ)_t$. Standard errors are of correction of \cite{newey1994}.

Panel A of Table~\ref{tab:ciq_factors_regressions} shows that the lower-tail factor is strongly mean-reverting ($\beta_1<0$) and, more importantly, is systematically related to financial conditions. A deterioration in intermediary capital predicts a decline in $\Delta CIQ^{LT}$ (i.e., worse downside tail conditions), and higher $VIX^2$ similarly predicts declines. Liquidity conditions add incremental predictive content: increases in average illiquidity predict declines in $\Delta CIQ^{LT}$, and this effect is concentrated in sell-side illiquidity ($\Delta Avg(ILLQ)^-$), consistent with an inherently asymmetric mechanism. In addition, increases in illiquidity dispersion predict declines in $\Delta CIQ^{LT}$, indicating that downside tail conditions worsen when liquidity stress becomes uneven and widespread. The interaction terms imply state dependence: the predictive effect of weak intermediary capital is amplified when average illiquidity or illiquidity dispersion increases.

\begin{table}[t!]
\caption{Predicting the $\Delta$CIQ Factors} 
\centering
\scriptsize
\begin{minipage}{\textwidth} 
The table reports predictive regressions of $\Delta CIQ$ factors on its lagged value, intermediary capital factor of \cite{HeKellyManela2017} ($ICF$), squared CBOE volatility index ($VIX^2$), and various cross-sectional measures of illiquidity based on \cite{AMIHUD200231}: average illiquidity ($Avg(ILLQ)$), variance of illiquidity ($Var(ILLQ)$), and downside (upside) illiquidity averaged by using only stocks that experienced negative (positive) monthly returns ($Avg(ILLQ)^-$ and $Avg(ILLQ)^+$), respectively). Explanatory variables enter the regressions at time $t$, while the dependent variables are in time $t+1$. $t$-statistics (in parentheses) are based on standard errors of \cite{newey1994}. The $\Delta$CIQ factors and the illiquidity measures cover the period from February 1968 to December 2024, $ICF$ covers the period between February 1970 to December 2018, $VIX^2$ data cover the period between February 1990 and December 2024.
\end{minipage}
\vspace{1em}
%\resizebox{\textwidth}{!}{%
\adjustbox{max height=\dimexpr\textheight-5.5cm\relax,
           max width=\textwidth}{
\begin{tabular}{lccccccccc}
  \toprule
  \midrule
 & (1) & (2) & (3) & (4) & (5) & (6) & (7) & (8) & (9) \\
  \midrule
  \textbf{Panel A:} &&&&&& \\
\textit{Dependent variable:} $\Delta CIQ_{t+1}^{LT}$ &&&&&& \\
\midrule
	$\Delta CIQ_{t}$ & -0.44 & -0.41 & -0.51 & -0.50 & -0.50 & -0.51 & -0.50 & -0.50 & -0.50 \\ 
   & (-13.20) & (-11.72) & (-19.25) & (-15.61) & (-15.33) & (-14.21) & (-15.17) & (-16.21) & (-15.45) \\ 
  $ICF_{t}$ &  & 0.52 &  & 0.67 & 0.63 & 0.69 & 0.66 & 0.65 & 0.67 \\ 
   &  & (4.39) &  & (7.33) & (6.69) & (6.65) & (7.11) & (6.79) & (7.21) \\ 
  $VIX^2_{t}$ &  &  & -1.11 & -0.71 & -0.71 & -0.70 & -0.71 & -0.61 & -0.71 \\ 
   &  &  & (-5.78) & (-4.66) & (-4.46) & (-4.28) & (-4.68) & (-3.10) & (-4.56) \\ 
   \textit{Cross-sectional liquidity measures} &&&&&&&&& \\
   \cmidrule(lr){1-1}
  $\Delta Avg(ILLQ)_t$ &  &  &  &  & -0.09 &  &  & -0.12 &  \\ 
   &  &  &  &  & (-3.05) &  &  & (-3.50) &  \\ 
  $\Delta Avg(ILLQ)_t^-$ &  &  &  &  &  & -0.06 &  &  &  \\ 
   &  &  &  &  &  & (-3.06) &  &  &  \\ 
  $\Delta Avg(ILLQ)_t^+$ &  &  &  &  &  & 0.20 &  &  &  \\ 
   &  &  &  &  &  & (1.37) &  &  &  \\ 
  $\Delta Var(ILLQ)_t$ &  &  &  &  &  &  & -0.07 &  & -0.10 \\ 
   &  &  &  &  &  &  & (-5.83) &  & (-10.93) \\
   \textit{Interactions with capital factor} &&&&&&&&& \\
   \cmidrule(lr){1-1}
  $ICF_{t} \times \Delta Avg(ILLQ)_t$ &  &  &  &  &  &  &  & 0.97 &  \\ 
   &  &  &  &  &  &  &  & (3.62) &  \\ 
  $ICF_{t} \times \Delta Var(ILLQ)_t$ &  &  &  &  &  &  &  &  & 1.10 \\ 
   &  &  &  &  &  &  &  &  & (10.25) \\
   %\midrule
   Intercept & -0.00 & -0.01 & 0.04 & 0.02 & 0.02 & 0.02 & 0.02 & 0.02 & 0.02 \\ 
   & (-0.75) & (-0.97) & (3.17) & (1.41) & (1.43) & (1.38) & (1.43) & (1.23) & (1.46) \\ 
  %R2 & 19.41 & 22.70 & 28.51 & 34.31 & 34.68 & 35.03 & 34.45 & 35.00 & 34.61 \\ 
  $R_{adj}^2$ & 19.29 & 22.43 & 28.17 & 33.74 & 33.92 & 34.08 & 33.68 & 34.05 & 33.65 \\ 
  $T$ & 683 & 587 & 419 & 347 & 347 & 347 & 347 & 347 & 347 \\ 
  \midrule
  \textbf{Panel B:} &&&&&& \\
\textit{Dependent variable:} $\Delta CIQ_{t+1}^{UT}$ &&&&&& \\
\midrule
	$\Delta CIQ_{t}$ & -0.43 & -0.42 & -0.45 & -0.45 & -0.44 & -0.44 & -0.45 & -0.44 & -0.45 \\ 
   & (-14.94) & (-10.67) & (-11.41) & (-9.62) & (-9.53) & (-10.56) & (-9.57) & (-9.57) & (-9.53) \\ 
  $ICF_{t}$ &  & -0.51 &  & -0.34 & -0.37 & -0.39 & -0.34 & -0.39 & -0.35 \\ 
   &  & (-3.58) &  & (-1.69) & (-1.76) & (-2.01) & (-1.69) & (-1.84) & (-1.69) \\ 
  $VIX^2_{t}$ &  &  & 0.61 & 0.26 & 0.25 & 0.25 & 0.26 & 0.11 & 0.26 \\ 
   &  &  & (3.32) & (1.48) & (1.47) & (1.51) & (1.47) & (0.43) & (1.47) \\
   \textit{Cross-sectional liquidity measures} &&&&&&&&& \\
      \cmidrule(lr){1-1}
  $\Delta Avg(ILLQ)_t$ &  &  &  &  & -0.06 &  &  & -0.02 &  \\ 
   &  &  &  &  & (-0.84) &  &  & (-0.41) &  \\ 
  $\Delta Avg(ILLQ)_t^-$ &  &  &  &  &  & 0.01 &  &  &  \\ 
   &  &  &  &  &  & (0.31) &  &  &  \\ 
  $\Delta Avg(ILLQ)_t^+$ &  &  &  &  &  & -0.30 &  &  &  \\ 
   &  &  &  &  &  & (-1.44) &  &  &  \\ 
  $\Delta Var(ILLQ)_t$ &  &  &  &  &  &  & -0.03 &  & -0.03 \\ 
   &  &  &  &  &  &  & (-1.67) &  & (-2.28) \\
   \textit{Interactions with capital factor} &&&&&&&&& \\
    \cmidrule(lr){1-1}
  $ICF_{t} \times \Delta Avg(ILLQ)_t$ &  &  &  &  &  &  &  & -1.45 &  \\ 
   &  &  &  &  &  &  &  & (-1.25) &  \\ 
  $ICF_{t} \times \Delta Var(ILLQ)_t$ &  &  &  &  &  &  &  &  & -0.10 \\ 
   &  &  &  &  &  &  &  &  & (-0.26) \\
   Intercept & -0.01 & -0.01 & -0.04 & -0.02 & -0.02 & -0.02 & -0.02 & -0.02 & -0.02 \\ 
   & (-1.45) & (-1.26) & (-2.66) & (-1.34) & (-1.32) & (-1.29) & (-1.33) & (-1.09) & (-1.33) \\ 
  %R2 & 17.80 & 21.19 & 20.17 & 20.91 & 21.03 & 21.34 & 20.92 & 21.50 & 20.92 \\ 
  $R_{adj}^2$ & 17.68 & 20.92 & 19.78 & 20.21 & 20.11 & 20.19 & 20.00 & 20.35 & 19.76 \\ 
  $T$ & 683 & 587 & 419 & 347 & 347 & 347 & 347 & 347 & 347 \\ 
  \midrule
   \bottomrule
\end{tabular}
}
\label{tab:ciq_factors_regressions}
\end{table}

Panel B reports analogous results for the upper-tail factor. While $\Delta CIQ^{UT}$ is also mean-reverting, its relationship with intermediary capital and illiquidity is markedly weaker and less robust. In particular, the illiquidity variables that strongly predict the lower-tail factor have little explanatory power for the upper-tail factor, and interaction effects are negligible.

Taken together, the results reveal a pronounced asymmetry between downside and upside idiosyncratic dynamics. Innovations in the lower-tail CIQ factor are closely linked to intermediary balance-sheet conditions, aggregate uncertainty, sell-side liquidity stress and general cross-sectional dispersion of illiquidity. In contrast, upside CIQ innovations exhibit substantially weaker connections to liquidity and intermediary constraints. This asymmetry suggests that the lower-tail CIQ factor captures endogenous, liquidity-driven tail risk that emerges when intermediary capacity is strained, whereas upside tail realizations are less constrained by financial frictions.

Such asymmetry is consistent with the inherently one-sided nature of balance-sheet constraints. When intermediaries are constrained, selling pressure is amplified into large negative price movements, while there is no comparable mechanism generating forced buying. As a result, intermediation stress manifests itself primarily through a deterioration of the lower tail of the cross-sectional distribution of idiosyncratic returns rather than through changes of the upper tail. This asymmetry is a direct implication of intermediary-based asset pricing models and motivates our focus on downside CIQ innovations.

In the next section, we test the hypothesis that the exposure to the lower-tail CIQ factor captures priced risk by examining the cross-section of stock returns.

%------------------------------------------------------------------------------------------------------------------------------------------------------%

\section{Pricing the CIQ($\tau$) Risks in the Cross-Section}
\label{sec:pricing}

The central prediction implied by the preceding analysis is that only exposure to common downside idiosyncratic tail risk commands a risk premium, whereas exposure to upside or central tail movements should not. We test this prediction by documenting heterogeneity in the prices of risk associated with the lower, central, and upper portions of the cross-sectional distribution of idiosyncratic returns. Consistent with the asymmetric economic mechanisms highlighted above, we show that only exposure to common lower-tail CIQ innovations is systematically priced and is distinct from existing volatility, intermediary, and liquidity factors. Importantly, this pricing effect cannot be accounted for by tail-based proxies, either.

To measure stock’s sensitivities to the lower-tail ($\Delta CIQ^{LT}$), central ($\Delta CIQ^{C}$), and upper-tail ($\Delta CIQ^{UT}$) factors, each month $m$, we separately estimate factor loadings using time-series regressions of excess returns on the corresponding CIQ innovations as
\begin{align}
\label{eq:ciq_betas}
r_{i,t} = a_{i, m} + \beta_{i,j,m}^{CIQ} , \Delta CIQ_{t,m}^{j} + v_{i,t,m}, \quad t = m-59,\ldots,m,
\end{align}
for each $ j \in {LT, C, UT}$ where $\beta_{i,j,m}^{CIQ}$ captures stock $i$’s exposure to quantile-specific innovations in common idiosyncratic risk.\footnote{Although the quantile factor model of \cite{chen2021} delivers quantile-specific loadings $\gamma_{i,m}(\tau)$ as part of the cross-sectional estimation, we do not interpret these objects as pricing betas. The $\gamma_{i,m}(\tau)$ coefficients characterize how individual returns contribute to the conditional cross-sectional quantile at a given $\tau$, and are therefore cross-sectional objects tied to the level of the quantile factor. In contrast, asset pricing requires a measure of exposure to innovations in systematic risk. Since our analysis focuses on $\Delta CIQ(\tau)$—which capture changes in tail conditions rather than persistent distributional states—the economically relevant exposure is a stock’s time-series covariance with these innovations. Estimating betas via time-series regressions therefore provides a direct measure of sensitivity to quantile-specific risk shocks that enter the stochastic discount factor, whereas the $\gamma_{i,m}(\tau)$ loadings need not coincide with exposure to time-series innovations in tail risk. In this sense, the time-series betas and the quantile factor loadings capture distinct economic concepts, and only the former is appropriate for pricing tests based on risk innovations.} Betas are estimated over the same 60-month rolling window used to construct the CIQ factors, including stocks with at least 48 monthly observations. Factor realizations and betas estimated using information up to time \(m\) are used to predict returns in month \(m+1\), ensuring no overlap between estimation and prediction periods. Unless stated otherwise, the dependent variable is the one-month-ahead out-of-sample return. We also examine longer-horizon returns using portfolios to assess the persistence of $\Delta CIQ$ exposures and, indirectly, the role of trading frictions.

The data cover the usual cross-sectional asset pricing period between January 1963 and December 2024. The first returns that we predict pertain to January 1968. In total, our baseline dataset without penny stocks consists of 2,082,857 stock-month observations.

\subsection{Univariate Portfolio Sorts}

We start our analysis by investigating the performance of portfolios formed on the basis of the $\Delta CIQ$ betas. Each month, stocks are sorted into five or ten portfolios according to their $\Delta$CIQ betas estimated using information available up to time \(m\).\footnote{In the baseline analysis, portfolio breakpoints are constructed using the full cross-section of stocks. In Table~\ref{tab:robustness}, we demonstrate that the results are robust to using breakpoints computed solely from the NYSE sample.} The portfolios are formed at the end of month \(m\) and held during month \(m+1\), with returns computed using either equal- or value-weighted schemes based on market capitalization at formation. Portfolios are rebalanced monthly as betas are re-estimated using a rolling window. These univariate portfolio results are summarized in Table~\ref{tab:portfolios_univariate}. In Panel~A, we report decile sorts. For the lower-tail $\Delta CIQ$ factor, we observe a monotonic increase in returns from low- to high-exposure portfolios, whereas no significant return spread is observed for portfolios sorted on the central or upper-tail $\Delta CIQ$ factors.

\begin{table}[t!]
\caption{Portfolios Sorted on Exposures to the $\Delta$CIQ Factors} 
\centering
\scriptsize
\begin{minipage}{\textwidth} 
The table reports the annualized out-of-sample excess returns of portfolios sorted on the exposure to the lower-tail ($\tau = 0.2$), central ($\tau = 0.5$) and upper-tail ($\tau = 0.8$) $\Delta$CIQ factors. We also report the returns of zero-cost portfolios obtained by buying the high-exposure portfolio and selling the low-exposure portfolio (High - Low). The corresponding $t$-statistics (in parentheses) are computed using the robust standard errors suggested by \cite{newey1987} with six lags. Stocks are either sorted into ten portfolios in Panel A or into five portfolios in Panel B, with portfolio returns obtained by either equally weighting stock returns (EW) or value weighting by their market capitalization (VW). The portfolios are formed each month based on the sensitivity to the $\Delta$CIQ factors estimated using time-series regression over the previous 60 months. The corresponding next-period out-of-sample return is then recorded for each portfolio. We also report time-series averages of equally-weighted stock-level $\Delta$CIQ betas within portfolios during the formation. The return sample covers period between January 1968 and December 2024. Each month, we use all the CRSP stocks for which at least 48 monthly observations are available over the last 60 months and exclude penny stocks with prices below \$1.
\end{minipage}
\vspace{1em}
\adjustbox{max height=\dimexpr\textheight-5.5cm\relax,
           max width=\textwidth}{
%\resizebox{\textwidth}{!}{%

\begin{tabular}{lccccccccc}
  \toprule
  \midrule
  & \multicolumn{3}{c}{Lower-Tail} & \multicolumn{3}{c}{Central} & \multicolumn{3}{c}{Upper-Tail} \\
  \cmidrule(lr){2-4}\cmidrule(lr){5-7}\cmidrule(lr){8-10}
 & $\beta^{CIQ}_{LT}$ & EW & VW & $\beta^{CIQ}_{C}$ & EW & VW & $\beta^{CIQ}_{UT}$ & EW & VW \\ 
  \midrule
  \midrule
  %\multicolumn{10}{l}{\textit{\textbf{Panel A:}} Decile Sorts} \\
  \textit{\textbf{Panel A:}} Decile Sorts &&&&&&&&& \\
  \midrule
  Low $\beta^{CIQ}$ & -0.29 & 4.36 & 3.36 & -0.03 & 6.90 & 5.02 & -0.13 & 8.86 & 7.23 \\ 
  2 & -0.17 & 7.94 & 7.28 & -0.01 & 9.06 & 5.22 & -0.04 & 10.10 & 9.12 \\ 
  3 & -0.12 & 9.46 & 6.66 & -0.01 & 10.34 & 6.62 & -0.01 & 10.45 & 8.15 \\ 
  4 & -0.09 & 9.63 & 6.57 & -0.01 & 9.05 & 6.58 & 0.01 & 9.61 & 7.06 \\ 
  5 & -0.07 & 9.45 & 7.16 & -0.00 & 10.23 & 7.52 & 0.03 & 9.74 & 8.13 \\ 
  6 & -0.04 & 10.37 & 8.12 & 0.00 & 9.86 & 8.43 & 0.05 & 9.70 & 7.83 \\ 
  7 & -0.02 & 10.41 & 7.97 & 0.00 & 10.32 & 8.15 & 0.07 & 9.91 & 6.81 \\ 
  8 & 0.01 & 10.74 & 8.22 & 0.01 & 10.23 & 7.73 & 0.09 & 8.98 & 6.36 \\ 
  9 & 0.04 & 11.40 & 9.85 & 0.01 & 10.30 & 6.49 & 0.13 & 9.75 & 5.88 \\ 
  High $\beta^{CIQ}$ & 0.14 & 11.77 & 10.73 & 0.03 & 9.26 & 9.39 & 0.21 & 8.43 & 6.35 \\ 
  High - Low &  & 7.41 & 7.36 &  & 2.36 & 4.37 &  & -0.43 & -0.88 \\ 
  $t$-statistic &  & (4.30) & (2.73) &  & (1.36) & (1.82) &  & (-0.26) & (-0.35) \\
\midrule
  %\multicolumn{10}{l}{\textit{\textbf{Panel B:}} Quintile Sorts} \\
    \textit{\textbf{Panel B:}} Quintile Sorts &&&&&&&&& \\
    \midrule
  Low $\beta^{CIQ}$ & -0.23 & 6.15 & 5.89 & -0.02 & 7.98 & 5.08 & -0.09 & 9.48 & 8.34 \\ 
  2 & -0.11 & 9.54 & 6.62 & -0.01 & 9.69 & 6.54 & -0.00 & 10.03 & 7.74 \\ 
  3 & -0.06 & 9.91 & 7.69 & -0.00 & 10.04 & 7.89 & 0.04 & 9.72 & 7.75 \\ 
  4 & -0.01 & 10.57 & 7.96 & 0.01 & 10.28 & 8.04 & 0.08 & 9.45 & 6.61 \\ 
  High $\beta^{CIQ}$ & 0.09 & 11.59 & 9.98 & 0.02 & 9.78 & 7.57 & 0.17 & 9.09 & 6.35 \\ 
  High - Low &  & 5.44 & 4.09 &  & 1.80 & 2.49 &  & -0.39 & -1.99 \\ 
  $t$-statistic &  & (3.76) & (2.04) &  & (1.36) & (1.44) &  & (-0.29) & (-1.12) \\
   \midrule
   \bottomrule
\end{tabular}%
}
\label{tab:portfolios_univariate}
\end{table}

Moreover, to formally assess the presence of compensation for bearing the CIQ risks, we present the returns of the high-minus-low portfolios obtained as the difference between the returns of portfolios with the highest $\Delta$CIQ betas and those of portfolios with the lowest $\Delta$CIQ betas. These returns correspond to the investment strategy that buys stocks with high exposures and sells stocks with low exposures to the $\Delta$CIQ factors. These portfolios are zero-cost portfolios and capture the risk premium associated with the specific joint movements of the idiosyncratic returns. We observe a significant positive premium for the difference portfolio only for the lower-tail $\Delta$CIQ factor. This premium is both economically and statistically significant. In the case of the decile equal-weighted portfolio, the premium reaches 7.41\% on an annual basis with a robust $t$ statistic on the basis of \cite{newey1987} standard error correction with six lags of 4.30. The premium of the value-weighted portfolio achieves a similar performance of 7.36\% p.a. with a $t$ statistic of 2.73. The stocks that hedge the lower-tail CIQ movements possess particularly low excess returns. Investors clearly value this feature, which pushes up the price of these stocks and lowers their expected returns.

We also observe sizeable spread in the exposures across the portfolios captured by the time series average of equally weighted stock-level betas within portfolios during the formation. The low-exposure portfolio has an average $\beta^{CIQ}_{LT}$ of -0.29, which is the greatest difference with respect to a neighborhood portfolio across all the portfolios (the second-lowest-exposure portfolio has an average $\beta^{CIQ}_{LT}$ of -0.17, a difference of 0.12).

To show that the premium is not driven by a particular choice of the sorting scheme, in Panel B, we report the results from sorting the stocks into quintiles. Although the premium is smaller than in the case of decile sorts, it remains both economically and statistically significant at 5.44\% p.a. with a $t$ statistic of 3.76 in the case of an equally-weighted portfolio and 4.09\% p.a. with a $t$ statistic of 2.04 in the case of a value-weighted portfolio. This slightly lower significance in the case of the value-weighted portfolio may be partially caused by the fact that the value-weighted portfolios possess a higher concentration, which leads to more volatile returns.

On the other hand, if we consider the returns of the zero-cost portfolios associated with either central or upper-tail $\Delta$CIQ factors, these portfolios yield premia indistinguishable from zero. This observation holds across weighting and sorting schemes. Investors clearly value exposures to the downside and upside idiosyncratic events asymmetrically, and they exhibit a clear preference to hedge against the former but not the latter. The fact that only the exposures to the lower-tail common movements yield a premium suggests that the $\Delta$CIQ risks are not driven by the effect of the common volatility. If volatility was the main driver of the factors, we would observe symmetrical compensation for the exposures to both the downside and upside factors, which we do not observe here. To illustrate this point in further detail, in Appendix \ref{sec:simul}, we provide evidence indicating that, in an economy that specifically rewards holding assets with exposure to a common volatility factor, the upside and downside factors are priced symmetrically. On the basis of this result, we argue that the typical location-scale model is not consistent with the premia that we observe here.

\begin{figure}[t!]
  \caption{Performance of the $\Delta$CIQ Portfolios}
  \centering
  \scriptsize
  \begin{minipage}{\textwidth} 
  The figure depicts performance of a strategy that buy stocks with high exposure to the $\Delta$CIQ factors and sell stocks with low exposure. It plots cumulative log-return obtained from sorting the stocks into decile portfolios with equal-weighting the stocks. The return sample covers period between January 1968 and December 2024. Each month, we use all the CRSP stocks with at least 48 monthly observations over the last 60 months and exclude penny stocks with prices below \$1. The shaded areas represent NBER recessions. We include the performance of the excess market return as a comparison.
  \end{minipage}
  \vspace{1em}
  \includegraphics[scale=0.45]{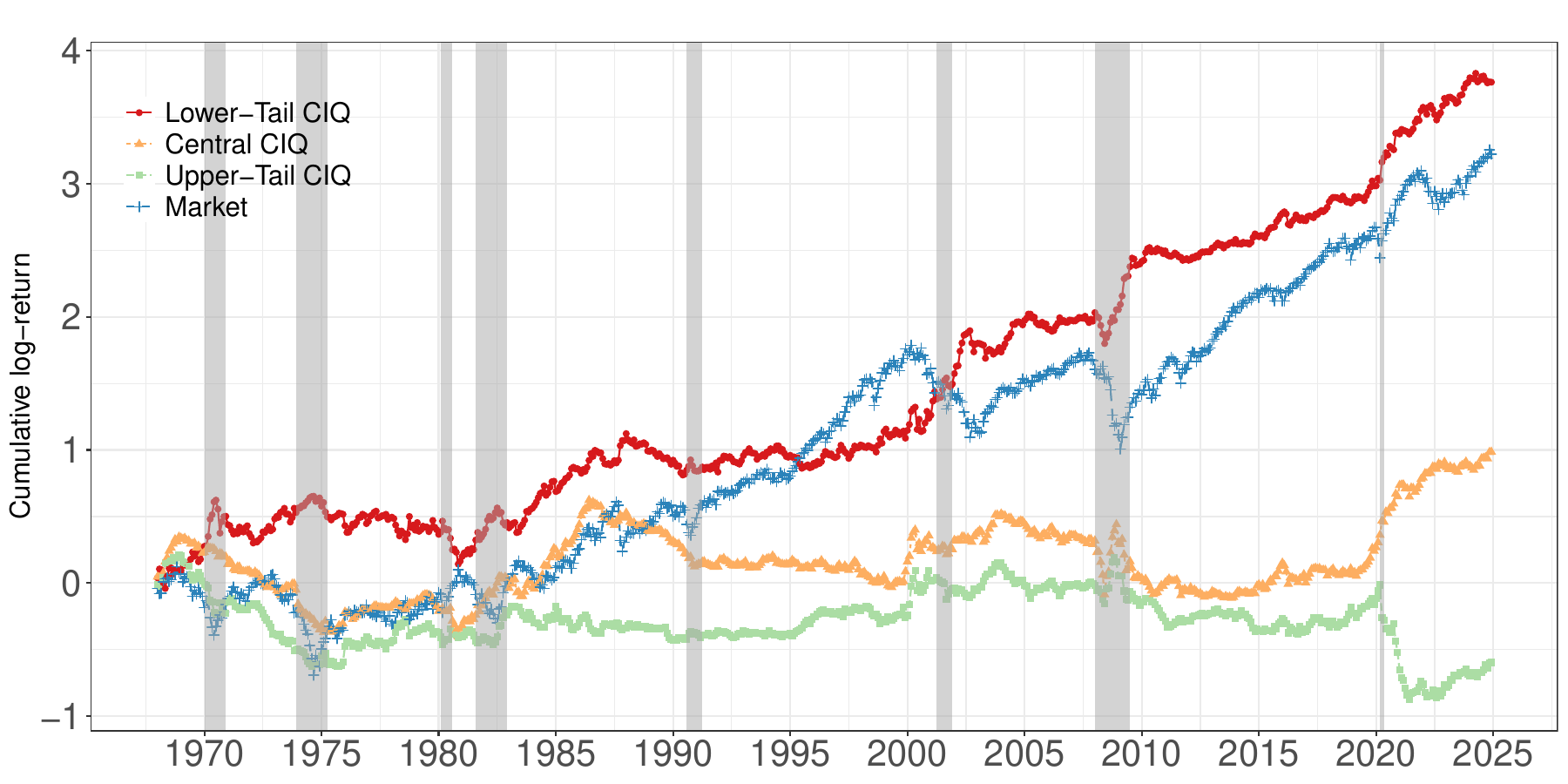}
  \label{fig:ciq_fac_traded}
\end{figure}

To visually inspect the performance of the portfolios sorted on the basis of the exposure to the $\Delta$CIQ factors, we present in Figure \ref{fig:ciq_fac_traded} the cumulative log-returns of the equal-weighted high-minus-low portfolios. Consistent with the numerical portfolio results, only the portfolio based on the lower-tail $\Delta$CIQ factor provides strong performance during the sample period that is no worse than the performance of the aggregate market as measured by the value-weighted return of all CRSP firms.

The newly discovered lower-tail $\Delta$CIQ premium raises the question of whether it simply mirrors previously reported risks associated with a particular factor model. Accordingly, we regress the returns of the high-minus-low portfolios on various sets of factors and report estimated intercepts--alphas--from this exercise. We summarize the results in Table \ref{tab:alphas}. We report the annualized alphas for both the equal- and value-weighted portfolios that are sorted either in quintiles or deciles. We start the investigation by regressing the returns on the market factor (CAPM). We observe that the market factor is not successful in explaining the abnormal returns across the specifications. We repeat the regressions on the basis of the three-factor model of \cite{FAMA19933} (FF3), the five-factor model of \cite{FAMA20151} (FF5), and its extension incorporating the momentum factor of \cite{FAMA2018234} (FF6). The alphas remain economically large and generally statistically significant, so standard Fama-French models do not fully span the lower-tail $\Delta$CIQ premium.

\begin{table}[t!]
\caption{Alphas of the Zero-Cost Lower-Tail $\Delta$CIQ Portfolios} 
\centering
\scriptsize
\begin{minipage}{\textwidth}
The table reports annualized abnormal returns of zero-cost portfolios obtained from buying high-exposure and selling low-exposure stocks with respect to the lower-tail $\Delta$CIQ factor. Each month, stocks are sorted on the exposure estimated from the last 60 months into decile or quintile portfolios. Returns within a portfolio are either equal- or value-weighted based on the market capitalization at the time of the portfolio formation. Zero-cost portfolio return is calculated as a difference between high-beta and low-beta portfolio. The portfolio is then held for one month and subsequently rebalanced. We report estimated intercepts (alphas) from regressing the portfolio out-of-sample returns on various sets of asset pricing factors: market (CAPM), three factors of \cite{FAMA19933} (FF3), five factors of \cite{FAMA20151} (FF5) and its extension with the momentum factor of \cite{FAMA2018234} (FF6), two factors of \cite{HeKellyManela2017} (Intermediary), five factors of \cite{q5factor} (Q5), and four factors of \cite{mispricing} (M4). Moreover, we formulate an ad-hoc model based on four factors of \cite{carhart1997}, CIV shocks of \cite{HERSKOVIC2016}, and BAB factor of \cite{FRAZZINI20141} (FF alt.) and augment it with either short-term reversal factor (Rev.) or liquidity factor of \cite{pastor2003} (Liquid.). The corresponding $t$-statistics (in parentheses) are computed using the robust standard errors suggested by \cite{newey1987} with six lags. The return sample covers period between January 1968 and December 2024, except for M4, which is only available up to December 2016, and Intermediary, which covers period between January 1970 and November 2018. Each month, we use all the CRSP stocks for which at least 48 monthly observations are available over the last 60 months and exclude penny stocks with prices below \$1.
\end{minipage}
\vspace{1em}
%\adjustbox{max height=\dimexpr\textheight-5.5cm\relax,
%           max width=\textwidth}{
%\resizebox{\textwidth}{!}{%
\begin{tabular}{lcccc}
  \toprule
  \midrule
  & \multicolumn{2}{c}{Decile} & \multicolumn{2}{c}{Quintile} \\
  \cmidrule(lr){2-3}\cmidrule(lr){4-5}
 & EW & VW & EW & VW \\ 
  \midrule
  \midrule
  CAPM & 9.72 & 9.31 & 7.35 & 5.78 \\ 
   & (6.39) & (3.65) & (5.73) & (3.04) \\ 
  FF3 & 8.22 & 6.74 & 6.08 & 3.44 \\ 
   & (5.52) & (2.83) & (5.01) & (2.04) \\ 
  FF5 & 7.59 & 6.71 & 6.14 & 3.57 \\ 
   & (4.87) & (2.56) & (4.91) & (2.05) \\ 
  FF6 & 8.31 & 7.85 & 6.83 & 4.85 \\ 
   & (4.97) & (2.84) & (5.11) & (2.72) \\ 
  FF alt. & 8.51 & 8.39 & 6.75 & 5.21 \\ 
   & (4.47) & (2.91) & (4.43) & (2.72) \\ 
  FF alt. + Rev. & 8.47 & 8.64 & 6.72 & 4.88 \\ 
   & (4.15) & (2.84) & (4.05) & (2.37) \\ 
  FF alt. + Liquid. & 8.28 & 8.17 & 6.57 & 5.07 \\ 
   & (4.36) & (2.77) & (4.32) & (2.65) \\
   Intermediary & 8.07 & 9.34 & 6.11 & 5.53 \\ 
   & (5.24) & (3.82) & (4.70) & (3.03) \\
  Q5 & 7.18 & 8.06 & 6.09 & 5.58 \\ 
   & (3.59) & (2.70) & (3.55) & (2.53) \\ 
  M4 & 5.94 & 7.61 & 5.52 & 5.18 \\ 
   & (3.17) & (2.85) & (3.79) & (2.73) \\
   \midrule
   \bottomrule
\end{tabular}%
%}
\label{tab:alphas}
\end{table}

Next, we form an extension of the four-factor model of \cite{carhart1997} by augmenting it with the $\Delta$CIV factor of \cite{HERSKOVIC2016} and the betting-against-beta (BAB) factor of \cite{FRAZZINI20141}. We then add either the short-term reversal (Rev.) or liquidity factor of \cite{pastor2003} (Liquid.). Similarly, we observe that the premia associated with these risks do not subsume the abnormal returns of the lower-tail $\Delta$CIQ portfolio.

We also verify that the premium does not mirror risk premium associated with benchmark two-factor intermediary pricing model of \cite{HeKellyManela2017}, which includes the market factor and traded capital factor. We observe that neither this model captures complex risk associated with our lower-tail idiosyncratic factor. 

Finally, we also regress the portfolio returns on the five-factor model of \cite{q5factor} (Q5) and the four factors of \cite{mispricing}, which include two mispricing factors (M4). The results also suggest that these models do not span the abnormal premium associated with the lower-tail $\Delta$CIQ risk.

\subsection{Bivariate Portfolio Sorts}

Although none of the factor models considered here can explain the lower-tail $\Delta$CIQ premium, it may be the case that some stock-level characteristics do. For this reason, we investigate how stock-level lower-tail exposures relate to other stock-specific characteristics and exposures by performing dependent bivariate sorts. Every month, we first sort the stocks into ten (five) portfolios on the basis of the value of their control variable. Then, within each of the control-variable-sorted portfolios, we sort the stocks into another ten (five) portfolios on the basis of their lower-tail $\Delta$CIQ betas. Finally, we form each $\Delta$CIQ portfolio by collapsing all the corresponding control-variable-sorted portfolios into one $\Delta$CIQ portfolio. This procedure yields single-sorted portfolios that vary in terms of their exposure to the lower-tail $\Delta$CIQ factor but exhibit approximately equal values of the control variable.

The obtained results are summarized in Table \ref{tab:portfolios_bivariate}. We focus here on some of the most researched stock characteristics and most related risk measures.\footnote{We provide information on how we obtain the control variables in Appendix \ref{sec:definitions}. We tried to adhere to the originally proposed specifications to the greatest extent possible. Stock characteristics are obtained following the specifications of \cite{LANGLOIS2020399}.} We start the investigation by using the market beta, firm size, book-to-price ratio, momentum and short term reversal. We observe that the zero-cost portfolio returns remain both economically and statistically significant when either decile or quintile sorts are used, with an annual premium of at least 4.65\% ($t$ statistic of 3.02) in the case of decile sorts and 3.66 ($t$ statistic of 2.79) in the case of quintile sorts. Moreover, we report the six-factor alphas with respect to the model of \cite{FAMA2018234} for the portfolios and observe that the premia are not subsumed by the model.

\begin{landscape}
\begin{table}[t!]
\caption{Dependent Bivariate Portfolio Sorts}
\centering

\tiny
\setlength{\tabcolsep}{3.5pt}        % efficient column spacing
\renewcommand{\arraystretch}{1.05}   % improve readability

\begin{minipage}{0.95\linewidth}
The table reports annualized out-of-sample excess returns of portfolios double-sorted on the exposure to the lower-tail $\Delta$CIQ factor and a control variable. In Panel A, the portfolios are constructed by first sorting the stocks into deciles based on a control variable and then within each portfolio we sort stocks into deciles based on the exposure to the lower-tail $\Delta$CIQ factor. Final portfolio returns are calculated by averaging the returns across the control deciles for every decile of the $\beta^{CIQ}_{LT}$. This procedure yields spread in the exposure to the lower-tail $\Delta$CIQ factor, while holding control variable approximately constant across portfolios. In Panel B, we repeat the procedure by sorting stocks into quintiles. The portfolios are formed every month and returns within portfolios are equally weighted. We also report returns of the high minus low portfolios, annualized alphas with respect to the six-factor model of \cite{FAMA2018234} and their $t$-statistics using the robust standard errors suggested by \cite{newey1987} with six lags. The return sample covers period between January 1968 and December 2024, with the exception of the case of $\beta^{VIX}$, which begins in February 1990. Each month, we use all the CRSP stocks for which at least 48 monthly observations are available over the last 60 months and exclude penny stocks with prices below \$1.
\end{minipage}

\vspace{1em}

\begin{tabularx}{0.95\linewidth}{
    l *{14}{>{\centering\arraybackslash}X}
}
\toprule
\midrule
& $\beta^{CAPM}$ & Size & B/P & MOM & STR & IVOL & ISKEW & CSK & CKT
& $\beta^{down}$ & $\beta^{PCA\text{-}SQ}$ & $\beta^{CIV}$ & $\beta^{tail}$ & $\beta^{VIX}$ \\
\midrule
\textit{\textbf{Panel A:}} & & & & & & & & & & & & & & \\
Decile Sorts & & & & & & & & & & & & & & \\
\midrule
\midrule
  Low & 6.82 & 4.85 & 5.99 & 5.51 & 5.66 & 6.71 & 4.54 & 4.64 & 5.27 & 5.89 & 6.00 & 6.34 & 5.55 & 6.48 \\ 
  2 & 8.59 & 7.53 & 7.54 & 8.11 & 8.94 & 7.77 & 8.09 & 8.23 & 7.53 & 8.20 & 8.14 & 9.13 & 7.78 & 9.32 \\ 
  3 & 9.38 & 9.66 & 9.86 & 8.80 & 9.44 & 8.65 & 8.87 & 8.98 & 9.43 & 9.09 & 8.88 & 9.00 & 9.06 & 9.97 \\ 
  4 & 9.21 & 9.37 & 9.19 & 9.57 & 9.52 & 9.40 & 9.57 & 9.58 & 9.92 & 9.30 & 9.84 & 9.68 & 9.41 & 10.97 \\ 
  5 & 9.40 & 9.88 & 9.37 & 9.55 & 10.38 & 9.34 & 10.14 & 9.60 & 9.20 & 9.64 & 9.24 & 9.68 & 9.63 & 10.44 \\ 
  6 & 9.66 & 10.19 & 9.89 & 10.24 & 9.59 & 9.61 & 10.54 & 10.46 & 10.06 & 9.21 & 10.29 & 10.30 & 10.67 & 10.98 \\ 
  7 & 9.88 & 10.75 & 10.14 & 10.65 & 10.63 & 10.97 & 10.03 & 9.97 & 10.47 & 10.42 & 10.05 & 10.46 & 10.61 & 12.06 \\ 
  8 & 10.48 & 11.16 & 10.56 & 9.88 & 9.87 & 10.21 & 11.13 & 10.34 & 10.86 & 10.58 & 10.47 & 10.41 & 9.98 & 12.37 \\ 
  9 & 11.22 & 11.60 & 11.31 & 11.35 & 11.19 & 11.16 & 11.34 & 11.88 & 11.56 & 11.41 & 11.09 & 10.40 & 11.72 & 13.77 \\ 
  High & 11.69 & 10.62 & 11.74 & 11.87 & 10.31 & 11.76 & 11.35 & 11.83 & 11.24 & 11.76 & 11.56 & 11.02 & 11.14 & 14.34 \\ 
  High - Low & 4.87 & 5.76 & 5.75 & 6.36 & 4.65 & 5.05 & 6.81 & 7.19 & 5.97 & 5.87 & 5.55 & 4.68 & 5.60 & 7.85 \\ 
  $t$-statistic & (3.64) & (3.45) & (3.77) & (4.29) & (3.02) & (3.79) & (4.02) & (4.34) & (3.86) & (4.25) & (3.91) & (3.19) & (3.63) & (3.99) \\ 
  $\alpha^{FF6}$ & 6.11 & 6.52 & 7.36 & 7.70 & 5.24 & 5.99 & 7.64 & 8.07 & 6.75 & 7.23 & 6.06 & 6.76 & 6.93 & 7.20 \\ 
  $t$-statistic & (4.20) & (4.39) & (4.73) & (5.71) & (3.73) & (5.07) & (4.84) & (5.04) & (4.66) & (4.84) & (4.17) & (4.87) & (4.74) & (3.74) \\

\midrule
\textit{\textbf{Panel B:}} & & & & & & & & & & & & & & \\
Quintile Sorts & & & & & & & & & & & & & & \\
\midrule

  Low & 7.60 & 6.09 & 6.79 & 6.66 & 7.10 & 7.22 & 6.28 & 6.31 & 6.33 & 6.87 & 7.17 & 7.60 & 6.57 & 7.86 \\ 
  2 & 9.17 & 9.71 & 9.55 & 9.46 & 9.54 & 9.08 & 9.20 & 9.34 & 9.49 & 9.33 & 8.75 & 9.28 & 9.11 & 10.29 \\ 
  3 & 9.59 & 9.92 & 9.55 & 9.70 & 9.96 & 9.39 & 10.12 & 10.13 & 9.79 & 9.56 & 10.23 & 9.85 & 10.29 & 11.05 \\ 
  4 & 10.31 & 10.83 & 10.46 & 10.33 & 10.40 & 10.62 & 10.72 & 10.18 & 10.75 & 10.35 & 10.27 & 10.60 & 10.42 & 11.92 \\ 
  High & 11.49 & 11.23 & 11.46 & 11.60 & 10.76 & 11.46 & 11.46 & 11.78 & 11.38 & 11.63 & 11.35 & 10.87 & 11.37 & 14.20 \\ 
  High - Low & 3.88 & 5.14 & 4.68 & 4.93 & 3.66 & 4.24 & 5.18 & 5.47 & 5.05 & 4.75 & 4.18 & 3.27 & 4.80 & 6.34 \\ 
  $t$-statistic & (3.46) & (3.78) & (3.60) & (3.96) & (2.79) & (3.77) & (3.73) & (3.94) & (3.76) & (4.11) & (3.79) & (2.74) & (3.72) & (4.08) \\ 
  $\alpha^{FF6}$ & 5.12 & 6.23 & 6.26 & 6.41 & 4.69 & 5.24 & 6.47 & 6.83 & 6.20 & 6.02 & 4.70 & 5.12 & 6.28 & 6.26 \\ 
  $t$-statistic & (4.31) & (5.19) & (4.82) & (5.67) & (3.94) & (5.19) & (5.16) & (5.17) & (4.95) & (5.01) & (4.24) & (4.87) & (5.27) & (4.08) \\
\midrule
\bottomrule
\end{tabularx}

\label{tab:portfolios_bivariate}
\end{table}
\end{landscape}

Next, we focus on risk measures related to idiosyncratic risk and examine nonsystematic measures of idiosyncratic volatility and idiosyncratic skewness. We observe that even these measures do not erase the economic and statistical significance of the lower-tail $\Delta$CIQ risk premium. The premium remains approximately 5\% for decile sorts and 4\% for quintile sorts, with $t$ statistics above 3 in every case.

We then investigate the cases in which we control for nonlinear systematic risk measures. We start with coskewness and cokurtosis and observe that very significant risk premia remain after we control for these measures. Then, we consider three measures that aim to capture exposure to systematic common volatility. First, we investigate exposure to the $\Delta$PCA-SQ factor. If the performance of the lower-tail $\Delta$CIQ portfolio is due to the effect of the common idiosyncratic volatility, the exposure to the $\Delta$PCA-SQ factor should erase its significance. However, we do not observe that and still observe a very sizeable premium, which is not even explained by the six-factor model, with alphas equal to 6.06\% p.a. ($t$ statistic of 4.17) and 4.70\% p.a. ($t$ statistic of 4.24) with respect to the decile and quintile sorts, respectively.

Our simulation results in Appendix \ref{sec:simul}, which are based on the stock returns obtained from a location-scale model with a symmetrical common volatility factor, show that exposure to the $\Delta$PCA-SQ factor would yield a significant premium in that case. Therefore, if the returns are generated by the location-scale model, the quantile risk is captured by this volatility risk.

We then control for the exposure to the $\Delta$CIV factor of \cite{HERSKOVIC2016}. We observe that the premium remains solid at 4.68\% p.a. ($t$ statistic of 3.19) and 3.27\% p.a. ($t$ statistic of 2.74) for the decile and quintile sorts, respectively. Moreover, the remaining premium is not explainable by the six-factor model. The last volatility-related systemic measure that we explore is the VIX beta of \cite{ang2006vol}. Once again, we observe that the premium is not subsumed by this exposure. Finally, we include the tail-risk beta of \cite{kelly2014} and draw the same conclusion.

\subsection{Fama-MacBeth Regressions}

In the next step, we perform two-stage \cite{famamacbeth} predictive regressions. In contrast to portfolio sorting, this type of asset pricing test conveniently allows for simultaneous estimation of many risk premia associated with various stock-level characteristics. This means that we can estimate the risk premium associated with the $\Delta$CIQ$_j$ factors $j \in \{LT, C, UT\}$ while controlling for other risk measures that have previously been proposed in the literature. Specifically, for each time $t=1,\ldots,T-1$ using all of the stocks $i=1,\ldots,N$ available at time $t$ and $t+1$,\footnote{A stock is identified as available if it possesses at least 48 monthly return observations during the last 60-month window up to time $t$ and an observation at time $t+1$.} we cross-sectionally regress all the returns at time $t+1$ on the betas estimated using only the information available up to time $t$. This procedure yields estimates of prices of risk $\lambda_{t+1, j}^{CIQ}$ while controlling for the most widely used measures of risk. More specifically, we use variations of the following cross-sectional regressions:
\begin{align}
	r_{i, t+1} = \alpha_{t+1} + \sum_j\beta_{i, t, j}^{CIQ} \lambda_{t+1, j}^{CIQ} + Z'_{i,t} \lambda_{t+1}^{Z} + e_{i, t+1}%, \\ &i = 1,\ldots,N_t, \, t=1,\ldots,T.
\end{align}
where $Z_{i,t}$ is a vector of control variables and $\lambda_{t+1}^{Z}$ is a vector of corresponding prices of risk. Using $T-1$ cross-sectional estimates of the prices of risk, we compute the average price of risk associated with $\lambda_{j}^{CIQ}$ as
\begin{align}
\widehat{\lambda}_j^{CIQ} = \frac{1}{T-1} \sum_{t=2}^T \widehat{\lambda}_{t, j}^{CIQ}, \quad j = LT, C, UT
\end{align}
and report them alongside their $t$ statistics on the basis of the \cite{newey1987} robust standard errors with six lags. Moreover, unlike in the case of portfolio sorts, this setup enables us to jointly evaluate the risk premia of the $\Delta$CIQ factors by including them simultaneously in the regressions.

We summarize the first set of results in Panel A of Table \ref{tab:fm_gen}, where we report the estimation outcomes of the regressions with general risk measures. First, by considering the settings featuring all three $\Delta$CIQ exposures, we observe that only the lower-tail exposure significantly predicts future returns, with a coefficient of 1.26 ($t$ statistic of 2.78). Adding idiosyncratic volatility, total and idiosyncratic skewness and the market beta to the regression does not alter the results, and the coefficient for lower-tail exposure is 0.82 ($t$ statistic of 2.68) in the full setting. On the other hand, exposures to the central or upper-tail $\Delta$CIQ factors remain unpriced.

\begin{table}[t!]
\caption{Fama--MacBeth Regressions with General Characteristics} 
\centering
\scriptsize
\begin{minipage}{\textwidth} 
The table shows estimated prices of risk and their $t$-statistics from Fama-MacBeth predictive regressions. Each month, we cross-sectionally regress next-month stock returns on current-month estimate of the exposures to the $\Delta$CIQ factors while controlling for various stock and firm characteristics. In Panel A, we control for idiosyncratic volatility (IVOL) and skewness (ISKEW), total skewness (SKEW) and market beta ($\beta^{CAPM}$). In Panel B, we focus on lower-tail $\Delta$CIQ exposure and various firm-specific characteristics. The resulting coefficients are calculated as averages of the monthly estimated coefficients and corresponding $t$-statistics are based on the robust standard errors suggested by \cite{newey1987} with six lags. The return sample covers period between January 1968 and December 2024. Each month, we use all the CRSP stocks with at least 48 monthly observations over the last 60 months, and exclude penny stocks with prices below \$1. Note the coefficients are multiplied by 100 to ensure the clarity of the presentation.
\end{minipage}
\vspace{1em}
\begin{tabularx}{0.85\linewidth}{l *{9}{>{\centering\arraybackslash}X}}
\toprule
\midrule
\multicolumn{10}{l}{\textbf{\textit{Panel A}}: Risk Characteristics} \\
\midrule
 & (1) & (2) & (3) & (4) & (5) & (6) & (7) & (8) & (9) \\ 
\midrule
	$\beta^{CIQ}_{LT}$ & 1.32 &  &  & 1.26 & 1.15 & 1.25 & 1.26 & 0.81 & 0.82 \\ 
   & (3.38) &  &  & (2.78) & (2.82) & (2.77) & (2.79) & (2.39) & (2.68) \\ 
  $\beta^{CIQ}_{C}$ &  & 3.32 &  & 1.16 & -0.06 & 0.93 & 0.94 & 1.60 & 0.27 \\ 
   &  & (1.20) &  & (0.39) & (-0.02) & (0.31) & (0.32) & (0.56) & (0.11) \\ 
  $\beta^{CIQ}_{UT}$ &  &  & -0.15 & -0.48 & -0.36 & -0.44 & -0.45 & -0.07 & -0.01 \\ 
   &  &  & (-0.34) & (-1.09) & (-0.91) & (-1.02) & (-1.04) & (-0.18) & (-0.04) \\ 
  IVOL &  &  &  &  & -16.70 &  &  &  & -15.80 \\ 
   &  &  &  &  & (-4.51) &  &  &  & (-4.68) \\ 
  SKEW &  &  &  &  &  & 0.00 &  &  & -0.14 \\ 
   &  &  &  &  &  & (-0.13) &  &  & (-2.15) \\ 
  ISKEW &  &  &  &  &  &  & 0.01 &  & 0.14 \\ 
   &  &  &  &  &  &  & (0.32) &  & (2.14) \\ 
  $\beta^{CAPM}$ &  &  &  &  &  &  &  & -0.19 & -0.14 \\ 
   &  &  &  &  &  &  &  & (-1.59) & (-1.14) \\
   Intercept & 0.82 & 0.80 & 0.80 & 0.83 & 1.12 & 0.81 & 0.81 & 0.92 & 1.15\\
 & (3.77) & (3.44) & (3.67) & (4.07) & (6.39) & (4.06) & (4.04) & (5.18) & (7.55)\\
	$R_{adj}^2$ & 0.66 & 0.41 & 0.43 & 1.50 & 2.62 & 1.65 & 1.64 & 3.14 & 4.23\\
	$\bar{n}$ & 3045 & 3045 & 3045 & 3045 & 3043 & 3043 & 3043 & 3043 & 3043\\
	$T$ & 684 & 684 & 684 & 684 & 684 & 684 & 684 & 684 & 684\\
   \midrule
\end{tabularx}
%\vspace{0em}
\begin{tabularx}{0.85\linewidth}{l *{8}{>{\centering\arraybackslash}X}}
\multicolumn{9}{l}{\textbf{\textit{Panel B}}: Firm Characteristics} \\
\midrule
 & (1) & (2) & (3) & (4) & (5) & (6) & (7) & (8) \\ 
\midrule
  $\beta^{CIQ}_{LT}$ & 1.34 & 1.30 & 1.23 & 1.20 & 1.32 & 1.32 & 1.19 & 1.08 \\ 
   & (3.48) & (3.47) & (3.48) & (3.12) & (3.30) & (3.36) & (3.14) & (3.17) \\ 
  Size & -0.02 &  &  &  &  &  &  & -0.01 \\ 
   & (-1.91) &  &  &  &  &  &  & (-1.71) \\ 
  Book-to-price &  & 0.13 &  &  &  &  &  & 0.12 \\ 
   &  & (2.33) &  &  &  &  &  & (2.00) \\ 
  Net payout yield &  &  & 1.16 &  &  &  &  & 0.84 \\ 
   &  &  & (1.59) &  &  &  &  & (1.18) \\ 
  Turnover &  &  &  & -0.08 &  &  &  & -0.08 \\ 
   &  &  &  & (-1.48) &  &  &  & (-1.84) \\ 
  Illiquidity &  &  &  &  & 1.49 &  &  & 1.92 \\ 
   &  &  &  &  & (1.27) &  &  & (1.22) \\ 
  Profit &  &  &  &  &  & 0.45 &  & 0.48 \\ 
   &  &  &  &  &  & (3.75) &  & (3.87) \\ 
  Investment &  &  &  &  &  &  & -0.43 & -0.38 \\ 
   &  &  &  &  &  &  & (-6.62) & (-6.38) \\
   Intercept & 0.83 & 0.68 & 0.79 & 0.83 & 0.80 & 0.68 & 0.87 & 0.60\\
 & (3.73) & (3.33) & (3.55) & (3.88) & (3.73) & (3.21) & (4.03) & (2.99)\\
	$R^2_{adj}$ & 0.99 & 1.45 & 1.25 & 1.52 & 1.34 & 1.19 & 1.06 & 3.26\\
	$\bar{n}$ & 3045 & 2897 & 2900 & 2893 & 2871 & 2887 & 2887 & 2741\\
	$T$ & 684 & 684 & 684 & 684 & 684 & 684 & 684 & 684\\
   \midrule
   \bottomrule
\end{tabularx}
\label{tab:fm_gen}
\end{table}

We also investigate whether firm-specific characteristics based on accounting and trading information can capture the lower-tail premium.\footnote{We construct the variables in the same vein as in \cite{LANGLOIS2020399}.}. In Panel B, we predict the returns using the exposure to the lower-tail $\Delta$CIQ factor and size, book-to-price ratio, net payout yield, turnover, illiquidity, profit and investment. All these specifications do not affect the significance and magnitude of the effect of $\beta_{LT}^{CIQ}$ on the expected returns, and we observe that the exposure to the common idiosyncratic lower-tail events is robustly compensated in the cross-section of stock returns. For example, in the setting featuring all the characteristics, the estimated price of risk is equal to 1.08 ($t$ statistic of 3.17).

Finally, we focus on nonlinear risk measures that may be correlated with exposure to common idiosyncratic lower-tail movements. We report the results from the regressions in Table \ref{tab:fm_nonlinear}. We start by considering coskewness and cokurtosis, following the specifications of \cite{harvey2000conditional} and \cite{dittmar2002}, respectively, and control for these measures simultaneously. We show that these measures do not drive out the significance of the lower-tail $\Delta$CIQ betas and that both factors are nonsignificant in this setting. Next, we consider the effect of a stock's lagged return and its momentum (cumulative return over the last twelve months while excluding the most recent return). Both control variables are highly significant, but the lower-tail $\Delta$CIQ exposure remains significant, with a slightly diminished coefficient of 0.94 ($t$ statistic of 2.67).

\begin{table}[t!]
\caption{Fama--MacBeth Regressions with Non-Linear Risk Measures} 
\centering
\scriptsize
\begin{minipage}{\textwidth} 
The table shows estimated prices of risk and their $t$-statistics from Fama-MacBeth predictive regressions. Each month, we cross-sectionally regress next-month stock returns on current-month estimates of exposure to the lower-tail $\Delta$CIQ factor and other risk measure that captures firm's risk exposure. We control for co-skewness (CSK), co-kurtosis (CKT), previous-month return (STR), cumulative return over $t-11$ to $t-1$ months (MOM), exposure to the $\Delta$PCA-SQ factor ($\beta^{PCA-SQ}$), exposure to the $\Delta$CIV factor ($\beta^{CIV}$), exposure to the $\Delta VIX$ ($\beta^{VIX}$), tail-risk beta ($\beta^{tail}$), downside-risk beta ($\beta^{down}$), hybrid tail covariance risk (HTCR), multivariate crash risk (MCRASH) and predicted co-skewness (PSS). The resulting coefficients are calculated as averages of the monthly estimated coefficients and corresponding $t$-statistics are based on the robust standard errors suggested by \cite{newey1987} with six lags. The return sample covers period between January 1968 and December 2024, with the exception of the case of $\beta^{VIX}$, which begins in February 1990. Each month, we use all the CRSP stocks with at least 48 monthly observations over the last 60 months, and exclude penny stocks with prices below \$1. Note the coefficients are multiplied by 100 to ensure the clarity of the presentation.
\end{minipage}
\vspace{1em}
%\resizebox{\textwidth}{!}{%
%\adjustbox{max height=\dimexpr\textheight-5.5cm\relax,
%           max width=\textwidth}{

\begin{tabular}{lcccccccc}
\toprule
\midrule
 & (1) & (2) & (3) & (4) & (5) & (6) & (7) & (8) \\ 
  \midrule
	$\beta^{CIQ}_{LT}$ & 1.32 & 1.22 & 0.94 & 1.32 & 1.02 & 1.35 & 1.48 & 1.25 \\ 
   & (3.38) & (3.25) & (2.67) & (3.47) & (3.14) & (3.52) & (3.43) & (3.44) \\ 
  CSK &  & -0.30 &  &  &  &  &  &  \\ 
   &  & (-1.10) &  &  &  &  &  &  \\ 
  CKT &  & -0.09 &  &  &  &  &  &  \\ 
   &  & (-1.28) &  &  &  &  &  &  \\ 
  STR &  &  & -3.86 &  &  &  &  &  \\ 
   &  &  & (-9.11) &  &  &  &  &  \\ 
  MOM &  &  & 0.64 &  &  &  &  &  \\ 
   &  &  & (4.28) &  &  &  &  &  \\ 
  $\beta^{PCA\text{-}SQ}$ &  &  &  & 19.07 &  &  &  &  \\ 
   &  &  &  & (0.64) &  &  &  &  \\ 
  $\beta^{CIV}$ &  &  &  & -0.23 &  &  &  &  \\ 
   &  &  &  & (-2.12) &  &  &  &  \\ 
  $\beta^{VIX}$ &  &  &  & -6.25 &  &  &  &  \\ 
   &  &  &  & (-1.62) &  &  &  &  \\ 
  $\beta^{down}$ &  &  &  &  & -0.04 &  &  &  \\ 
   &  &  &  &  & (-0.42) &  &  &  \\ 
  $\beta^{tail}$ &  &  &  &  & 0.09 &  &  &  \\ 
   &  &  &  &  & (1.24) &  &  &  \\ 
  HTCR &  &  &  &  &  & 1.02 &  &  \\ 
   &  &  &  &  &  & (2.75) &  &  \\ 
  MCRASH &  &  &  &  &  &  & 2.39 &  \\ 
   &  &  &  &  &  &  & (2.51) &  \\ 
  PSS &  &  &  &  &  &  &  & -2.53 \\ 
   &  &  &  &  &  &  &  & (-1.48) \\
	
Intercept & 0.82 & 0.86 & 0.71 & 0.91 & 0.81 & 0.90 & 0.57 & 2.11\\
 & (3.77) & (3.37) & (3.23) & (3.48) & (4.44) & (4.52) & (2.40) & (2.13)\\
$R^2_{adj}$ & 0.66 & 1.87 & 2.64 & 1.34 & 2.48 & 1.56 & 1.43 & 2.09\\
$\bar{n}$ & 3045 & 3043 & 3042 & 3009 & 3043 & 3044 & 2039 & 2739\\
$T$ & 684 & 684 & 684 & 418 & 684 & 684 & 684 & 684\\
   \midrule
   \bottomrule
\end{tabular}
%}
\label{tab:fm_nonlinear}
\end{table}

Extending the bivariate portfolio results, we consider the simultaneous effect of the exposures to three systematic volatility measures- the $\Delta$PCA-SQ factor, the $\Delta$CIV factor of \cite{HERSKOVIC2016} and the $\Delta$VIX factor--on the lower-tail premium. We can conclude that the lower-tail factor extracts different pricing information, as the volatility exposures do not erase either its magnitude or significance. One must investigate the common distribution in further depth if one wants to identify priced information regarding the common distributional movements.

Similarly, we jointly consider exposure to the tail-risk factor of \cite{kelly2014} and the downside beta of \cite{ang2006down}. The results show that these two measures do not drive out the effect of the lower-tail $\Delta$CIQ betas, which remains significant, similar to the univariate specification.

Finally, we consider the following three measures separately: the hybrid tail covariance risk (HTCR) proposed by \cite{htcr2014}, the multivariate crash risk (MCRASH) of \cite{chabi2022multivariate} and the predicted systematic skewness (PSS) of \cite{LANGLOIS2020399}. Although the HTCR and MCRASH are highly significant predictors of expected returns, these factors do not alter the effect of CIQ$(\tau)$ risks.

Using portfolio sorts and firm-level cross-sectional regressions, we showed that exposure to common downside idiosyncratic tail risk is significantly priced in the cross-section of stock returns. Assets with higher exposure to lower-tail $\Delta CIQ$ innovations earn higher expected returns, while exposure to upper-tail or central CIQ innovations carries no significant premium. Importantly, this pricing effect is not subsumed by common volatility or other existing measures of aggregate risk, reinforcing the conclusion that downside CIQ captures economically distinct information.

Further, consistent with the intuition of assets hedging the states in which liquidity provision is costly and intermediary capital is scarce command higher prices and lower expected returns, we examine whether the lower-tail CIQ premium is related to forward-looking measures of risk.\footnote{See Table \ref{tab:ciq_port_regressions} in Appendix \ref{sec:pred_ciq_premium}.} Regressions of portfolio returns on expected variance and the variance risk premium show that the CIQ premium is primarily associated with variation in the variance risk premium, rather than expected variance. This distinction is important: while expected variance reflects changes in aggregate uncertainty, the variance risk premium is commonly interpreted as capturing fluctuations in risk-bearing capacity and liquidity supply. The results therefore reinforce the view that lower-tail CIQ risk is priced because it loads on states in which liquidity provision is costly, rather than on periods of elevated uncertainty alone.

%------------------------------------------------------------------------------------------------------------------------------------------------------%

\section{Robustness Checks}
\label{sec:robustness}

In this section, we investigate how the lower-tail $\Delta$CIQ premium holds across specifications. First, we analyze how the CIQ premium varies over a finer grid of quantiles. Secondly, we explore the alteration of the data that we use and the specification of the model that lies behind the abnormal returns. Then, we introduce several new approaches for estimating the exposure to the $\Delta$CIQ factors.

\subsection{CIQ Premium across Distribution}

\begin{figure}[t!]
\caption{CIQ premia} 
\centering
\scriptsize
\begin{minipage}{\textwidth} 
The figure contains premia related to the exposures to the $\Delta$CIQ factors. We report average annual returns of high-minus-low equal-weighted portfolios from decile and quintile sorts and their alphas with respect to the six-factor model of \cite{FAMA2018234}. We also report $t$-statistics based on the correction of \cite{newey1994}. The data come from the CRSP and cover the period from January 1968 to December 2024. We exclude penny stocks with prices below \$1.
\end{minipage}
\vspace{1em}
\includegraphics[scale=0.45]{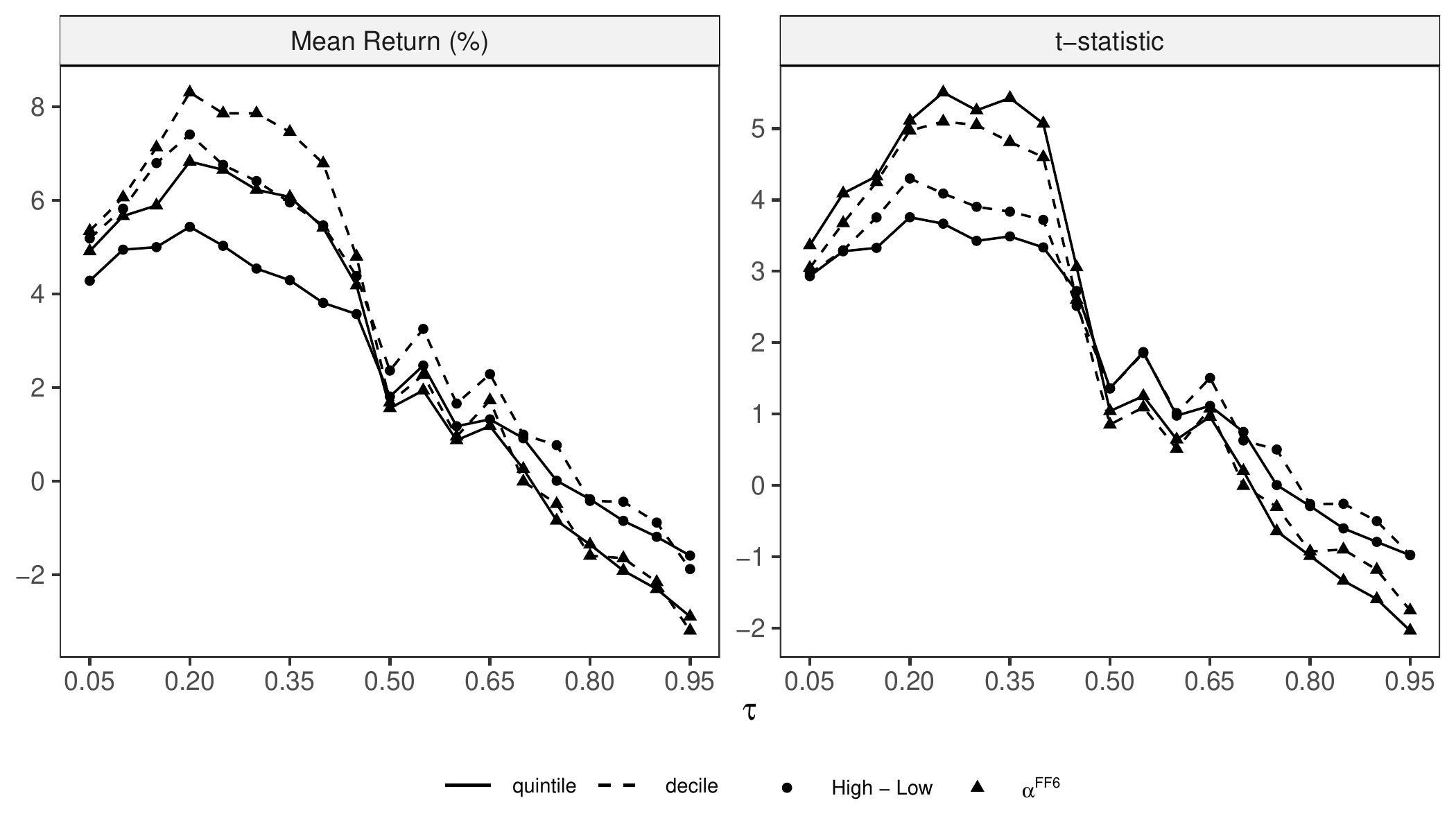}
\label{fig:ciq_premia}
\end{figure}

Figure~\ref{fig:ciq_premia} reports estimated risk premia for CIQ factors across quantile levels $\tau \in [0.05,0.95]$ in increments of 0.05.\footnote{Full results across specifications are reported in Appendix~\ref{sec:ciq_all}.} The pricing pattern varies systematically across quantiles: only lower-tail exposures ($\tau<0.5$) command a significant premium, with the largest premia concentrated around $\tau \approx 0.2$. This result reinforces the central theme of the paper that downside idiosyncratic tail risk, rather than symmetric dispersion or upside potential, is the primary priced dimension of the common idiosyncratic risk.

To better understand the economic forces driving the cross-quantile variation in premia, we examine how differences in portfolio returns across quantiles vary over time. Specifically, for two quantile levels $\tau_1 < \tau_2$, we construct the return differential between the corresponding high-minus-low portfolios from equal-weighted decile sorts,
\[
r_t^{diff(\tau_1,\tau_2)} \equiv r_t^{CIQ(\tau_1)} - r_t^{CIQ(\tau_2)}.
\]
This difference is positive when exposure to more extreme lower tail is compensated more than exposure to less severe states. We then compute the correlation between these return differentials and the lagged variance risk premium (VRP), which proxies for fluctuations in risk aversion and risk-bearing capacity.

Panel~A of Table~\ref{tab:ciq_diff_vrp} shows that these correlations are predominantly positive and increase as the distance between $\tau_1$ and $\tau_2$ widens. For example, the return differential between portfolios associated with $\tau=0.05$ and $\tau=0.35$ is strongly positively correlated with the lagged VRP. This pattern implies that when risk aversion is elevated, compensation for exposure to more extreme downside tail risk increases disproportionately relative to milder tail exposures. In other words, investors require especially high compensation for assets that perform poorly in states associated with severe common downside realizations.

This evidence is consistent with models of quantile-dependent preferences, in which agents focusing on lower quantiles behave in a more risk-averse manner \citep{decastro2019}. Importantly, repeating the same analysis using conditional variance instead of the VRP yields no comparable pattern, indicating that the cross-quantile variation in premia is driven by risk aversion and risk-bearing capacity rather than by fluctuations in aggregate uncertainty. This finding complements the results from the previous section and further supports the interpretation of lower-tail CIQ risk as a priced downside state variable.

\begin{table}[t!]
\caption{State-dependent CIQ Premium}
\centering
\scriptsize
\begin{minipage}{\textwidth} 
The table reports forward correlations between differences in CIQ portfolio returns and lagged measures of market volatility. The CIQ return differential is defined as the month-$t$ return difference between two high-minus-low portfolios constructed using exposures to the $\Delta CIQ(\tau)$ factor at quantile levels $\tau_1$ and $\tau_2$, with $\tau_1 < \tau_2$. CIQ portfolios are equal-weighted decile-sorted portfolios based on $\Delta CIQ(\tau)$ betas. Volatility measures are observed at month $t-1$ and correspond to either the variance risk premium (VRP) in Panel~A or the conditional variance (CV) in Panel~B. The VRP and CV data are from \cite{BEKAERT2014181} and cover the period from February~1990 to January~2022.
\end{minipage}
\vspace{1em}
\begin{tabular}{lccccccc}
  \toprule
  \midrule
$\tau_1 / \tau_2$ & 0.1 & 0.15 & 0.2 & 0.25 & 0.3 & 0.35 & 0.4 \\ 
  \midrule
  \textbf{Panel A:} VRP &&&& \\
0.05 & 0.04 & 0.02 & 0.06 & 0.11 & 0.15 & 0.18 & 0.17 \\ 
  0.10 &  & -0.01 & 0.05 & 0.12 & 0.17 & 0.19 & 0.18 \\ 
  0.15 &  &  & 0.08 & 0.16 & 0.20 & 0.23 & 0.20 \\ 
  0.20 &  &  &  & 0.13 & 0.21 & 0.23 & 0.20 \\ 
  0.25 &  &  &  &  & 0.17 & 0.21 & 0.18 \\ 
  0.30 &  &  &  &  &  & 0.14 & 0.12 \\ 
  0.35 &  &  &  &  &  &  & 0.04 \\ 
  \midrule
  \textbf{Panel B:} CV &&&& \\
  0.05 & 0.00 & -0.02 & 0.00 & 0.01 & 0.01 & 0.03 & 0.02 \\ 
  0.10 &  & -0.03 & 0.00 & 0.02 & 0.01 & 0.04 & 0.02 \\ 
  0.15 &  &  & 0.04 & 0.05 & 0.04 & 0.07 & 0.04 \\ 
  0.20 &  &  &  & 0.02 & 0.01 & 0.05 & 0.02 \\ 
  0.25 &  &  &  &  & 0.00 & 0.06 & 0.02 \\ 
  0.30 &  &  &  &  &  & 0.08 & 0.02 \\ 
  0.35 &  &  &  &  &  &  & -0.04 \\
  \midrule
   \bottomrule
\end{tabular}
\label{tab:ciq_diff_vrp}
\end{table}

\subsection{Stability Checks}
\label{subsec:stability}

In this section, we investigate the stability of the premium associated with the lower-tail $\Delta$CIQ factor across various alternative data and model specifications. First, in Panel A of Table \ref{tab:robustness}, we further address the concern that the results are driven by volatility. To alleviate the risk that the driving force behind the premium is the firm-level time-varying idiosyncratic volatility, we standardize the idiosyncratic returns from Equation (\ref{eq:ret_ts}) by their estimate of time-varying volatility using the simple exponentially weighted moving average (EWMA) model and use the standardized returns to estimate the CIQ factors.\footnote{The EWMA volatility model for random variable $e_t$ is defined as $\sigma^2_{t+1} = \lambda \sigma^2_t + (1 - \lambda) e_{t}^2$, where we opt for the standard value of $\lambda = 0.94$.} We present the results of high-minus-low portfolio from the decile and quintile equal-weighted sorts. We observe that the premium remains highly significant, with annual returns of 6.92\% ($t$ statistic of 3.88) and 4.68\% ($t$ statistic of 3.32) in the cases of the decile and quintile sorts, respectively. Furthermore, the premium is not subsumed by the six-factor model of \cite{FAMA2018234}.

\begin{table}[t!]
\caption{Lower-Tail $\Delta$CIQ Premium across Different Specifications} 
\centering
\scriptsize
\begin{minipage}{\textwidth} 
The table contains risk premia associated with exposure to the lower-tail $\Delta$CIQ factor across various specifications. premia are obtained as differences between high- and low-exposure decile or quintile equal-weighted portfolio returns. We also report alphas with respect to the six-factor model of \cite{FAMA2018234} (FF6). In Panel A, firstly, we report results based on the lower-tail $\Delta$CIQ factor that is estimated by reference to idiosyncratic returns that are standardized by their time-varying volatility using the EWMA model. Second, we report results using $\Delta$CIQ betas that are estimated from multiple regression when controlling for the exposure to the $\Delta$PCA-SQ factor. Third, we report results using betas estimated on levels and AR(1) innovations of the lower-tail CIQ factor, respectively. In Panel B, we report results from models that use FF5 and FF6 model to compute the idiosyncratic returns. In Panel C, we report results using data that do not exclude stocks based on their price, as well as for dataset that includes stocks that are traded with price above \$5. In Panel D, first, we report results from sorting the stocks into portfolios based on the breakpoints obtained from NYSE stocks only. Second, we progressively exclude stocks with market capitalization below certain quantile level of NYSE traded companies. In Panel E, we report separately the premia across two disjoint time periods. Panel G captures returns that are obtained from average multi-period stock returns followed after the formation of the portfolios. The reported $t$-statistics are computed using \cite{newey1987} robust standard errors with six lags. The sample covers period between January 1968 and December 2024. Each month, we use all the CRSP stocks with at least 48 monthly observations over the last 60 months, and exclude penny stocks with prices below \$1 if not stated otherwise.
\end{minipage}
\vspace{1em}
%\resizebox{\textwidth}{!}{%
%\adjustbox{max height=\dimexpr\textheight-5.5cm\relax,
%           max width=\textwidth}{

\begin{tabular}{lcccccccc}
  \toprule
  \midrule
  & \multicolumn{4}{c}{Decile Sorts} & \multicolumn{4}{c}{Quintile Sorts}  \\ % align: l,c,r
  \cmidrule(lr){2-5}  \cmidrule(lr){6-9} 
 & Premium & $t$-stat & $\alpha^{FF6}$ & $t$-stat & Premium & $t$-stat & $\alpha^{FF6}$ & $t$-stat \\ 
  \midrule
  \textit{\textbf{Panel A:}} Volatility Control &&&&&&& \\
  \midrule
  TV volatility std. & 6.92 & 3.88 & 7.71 & 4.49 & 4.68 & 3.32 & 6.16 & 4.89 \\ 
  PCA-SQ control & 6.75 & 4.26 & 6.39 & 3.97 & 4.83 & 4.01 & 5.03 & 4.20 \\ 
  CIQ level & 8.01 & 2.88 & 9.74 & 4.86 & 5.87 & 2.57 & 7.37 & 4.56 \\
  AR(1) shocks & 8.70 & 3.56 & 9.82 & 5.12 & 6.54 & 3.23 & 7.75 & 5.12 \\
    \midrule
  \textit{\textbf{Panel B:}} Linear Specification &&&&&&& \\
  \midrule
  FF5 & 6.03 & 3.85 & 7.21 & 4.68 & 4.84 & 3.71 & 6.55 & 5.17 \\ 
  FF6 & 4.38 & 2.84 & 6.08 & 3.77 & 3.73 & 2.93 & 5.65 & 4.32 \\
  \midrule
  \textit{\textbf{Panel C:}} Stock Price &&&&&&& \\
  \midrule
  All stocks & 7.76 & 4.44 & 8.64 & 4.99 & 5.79 & 3.96 & 7.20 & 5.15 \\ 
  Price $>$ \$5 & 6.71 & 3.97 & 7.30 & 5.27 & 4.64 & 3.34 & 5.89 & 5.37 \\
  \midrule
  \textit{\textbf{Panel D:}} Firm Size &&&&&&& \\
  \midrule
  NYSE breakpoints & 6.03 & 3.80 & 7.36 & 5.12 & 4.64 & 3.43 & 6.05 & 5.18 \\ 
  Market cap $>$ $q_{NYSE}$(10\%) & 7.34 & 3.61 & 8.21 & 4.64 & 5.21 & 3.18 & 6.40 & 4.77 \\ 
  Market cap $>$ $q_{NYSE}$(20\%) & 6.75 & 3.28 & 7.85 & 4.41 & 4.83 & 2.87 & 6.10 & 4.39 \\ 
  Market cap $>$ $q_{NYSE}$(50\%) & 4.92 & 2.33 & 6.01 & 3.35 & 4.06 & 2.38 & 4.83 & 3.37 \\   
  \midrule
  \textit{\textbf{Panel E:}} Time Split &&&&&&& \\
  \midrule
  01/1968 - 12/1996 & 4.06 & 1.72 & 4.00 & 1.93 & 2.89 & 1.37 & 3.58 & 2.13 \\ 
  01/1997 - 12/2024 & 10.87 & 4.48 & 9.49 & 3.87 & 8.06 & 4.25 & 7.43 & 3.96 \\
  \midrule
  \textit{\textbf{Panel F:}} Multi-Period Returns &&&&&&& \\
  \midrule
  Next 3 months & 5.61 & 3.78 & 6.56 & 4.31 & 3.92 & 3.12 & 5.38 & 4.39 \\ 
  Next 6 months & 5.13 & 3.87 & 6.65 & 4.30 & 3.61 & 3.24 & 5.26 & 4.42 \\ 
  Next 12 months & 4.50 & 4.35 & 5.22 & 4.12 & 3.24 & 3.69 & 3.83 & 3.65 \\ 
  $t+2$ to $t+12$ & 4.30 & 4.05 & 5.22 & 4.20 & 3.04 & 3.39 & 3.90 & 3.76 \\
   \midrule
   \bottomrule
\end{tabular}
%}
\label{tab:robustness}
\end{table}

Next, to precisely measure the sensitivity of stocks to the lower-tail $\Delta$CIQ factor while controlling for the sensitivity to the changes in common idiosyncratic volatility, we estimate the lower-tail $\Delta$CIQ betas from Equation (\ref{eq:ciq_betas}) using multiple regression, in which we also include the $\Delta$PCA-SQ factor as a control. The results show that this alternative does not affect the premium associated with lower-tail $\Delta$CIQ risks, as the premia remain at 6.75\% ($t = 4.26$) and 4.83\% ($t = 4.01$) in the cases of the decile and quintile sorts, respectively.

We also include results in which the exposures (betas) are estimated with respect to \textit{levels} of the CIQ factors. The premium remains strong even with respect to this specification with values of 8.01\% ($t = 2.88$) and 5.87\% per annum for respective sorts. Finally, we also report results using AR(1) innovations of the lower-tail CIQ factor, which yield strong performance within this setting too.

In Panel B, we report the portfolio results using idiosyncratic returns with respect to the FF5 and FF6 models. While the FF5 specification does not quantitatively alter the results from our baseline FF3 specification, the results using the FF6 model slightly diminish the premium. As the only difference between the FF5 and FF6 models is the inclusion of the momentum factor in the linear specification, we can conclude that the momentum exposure is partially related to the common lower-tail events. The observation that the momentum returns are associated with extreme negative events, so-called momentum crashes, has been well documented; see, for example, \cite{DANIEL2016221} or \cite{BARROSO2015111}. This observation sheds further light on the drivers of the momentum risk premium capturing the risk of lower-tail events in the cross-section of stock returns. Moreover, the alphas associated with the lower-tail $\Delta$CIQ risk remain high and significant, with annualized values of 6.08\% ($t = 3.77$) and 5.65\% ($t = 4.32$) for the decile and quintile sorts, respectively.

Next, in Panel C, we vary the stocks that we consider when forming the portfolios. First, as we restrict our universe to stocks with prices above \$1 in our baseline specification, we report here the results using all the available stocks. We observe that the premia estimated using either decile or quintile sorts remain high and significant. This situation holds if we restrict our universe to stocks with prices above the \$5 threshold.

In Panel D, we first investigate how the results change if we sort stocks into portfolios based on New York Stock Exchange breakpoints. We see that both decile- and quintile-based high-minus-low portfolio results remain high and significant. These results show that the portfolio results are not driven by micro-cap stocks.

Next in the same panel, we restrict our investment universe to stocks with market capitalization above the $p$ percentile of the distribution of stocks traded on the New York Stock Exchange at a given time. We report the results for the stocks with market capitalization greater than the 10\%, 20\% and 50\% percentiles. We observe a stable premium that, even if we focus on the stocks with capitalization above the NYSE median, yields annual returns of 4.92\% ($t = 2.33$) and 4.06\% ($t=2.38$) for decile- and quintile-based premia, respectively.

In Panel E, we report separate results for two disjointed periods. The first period covers the time between January 1968 and December 1996, and the second period covers the time from January 1997 to December 2024. We observe that the premium is substantially lower for the first period---4.06\% and 2.89\% for the respective decile and quintile sorts. Figure \ref{fig:ciq_fac_traded} shows that this situation is caused mostly by the period between approximately 1974 and 1980, in which the strategy did not yield any significant premium. On the other hand, since that time, the strategy has experienced relatively steady growth, with a premium of 10.87\% ($t$ statistic of 4.48) for the decile-based high-minus-low strategy, and a premium of 8.06\% ($t$ statistic of 4.25) for the quintile-based strategy.

Finally, in Panel F, we consider the performance of the lower-tail $\Delta$CIQ-sorted portfolios as captured by the following multiperiod returns. This exercise helps us understand two points. First, we examine whether the investment strategy based on the lower-tail $\Delta$CIQ factor is feasible in terms of turnover by examining the returns of portfolios that are rebalanced with a lower than monthly frequency. Second, we can infer whether the premium associated with the exposure to the lower-tail $\Delta$CIQ factor is a compensation for risk and not just a reversal effect. If risk is the driving force underlying the abnormal returns, then multiperiod returns should remain economically and statistically significant. We proceed as follows. Each month, we form the portfolios as in the previous case. Instead of saving the next one-month return of the sorted portfolios, we record the three-, six- and twelve-month returns that follow after the formation period. We observe returns that are consistent with the results obtained using the one-month returns; for example, the high-minus-low portfolio rebalanced every six months yields a 5.13\% ($t = 3.87$) decile-based premium and a 3.61\% ($t=3.24$) quintile-based premium on an annual basis. These results suggest that an investor does not have to suffer the high turnover costs associated with the strategy to exploit the associated risk premium.

To mitigate the effect of return reversals, we also extend this analysis by working with one-year returns but excluding returns immediately following the formation period. We report the results in the last row of Panel F. The resulting returns are almost indistinguishable from the returns over the full one-year period, with premia of 4.30\% ($t = 4.05$) and 3.04\% ($t = 3.39$) for respective sorts.

\subsection{Beyond $\Delta$CIQ Betas}

In this subsection, we extend the $\Delta$CIQ betas to further demonstrate the importance of considering downside- and upside-specific risks and their heterogeneous implications. In particular, this approach has two objectives. First, we specifically capture additional information beyond the median dependence from the downside and upside parts of the distribution and define \textit{the relative} $\Delta$CIQ betas as follows:
\begin{align}
\begin{split}
    \beta_{i, j}^{CIQ, rel} &:= \beta_{i, j}^{CIQ} - \beta_{i, C}^{CIQ}, \quad j = LT, UT\\
   \end{split}
\end{align}
The results of the portfolio sorts on the basis of relative betas are summarized in Table \ref{tab:alternative_ciq}. These results are similar to the $\Delta$CIQ results presented above. The high-minus-low portfolio sorted on the lower-tail relative betas yields an annual excess return of 7.59\% ($t = 4.27$) with a six-factor $\alpha = 8.49$ ($t = 4.90$) for the equal-weighted portfolio. In the case of the value-weighted portfolio, we obtain an annual return of 6.04\% ($t = 2.22$) and $\alpha = 6.34$ ($t = 2.33$). Similarly, as in the previous section, we also present the results for the quintile sorts in Panel B and report significant abnormal returns and alphas. Moreover, the results suggest that the incremental upside exposure does not result in any premium with zero-cost portfolio returns being statistically indistinguishable from zero.

\begin{table}[t!]
\caption{Portfolios Sorted on Alternative Specifications of the $\Delta$CIQ Exposures} 
\centering
\scriptsize
\begin{minipage}{\textwidth} 
The table reports the annualized out-of-sample excess returns of portfolios sorted on exposure to the lower-tail and upper-tail $\Delta$CIQ($\tau$) factors. Relative betas are computed by subtracting central $\Delta$CIQ exposure from either lower- and upper-tail exposures. Combination betas are obtained as average rank of either lower- and upper-tai $\Delta$CIQ exposures. Relative combination exposures are obtained by first computing relative exposures and then by averaging across ranks of either lower- or upper-tail $\Delta$ exposures. We also report the returns of zero-cost portfolios obtained by buying a high-exposure portfolio and selling a low-exposure portfolio (High - Low). The corresponding $t$-statistics (in parentheses) are computed using the robust standard errors suggested by \cite{newey1987} with six lags. Stocks are either sorted into ten portfolios in Panel A or into five portfolios in Panel B, with portfolio returns obtained by either equally weighting stock returns or value weighting by their market capitalization. The portfolios are formed each month based on the sensitivity to the $\Delta$CIQ($\tau$) factors estimated using time-series regression over the previous 60 months. The return sample covers period between January 1968 and December 2024. Each month, we use all the CRSP stocks for which at least 48 monthly observations are available over the last 60 months and exclude penny stocks with prices below \$1.
\end{minipage}
\vspace{1em}
\adjustbox{max height=\dimexpr\textheight-5.5cm\relax,
           max width=\textwidth}{
%\resizebox{\textwidth}{!}{%
\begin{tabular}{lcccccccccccc}
  \toprule
  \midrule
  & \multicolumn{4}{c}{Relative} & \multicolumn{4}{c}{Combination} & \multicolumn{4}{c}{Rel. Comb.} \\
  \cmidrule(lr){2-5}\cmidrule(lr){6-9}\cmidrule(lr){10-13}
  & \multicolumn{2}{c}{Lower-Tail} & \multicolumn{2}{c}{Upper-Tail} & \multicolumn{2}{c}{Lower-Tail} & \multicolumn{2}{c}{Upper-Tail} & \multicolumn{2}{c}{Lower-Tail} & \multicolumn{2}{c}{Upper-Tail} \\
  \cmidrule(lr){2-3}\cmidrule(lr){4-5}\cmidrule(lr){6-7}\cmidrule(lr){8-9}\cmidrule(lr){10-11}\cmidrule(lr){12-13}
 & EW & VW & EW & VW & EW & VW &EW & VW & EW & VW & EW & VW \\ 
  \midrule
  %\multicolumn{10}{l}{\textit{\textbf{Panel A:}} Decile Sorts} \\
  \textit{\textbf{Panel A:}} Decile Sorts &&&&&&&&&&&& \\
  \midrule
  Low & 4.40 & 4.63 & 9.00 & 5.83 & 4.90 & 3.77 & 8.60 & 5.36 & 5.17 & 5.18 & 9.46 & 7.36 \\ 
  2 & 8.05 & 6.45 & 10.71 & 9.92 & 7.61 & 7.66 & 9.56 & 8.25 & 7.90 & 6.88 & 10.11 & 9.01 \\ 
  3 & 9.63 & 6.69 & 10.31 & 7.51 & 9.27 & 6.69 & 10.02 & 6.58 & 8.86 & 6.95 & 9.57 & 8.43 \\ 
  4 & 9.30 & 7.44 & 9.33 & 8.05 & 9.82 & 6.59 & 10.14 & 8.47 & 9.66 & 6.90 & 9.63 & 7.09 \\ 
  5 & 9.40 & 7.18 & 9.37 & 8.45 & 9.92 & 7.44 & 9.80 & 8.16 & 10.12 & 8.72 & 9.62 & 7.51 \\ 
  6 & 10.29 & 7.82 & 10.04 & 6.48 & 10.53 & 8.79 & 9.91 & 7.96 & 10.38 & 7.33 & 9.87 & 7.63 \\ 
  7 & 10.50 & 8.07 & 9.59 & 7.33 & 9.76 & 7.55 & 9.63 & 6.88 & 10.21 & 8.35 & 9.17 & 7.14 \\ 
  8 & 10.69 & 7.99 & 9.32 & 5.98 & 10.81 & 7.76 & 9.53 & 5.93 & 10.38 & 7.17 & 9.99 & 6.93 \\ 
  9 & 11.27 & 10.09 & 9.34 & 6.04 & 11.53 & 9.44 & 9.24 & 6.36 & 11.32 & 9.41 & 8.95 & 5.46 \\ 
  High & 11.99 & 10.68 & 8.51 & 6.03 & 11.38 & 10.80 & 9.09 & 7.01 & 11.54 & 10.82 & 9.16 & 6.55 \\ 
  High - Low & 7.59 & 6.04 & -0.49 & 0.20 & 6.48 & 7.04 & 0.49 & 1.65 & 6.36 & 5.64 & -0.30 & -0.81 \\ 
  $t$-statistic & (4.27) & (2.22) & (-0.29) & (0.08) & (3.83) & (2.73) & (0.29) & (0.74) & (3.65) & (2.23) & (-0.20) & (-0.33) \\ 
  $\alpha^{FF6}$ & 8.49 & 6.34 & -1.48 & 1.20 & 7.53 & 7.17 & -0.66 & 2.49 & 7.56 & 5.63 & -1.50 & -0.54 \\ 
  $t$-statistic & (4.90) & (2.33) & (-0.87) & (0.48) & (4.76) & (2.85) & (-0.35) & (1.04) & (4.52) & (2.29) & (-0.92) & (-0.20) \\
\midrule
  %\multicolumn{10}{l}{\textit{\textbf{Panel B:}} Quintile Sorts} \\
  \textit{\textbf{Panel B:}} Quintile Sorts &&&&&&&&&&&& \\
  \midrule
  Low & 6.23 & 5.77 & 9.85 & 8.63 & 6.26 & 6.22 & 9.08 & 7.30 & 6.54 & 6.32 & 9.78 & 8.37 \\ 
  2 & 9.47 & 7.10 & 9.82 & 7.94 & 9.54 & 6.62 & 10.08 & 7.59 & 9.26 & 6.90 & 9.60 & 7.72 \\ 
  3 & 9.85 & 7.56 & 9.71 & 7.36 & 10.22 & 8.11 & 9.85 & 8.06 & 10.25 & 7.94 & 9.74 & 7.55 \\ 
  4 & 10.60 & 7.96 & 9.46 & 6.73 & 10.28 & 7.57 & 9.58 & 6.53 & 10.29 & 7.56 & 9.58 & 6.91 \\ 
  High & 11.63 & 10.09 & 8.92 & 6.18 & 11.46 & 9.66 & 9.17 & 6.65 & 11.43 & 9.63 & 9.06 & 5.74 \\ 
  High - Low & 5.40 & 4.33 & -0.93 & -2.45 & 5.20 & 3.44 & 0.09 & -0.65 & 4.89 & 3.32 & -0.73 & -2.63 \\ 
  $t$-statistic & (3.71) & (2.09) & (-0.68) & (-1.36) & (3.77) & (1.83) & (0.07) & (-0.39) & (3.45) & (1.71) & (-0.59) & (-1.57) \\ 
  $\alpha^{FF6}$ & 6.67 & 4.59 & -1.93 & -2.74 & 6.85 & 4.27 & -0.92 & -0.85 & 6.45 & 4.02 & -1.69 & -2.61 \\ 
  $t$-statistic & (5.03) & (2.55) & (-1.42) & (-1.39) & (5.43) & (2.54) & (-0.65) & (-0.47) & (4.85) & (2.38) & (-1.36) & (-1.46) \\
   \midrule
   \bottomrule
\end{tabular}%
}
\label{tab:alternative_ciq}
\end{table}

Second, to show robustness with respect to the quantile level, $\tau$, which we choose to compute the lower- and upper-tail exposures and to provide a way to aggregate the information from the downside and upside parts, we define two compressed measures. To summarize the dependence across the entire lower or upper part of the factor structure, we define the \textit{combination} lower-tail and upper-tail $\Delta$CIQ betas as follows:
\begin{align}
\begin{split}
    \beta_{i, LT}^{CIQ, comb} &:=  \sum_{\tau \in \tau_{LT}} F\big(\beta_{i, \tau}^{CIQ}\big), \quad \tau_{LT} = \{ 0.05, 0.10,\ldots, 0.45 \} \\
    \beta_{i, UT}^{CIQ, comb} &:=  \sum_{\tau \in \tau_{UT}} F\big(\beta_{i, \tau}^{CIQ}\big), \quad \tau_{LT} = \{ 0.55, 0.60,\ldots, 0.95 \}
\end{split}
\end{align}
where $F\big(\beta_{i, \tau}^{CIQ}\big) = \frac{Rank(\beta_{i, \tau}^{CIQ})}{N_t + 1}$ and $\beta_{i, \tau}^{CIQ}$ is estimated using the $\Delta$CIQ factor for a given quantile level $\tau$, where we use a set of $\tau$s between 0.05 and 0.45 with 0.05 increments for the lower tail and a set of $\tau$s between 0.55 and 0.95 with same increments for the upper tail. We obtain the combination lower- and upper-tail $\Delta$CIQ betas as the average cross-sectional ranks of the $\Delta$CIQ betas for the lower- and upper-tail $\tau$s, respectively. The results of the portfolio sorts on the basis of those betas are also summarized in Table \ref{tab:alternative_ciq}. We observe that the long-short portfolios sorted on the basis of the combination lower-tail $\Delta$CIQ betas provide significant excess annual returns of 6.48\% ($t = 3.83$) and 7.04\% ($t = 2.73$) using decile sorting and equal- and value-weighted schemes, respectively. On the other hand, an investment strategy based on the combination upper-tail betas does not yield significant abnormal returns when either weighting approach is used.

Finally, to summarize the relative betas through the whole downside or upside parts of the joint structure, we introduce the \textit{relative combination} betas as follows:
\begin{align}
\begin{split}
    \beta_{i, LT}^{CIQ, rel-comb} &:=  \sum_{\tau \in \tau_{LT}} F\big(\beta_{i, \tau}^{CIQ, rel}\big), \quad \tau_{LT} = \{ 0.05, 0.10,\ldots, 0.45 \} \\
    \beta_{i, UT}^{CIQ, rel-comb} &:=  \sum_{\tau \in \tau_{UT}} F\big(\beta_{i, \tau}^{CIQ, rel}\big), \quad \tau_{LT} = \{ 0.55, 0.60,\ldots, 0.95 \}
\end{split}
\end{align}
which are obtained as a mean cross-sectional rank of the relative betas associated with the exposure to the lower- or upper-tail $\Delta$CIQ$(\tau)$ factors, respectively. We compute the relative betas with respect to the same quantiles as in the case of the combination betas. The associated returns are also summarized in Table \ref{tab:alternative_ciq}. Similarly, as in the case of the relative betas, the lower-tail relative combination betas provide an investment strategy with significant abnormal returns of 6.36\% ($t = 3.65$) and 5.64\% ($t = 2.23$) on an annual basis using the equal- or value-weighted returns, respectively. The returns of the portfolios based on the relative combination upper-tail betas remain nonsignificant.

The results presented here confirm the previous hypothesis that lower-tail idiosyncratic comovements present a priced risk factor in the economy. On the other hand, the upside idiosyncratic factor does not constitute a risk to which the average investor assigns a premium.

%------------------------------------------------------------------------------------------------------------------------------------------------------%

\section{Economic Origins and Pricing Implications of CIQ Risk}
\label{sec:origins}

In this section, we analyze the economic drivers and pricing implications of the $\Delta$CIQ risk premium. We first identify firm characteristics associated with exposure to common idiosyncratic tail movements, clarifying which firms are most vulnerable to $\Delta$CIQ risk and why this exposure commands compensation in the cross-section. We then examine whether the same state variable also governs time variation in the aggregate equity premium, assessing how innovations in $\Delta$CIQ risk map into fluctuations in market-wide expected returns.

\subsection{Cross-sectional Determinants}

Our cross-sectional pricing tests rely on firm-level heterogeneity in exposure to the lower-tail $\Delta CIQ$ factor. To understand the economic origins of this heterogeneity, we examine which firm characteristics are associated with sensitivity to common idiosyncratic tail movements. This analysis provides an economic interpretation of the $\Delta CIQ$ factor by identifying firm attributes—such as liquidity fragility, financial slack, and vulnerability—that are systematically related to CIQ exposure and, ultimately, to the associated risk premium.

We use a subset of firm characteristics from \cite{10.1093/rfs/hhz123} and \cite{10.1093/rfs/hhaa102}, reported in Table~\ref{tab:chars_list}, and organize them into economically motivated channels suggested by the intermediary-liquidity mechanism: (i) \textit{liquidity fragility and trading}, (ii) \textit{funding and financial slack}, (iii) \textit{risk controls}, and (iv) \textit{fundamental, vulnerability and investment controls}.\footnote{We are grateful to Andreas Neuhierl for sharing the data with us.} Within each channel, we select a parsimonious set of representative variables to limit redundancy among highly correlated characteristics. We estimate monthly Fama–MacBeth regressions of firm-level CIQ betas on these characteristics, after ranking each characteristic cross-sectionally and scaling it to the interval $(-1,+1)$. CIQ betas are similarly standardized, allowing coefficient magnitudes to be compared across specifications and across tails. Reported coefficients and $t$-statistics are obtained from the time series of monthly estimates using \cite{newey1994} corrections.

Table~\ref{tab:ciq_exp_lt_det} shows that exposure to lower-tail CIQ risk is concentrated in firms with fragile trading conditions and limited financial slack. Measures of liquidity fragility—such as low turnover depth ($\texttt{dto}$ and $\texttt{lturnover}$), high turnover volatility ($\texttt{std\_turn}$), and elevated unexplained volume ($\texttt{suv}$)—are strongly associated with higher CIQ betas. Firms with greater financial flexibility, as measured by cash holdings ($\texttt{c}$), exhibit lower CIQ exposure, whereas firms with higher net payouts ($\texttt{nop}$) exhibit higher exposure. Market beta ($\texttt{beta}$) enters with a negative sign, consistent with the view that CIQ exposure is higher for stocks whose risk is less hedgeable and therefore more costly for intermediaries to warehouse during stress. Firms trading closer to recent price highs ($\texttt{rel\_to\_high\_price}$) also exhibit greater lower-tail CIQ exposure, consistent with sharper valuation adjustments when downside risk materializes. Measures of fundamental vulnerability, such as industry-adjusted profit margins ($\texttt{pm\_adj}$), further indicate that CIQ exposure is concentrated among firms more susceptible to funding-driven sell pressure.\footnote{In Table \ref{tab:ciq_exp_det_full}, we provide regression results using all the characteristics considered in our investigation.}

Taken together, these patterns support the interpretation of CIQ as capturing exposure to idiosyncratic downside sell pressure amplified by constrained intermediation, rather than a generic volatility or value effect. CIQ-sorted portfolios tilt toward firms that are difficult to intermediate in stress states—stocks with fragile liquidity and limited financial slack—providing an economic foundation for the pricing results.

\begin{table}[t!]
\caption{Cross-Sectional Determinants of the Lower-Tail CIQ Exposure}
\centering
\scriptsize
\begin{minipage}{\textwidth}
This table reports Fama–MacBeth cross-sectional regressions explaining firm-level exposure to innovations in the lower-tail CIQ factor. The dependent variable is each stock’s rolling 60-month beta with respect to $\Delta CIQ^{LT}$. Firm characteristics are ranked cross-sectionally each month and linearly scaled to lie between $-1$ and $1$. Regressions are estimated monthly, and reported coefficients are time-series averages. $t$-statistics based on \cite{newey1994} are reported in parentheses. Firm-level characteristics are from \cite{10.1093/rfs/hhz123} and \cite{10.1093/rfs/hhaa102}. The sample covers the period from January 1968 to December 2018.
\end{minipage}
\vspace{0.8em}
\begin{tabular}{lccccc}
  \toprule
  \midrule
 & (1) & (2) & (3) & (4) & (5) \\
 & Liquidity & Funding & Risk & Vulnerability & Full \\ 
  \midrule
  \multicolumn{6}{l}{\textit{Liquidity fragility}}  \\
$\texttt{dto}$ & -0.02 &  &  &  & -0.01 \\ 
   & (-7.94) &  &  &  & (-6.85) \\ 
  $\texttt{lturnover}$ & -0.28 &  &  &  & -0.24 \\ 
   & (-11.47) &  &  &  & (-10.27) \\ 
  $\texttt{std\_turn}$ & 0.10 &  &  &  & 0.10 \\ 
   & (4.84) &  &  &  & (4.84) \\ 
  $\texttt{suv}$ & 0.02 &  &  &  & 0.01 \\ 
   & (6.35) &  &  &  & (3.63) \\
   \multicolumn{1}{l}{\textit{Funding and financial slack}} &&&&& \\
  $\texttt{c}$ &  & -0.06 &  &  & -0.04 \\ 
   &  & (-5.01) &  &  & (-3.70) \\ 
  $\texttt{nop}$ &  & 0.11 &  &  & 0.07 \\ 
   &  & (3.49) &  &  & (3.83) \\
   \multicolumn{6}{l}{\textit{Risk controls}} \\   
  $\texttt{beta}$ &  &  & -0.10 &  & -0.04 \\ 
   &  &  & (-13.05) &  & (-8.45) \\ 
  $\texttt{rel\_to\_high\_price}$ &  &  & 0.07 &  & 0.05 \\ 
   &  &  & (4.77) &  & (4.44) \\
\multicolumn{6}{l}{\textit{Vulnerability control}} \\   
  $\texttt{pm\_adj}$ &  &  &  & -0.06 & -0.04 \\ 
   &  &  &  & (-5.39) & (-6.42) \\ 
   Intercept & -0.00 & -0.00 & -0.00 & -0.00 & -0.00\\
 & (-1.07) & (-0.72) & (-1.20) & (-1.11) & (-1.13)\\
 %\midrule
$R_{adj}^2$ & 7.21 & 4.65 & 4.54 & 2.35 & 10.85\\
$\bar{n}$ & 2578 & 2578 & 2578 & 2578 & 2578\\
$T$ & 612 & 612 & 612 & 612 & 612\\
   \midrule
   \bottomrule
\end{tabular}
\label{tab:ciq_exp_lt_det}
\end{table}

The economic interpretation of CIQ further depends on whether these cross-sectional determinants are symmetric across return tails. If CIQ merely reflected generic order flow or trading activity, similar characteristics would explain both downside and upside exposures. In contrast, theories of constrained intermediation predict strong asymmetries: downside order flow is amplified by forced selling and balance-sheet constraints, whereas upside order flow reflects discretionary trading in favorable states. Consistent with this prediction, Fama–MacBeth regressions for upper-tail exposures in Table~\ref{tab:ciq_exp_ut_det} show that liquidity fragility and funding vulnerability lose explanatory power in the upper tail, while CIQ exposure instead loads positively on turnover depth and market beta. This sharp contrast across tails reinforces the interpretation of CIQ as a measure of downside liquidity risk arising from constrained intermediation rather than a symmetric volatility factor.

\subsection{Predictability of the Equity Premium}
\label{sec:predictability}

Having shown that innovations in the lower-tail $\Delta$CIQ factor reflect periods of strained intermediary balance sheets, elevated aggregate uncertainty, and sell-side liquidity stress, we now examine whether these states also carry information about the aggregate equity premium. If the lower-tail $\Delta$CIQ factor captures fluctuations in economy-wide risk-bearing capacity, it should predict subsequent market returns by identifying periods in which compensation for bearing risk is unusually high.

We investigate this hypothesis by examining the ability of the lower-tail $\Delta$CIQ factor to forecast short-horizon equity returns, measured as the value-weighted excess return of all CRSP firms. We show that the predictive content of the lower-tail $\Delta$CIQ factor is economically meaningful, robust to controls for existing predictors, and persists in an out-of-sample setting.

We begin with in-sample univariate predictive regressions of the form
\begin{align}
\label{eq:pred_univ}
	r_{m, t+1} = \gamma_0 + \gamma_1 \Delta CIQ_t^j + \epsilon_{t+1}, \quad j = LT, C, UT,
\end{align}
where $r_{m,t+1}$ denotes the monthly market excess return. Each month, CIQ factors are estimated using the preceding 60 months of data, and the most recent innovation is used to predict the subsequent market return. Coefficients are scaled to represent the effect of a one-standard-deviation change in the predictor on the annualized market return. Statistical inference is based on Newey--West standard errors with six lags.

The results, reported in Table~\ref{tab:time_series}, show that the lower-tail $\Delta$CIQ factor exhibits strong predictive power for the equity premium. A one-standard-deviation decrease in the factor—corresponding to a deterioration in common idiosyncratic downside conditions—predicts an increase of 5.49 percentage points in the subsequent annualized market return ($t=-2.26$). By contrast, the upper-tail $\Delta$CIQ factor displays substantially weaker and less stable univariate predictive power, with a smaller economic magnitude and lower explanatory power.  This asymmetry mirrors the earlier evidence that downside CIQ innovations are closely tied to binding intermediation constraints, whereas upside realizations are not.

\begin{table}[t!]
\caption{Market Return Predictability by Reference to $\Delta$CIQ Factors} 
\centering
\scriptsize
\begin{minipage}{\textwidth} 
The table reports results from various specifications of predictive regressions of the value-weighted return of all CRSP firms on the $\Delta$CIQ factors and control variables. We employ increments of the PCA-SQ factor ($\Delta PCA\text{-}SQ$), innovations of the CIV factor of \cite{HERSKOVIC2016} ($\Delta CIV$), tail risk factor of \cite{kelly2014} (TR), lagged market return (MKT$_{t-1}$), cross-sectional bivariate idiosyncratic volatility of \cite{Han_Li_2025} (CBIV), and short-term reversal factor. Coefficients are scaled to capture the effect of one-standard-deviation increase in the factor on the annualized market return in percentage points. The corresponding $t$-statistics reported in parentheses are computed using the \cite{newey1987} robust standard errors using six lags. We report both in-sample (IS) and out-of-sample (OOS) $R^2$s. We also truncate the predictions at zero following \cite{campbell2005} (CT) and report corresponding OOS $R^2$s. The data cover the period from January 1968 to December 2024, with the exception of the CBIV, which ends in December 2022.
\end{minipage}
\vspace{1em}
%\resizebox{\textwidth}{!}{%
\adjustbox{max height=\dimexpr\textheight-5.5cm\relax,
           max width=\textwidth}{
           
\begin{tabular}{lccccccccccccc}
  \toprule
  \midrule
 & (1) & (2) & (3) & (4) & (5) & (6) & (7) & (8) & (9) & (10) & (11) & (12) & (13) \\ 
  \midrule
  $\Delta CIQ^{LT}$ & -5.49 &  &  & -5.52 & -6.27 & -6.60 & -5.53 & -4.97 & -4.98 & -5.70 & -5.52 & -7.24 & -6.38 \\ 
   & (-2.26) &  &  & (-2.22) & (-2.11) & (-2.38) & (-2.24) & (-1.96) & (-1.96) & (-2.12) & (-2.19) & (-2.10) & (-1.80) \\ 
  $\Delta CIQ^{C}$ &  & -2.30 &  & -3.50 & -3.47 & -3.37 & -3.60 & -3.52 & -3.59 & -3.48 & -3.50 & -3.48 & -3.49 \\ 
   &  & (-0.96) &  & (-1.32) & (-1.29) & (-1.22) & (-1.35) & (-1.30) & (-1.33) & (-1.30) & (-1.31) & (-1.21) & (-1.21) \\ 
  $\Delta CIQ^{UT}$ &  &  & 3.75 & 5.98 & 6.84 & 7.09 & 6.22 & 5.96 & 6.08 & 6.07 & 5.98 & 8.30 & 7.89 \\ 
   &  &  & (1.71) & (2.49) & (2.39) & (2.81) & (2.60) & (2.43) & (2.48) & (2.47) & (2.50) & (2.74) & (2.54) \\ 
  $\Delta PCA\text{-}SQ$ &  &  &  &  & -1.51 &  &  &  &  &  &  & -1.53 & -1.35 \\ 
   &  &  &  &  & (-0.49) &  &  &  &  &  &  & (-0.48) & (-0.42) \\ 
  $\Delta CIV$ &  &  &  &  &  & -3.48 &  &  &  &  &  & -3.80 & -3.60 \\ 
   &  &  &  &  &  & (-0.99) &  &  &  &  &  & (-1.10) & (-1.02) \\ 
  TR &  &  &  &  &  &  & 4.83 &  & 2.27 &  &  & 4.92 & 2.61 \\ 
   &  &  &  &  &  &  & (2.34) &  & (0.59) &  &  & (2.36) & (0.66) \\ 
  CBIV &  &  &  &  &  &  &  & 4.87 & 2.89 &  &  &  & 2.63 \\ 
   &  &  &  &  &  &  &  & (2.21) & (0.70) &  &  &  & (0.62) \\ 
  MKT$_{t-1}$ &  &  &  &  &  &  &  &  &  & 0.43 &  & -0.41 & -0.75 \\ 
   &  &  &  &  &  &  &  &  &  & (0.21) &  & (-0.20) & (-0.34) \\ 
  STR &  &  &  &  &  &  &  &  &  &  & 0.00 & -0.53 & -0.40 \\ 
   &  &  &  &  &  &  &  &  &  &  & (0.00) & (-0.21) & (-0.15) \\ 
  $R^2$ IS & 1.00 & 0.17 & 0.46 & 2.00 & 2.04 & 2.34 & 2.77 & 2.56 & 2.60 & 2.01 & 2.00 & 3.18 & 2.97 \\ 
  %$R^2$ IS CT & 1.02 & 0.15 & 0.43 & 1.78 & 1.86 & 2.05 & 2.17 & 2.10 & 2.10 & 1.78 & 1.78 & 2.45 & 2.31 \\ 
  $R^2$ OOS & 0.62 & -0.17 & 0.01 & 0.95 & 0.20 & 0.27 & 1.67 & 1.30 & 0.87 & 0.65 & 0.33 & -0.70 & -1.65 \\ 
  $R^2$ OOS CT & 1.28 & -0.06 & 0.29 & 1.12 & 0.19 & 1.03 & 1.59 & 1.31 & 1.15 & 0.81 & 0.85 & 0.32 & -0.34 \\
   \midrule
   \bottomrule
\end{tabular}
}
\label{tab:time_series}
\end{table}

From an economic perspective, large negative innovations in the lower-tail $\Delta$CIQ factor identify periods in which intermediary risk-bearing capacity is impaired and sell-side liquidity is scarce. In such states, assets command higher expected returns, leading to an elevated subsequent equity premium. The predictability of the market return by the lower-tail $\Delta$CIQ factor therefore reflects time variation in the price of aggregate risk driven by fluctuations in liquidity provision and balance-sheet capacity.

We also evaluate the performance of the predictive regressions in an expanding-window out-of-sample framework. Specifically, we use data up to time $t$ to estimate a prediction model and then forecast a return at time $t+1$ (the first window contains 120 monthly periods to obtain sufficiently reasonable estimates). Then, the window is extended by one observation, the prediction model is re-estimated, and a new forecast is obtained. We repeat this procedure until the whole sample is exhausted. The corresponding $R^2$ value is computed by comparing the conditional forecast and historical mean computed using the available data up to time $t$, i.e., $1 - \sum_t (r_{m, t+1} - \widehat{r}_{m, t+1 | t})^2 / \sum_t (r_{m, t+1} - \bar{r}_{m, t})^2$, where $\widehat{r}_{m, t+1 | t}$ is the out-of-sample forecast of the $t+1$ return using data up to time $t$ and  $\bar{r}_{m, t}$ is the historical mean of the market return computed up to date $t$. Unlike the case of the in-sample $R^2$ value, the OOS $R^2$ value can attain negative values if the conditional forecasts perform worse than the historical mean forecast.

We observe that the lower-tail $\Delta$CIQ factor delivers positive and economically meaningful out-of-sample $R^2$ values, whereas the predictive performance of the upper-tail factor deteriorates substantially out of sample. Applying the truncation of \citep{campbell2005} further improves out-of-sample performance for the lower-tail factor, consistent with the notion that $\Delta$CIQ captures states in which the equity premium is elevated rather than negative.

Next, we examine whether the predictive power of the lower-tail $\Delta$CIQ factor is subsumed by existing predictors. We estimate multivariate regressions of the form
\begin{align}
\label{eq:pred_control}
	r_{m, t+1} = \gamma_0 + \sum_j \gamma_{1,j} \Delta CIQ_t^j + \gamma_2 f_t + \epsilon_{t+1},
\end{align}
where $f_t$ includes controls such as innovations in common idiosyncratic volatility (PCA-SQ and CIV), tail risk \citep{kelly2014}, cross-sectional bivariate idiosyncratic volatility \citep{Han_Li_2025}, lagged market returns, and short-term reversal. Across all specifications, the coefficient on the lower-tail $\Delta$CIQ factor remains economically large and statistically significant, while the inclusion of volatility-based controls does not eliminate its predictive content.

Importantly, although some control variables—such as tail risk and CBIV—also predict market returns, the lower-tail $\Delta$CIQ factor retains its predictive power even when these variables are included jointly. This evidence indicates that the predictive content of $\Delta$CIQ is not driven by common volatility or generic tail risk but instead reflects a distinct channel related to fluctuations in aggregate liquidity provision and intermediary constraints.

Taken together, these results show that innovations in the lower-tail $\Delta$CIQ factor forecast the equity premium because they capture aggregate states in which risk-bearing capacity is scarce and liquidity provision is costly. The predictability of the market return therefore complements the cross-sectional pricing evidence and reinforces the interpretation of $\Delta$CIQ as a state variable governing the price of risk in the economy.\footnote{In unreported results, we also control for a set of eleven macroeconomic variables investigated by \cite{goyal2007}, and the results remain identical. These results are available upon request.}

%------------------------------------------------------------------------------------------------------------------------------------------------------%

\section{Conclusion}
\label{sec:conclusion}

This paper examines the circumstances in which 'idiosyncratic' downside returns become systematically related across firms, and the reasons why this comovement is priced. Using a quantile factor model applied to idiosyncratic returns, we identify common idiosyncratic quantile factors that alter the lower, central and upper regions of the residual return distribution. Our central empirical finding is that exposure to innovations in the lower-tail CIQ factor commands a large premium, whereas exposure to the median and upper-tail CIQ factors is not priced. The downside CIQ premium is robust to standard factor models and a wide range of volatility- and tail-based risk controls, suggesting that it represents a unique source of systematic risk.

The paper makes three contributions. Firstly, it introduces a new return-based measure of systematic risk that explicitly considers the distribution and distinguishes between downside and upside idiosyncratic comovement. Secondly, it demonstrates that only the downside of common idiosyncratic tail comovement is priced, with premia that survive an extensive set of risk and characteristic controls, and which vary systematically across quantiles. Thirdly, it links the priced downside factor to intermediary constraints and liquidity conditions, demonstrating that the same state variable forecasts the aggregate equity premium. 

%The remainder of the paper describes the construction of CIQ factors, develops the economic mechanism and testable predictions, and presents evidence of cross-sectional pricing and robustness. It also studies the macro-financial content and return predictability of CIQ innovations.

%We interpret this evidence through the lens of constrained intermediation. 
When the risk-bearing capacity of intermediaries is impaired, funding and liquidity frictions amplify selling pressure and propagate losses across assets. This causes the residual downside returns of otherwise unrelated firms to move together. In line with this mechanism, downside $\Delta CIQ$ innovations are correlated with measures associated with intermediary constraints and uncertainty. Furthermore, firms that are more fragile in terms of liquidity and funding ex ante exhibit stronger exposure to downside $\Delta CIQ$. These patterns support the view that common downside idiosyncratic tail risk is a practical manifestation of limits to diversification in adverse conditions: what appears idiosyncratic in normal conditions becomes commonplace and priced in when balance sheet constraints take effect.

Our findings have implications for both asset pricing and risk management. In terms of asset pricing, they identify a state variable that links distributional shifts in residual returns to equilibrium risk premia, helping to explain why compensation for tail exposure is concentrated in the left tail. For practitioners and regulators, the results highlight that monitoring downside tail comovement in residual returns can provide information about latent fragility in market making and risk bearing capacity. Future work could investigate how these tail comovement states interact with market structure, dealer inventories and the propagation of shocks across asset classes at higher frequencies.

%------------------------------------------------------------------------------------------------------------------------------------------------------%

\newpage
\clearpage

\setlength{\bibsep}{3pt}
\bibliographystyle{chicago}
\bibliography{bibliography}

\begin{thebibliography}{}

\bibitem[\protect\citeauthoryear{Acharya, Engle, and Richardson}{Acharya
  et~al.}{2017}]{AcharyaEtAl2017}
Acharya, V.~V., R.~F. Engle, and M.~Richardson (2017).
\newblock {SRISK}: A conditional capital shortfall measure of systemic risk.
\newblock {\em Review of Financial Studies\/}~{\em 30\/}(1), 48--79.

\bibitem[\protect\citeauthoryear{Adrian and Brunnermeier}{Adrian and
  Brunnermeier}{2016}]{AdrianBrunnermeier2016}
Adrian, T. and M.~K. Brunnermeier (2016).
\newblock {CoVaR}.
\newblock {\em American Economic Review\/}~{\em 106\/}(7), 1705--1741.

\bibitem[\protect\citeauthoryear{Amengual and Sentana}{Amengual and
  Sentana}{2020}]{amengual2020}
Amengual, D. and E.~Sentana (2020).
\newblock Is a normal copula the right copula?
\newblock {\em Journal of Business \& Economic Statistics\/}~{\em 38\/}(2),
  350--366.

\bibitem[\protect\citeauthoryear{Amihud}{Amihud}{2002}]{AMIHUD200231}
Amihud, Y. (2002).
\newblock Illiquidity and stock returns: cross-section and time-series effects.
\newblock {\em Journal of Financial Markets\/}~{\em 5\/}(1), 31--56.

\bibitem[\protect\citeauthoryear{Ang, Chen, and Xing}{Ang
  et~al.}{2006}]{ang2006down}
Ang, A., J.~Chen, and Y.~Xing (2006, 03).
\newblock {Downside Risk}.
\newblock {\em The Review of Financial Studies\/}~{\em 19\/}(4), 1191--1239.

\bibitem[\protect\citeauthoryear{Ang, Hodrick, Xing, and Zhang}{Ang
  et~al.}{2006}]{ang2006vol}
Ang, A., R.~J. Hodrick, Y.~Xing, and X.~Zhang (2006).
\newblock The cross-section of volatility and expected returns.
\newblock {\em The Journal of Finance\/}~{\em 61\/}(1), 259--299.

\bibitem[\protect\citeauthoryear{Bai}{Bai}{2003}]{bai2003}
Bai, J. (2003).
\newblock Inferential theory for factor models of large dimensions.
\newblock {\em Econometrica\/}~{\em 71\/}(1), 135--171.

\bibitem[\protect\citeauthoryear{Bai and Ng}{Bai and
  Ng}{2002}]{bai2002determining}
Bai, J. and S.~Ng (2002).
\newblock Determining the number of factors in approximate factor models.
\newblock {\em Econometrica\/}~{\em 70\/}(1), 191--221.

\bibitem[\protect\citeauthoryear{Baker and Wurgler}{Baker and
  Wurgler}{2006}]{baker2006}
Baker, M. and J.~Wurgler (2006).
\newblock Investor sentiment and the cross-section of stock returns.
\newblock {\em The Journal of Finance\/}~{\em 61\/}(4), 1645--1680.

\bibitem[\protect\citeauthoryear{Baker and Wurgler}{Baker and
  Wurgler}{2007}]{baker2007}
Baker, M. and J.~Wurgler (2007, June).
\newblock Investor sentiment in the stock market.
\newblock {\em Journal of Economic Perspectives\/}~{\em 21\/}(2), 129--152.

\bibitem[\protect\citeauthoryear{Bali, Cakici, and Whitelaw}{Bali
  et~al.}{2014}]{htcr2014}
Bali, T.~G., N.~Cakici, and R.~F. Whitelaw (2014, 09).
\newblock {Hybrid Tail Risk and Expected Stock Returns: When Does the Tail Wag
  the Dog?}
\newblock {\em The Review of Asset Pricing Studies\/}~{\em 4\/}(2), 206--246.

\bibitem[\protect\citeauthoryear{Bali, Engle, and Murray}{Bali
  et~al.}{2016}]{bali2016empirical}
Bali, T.~G., R.~F. Engle, and S.~Murray (2016).
\newblock {\em Empirical asset pricing: The cross section of stock returns}.
\newblock John Wiley \& Sons.

\bibitem[\protect\citeauthoryear{Barroso and Santa-Clara}{Barroso and
  Santa-Clara}{2015}]{BARROSO2015111}
Barroso, P. and P.~Santa-Clara (2015).
\newblock Momentum has its moments.
\newblock {\em Journal of Financial Economics\/}~{\em 116\/}(1), 111--120.

\bibitem[\protect\citeauthoryear{Bekaert and Hoerova}{Bekaert and
  Hoerova}{2014}]{BEKAERT2014181}
Bekaert, G. and M.~Hoerova (2014).
\newblock The vix, the variance premium and stock market volatility.
\newblock {\em Journal of Econometrics\/}~{\em 183\/}(2), 181--192.
\newblock Analysis of Financial Data.

\bibitem[\protect\citeauthoryear{Bollerslev}{Bollerslev}{1986}]{BOLLERSLEV1986307}
Bollerslev, T. (1986).
\newblock Generalized autoregressive conditional heteroskedasticity.
\newblock {\em Journal of Econometrics\/}~{\em 31\/}(3), 307--327.

\bibitem[\protect\citeauthoryear{Brunnermeier and Pedersen}{Brunnermeier and
  Pedersen}{2009}]{BrunnermeierPedersen2009}
Brunnermeier, M.~K. and L.~H. Pedersen (2009).
\newblock Market liquidity and funding liquidity.
\newblock {\em Review of Financial Studies\/}~{\em 22\/}(6), 2201--2238.

\bibitem[\protect\citeauthoryear{Brunnermeier and Sannikov}{Brunnermeier and
  Sannikov}{2014}]{BrunnermeierSannikov2014}
Brunnermeier, M.~K. and Y.~Sannikov (2014).
\newblock A macroeconomic model with a financial sector.
\newblock {\em American Economic Review\/}~{\em 104\/}(2), 379--421.

\bibitem[\protect\citeauthoryear{Campbell and Thompson}{Campbell and
  Thompson}{2007}]{campbell2005}
Campbell, J.~Y. and S.~B. Thompson (2007, 11).
\newblock {Predicting Excess Stock Returns Out of Sample: Can Anything Beat the
  Historical Average?}
\newblock {\em The Review of Financial Studies\/}~{\em 21\/}(4), 1509--1531.

\bibitem[\protect\citeauthoryear{Carhart}{Carhart}{1997}]{carhart1997}
Carhart, M.~M. (1997).
\newblock On persistence in mutual fund performance.
\newblock {\em The Journal of Finance\/}~{\em 52\/}(1), 57--82.

\bibitem[\protect\citeauthoryear{Chabi-Yo, Huggenberger, and Weigert}{Chabi-Yo
  et~al.}{2022}]{chabi2022multivariate}
Chabi-Yo, F., M.~Huggenberger, and F.~Weigert (2022).
\newblock Multivariate crash risk.
\newblock {\em Journal of Financial Economics\/}~{\em 145\/}(1), 129--153.

\bibitem[\protect\citeauthoryear{Chen, Dolado, and Gonzalo}{Chen
  et~al.}{2021}]{chen2021}
Chen, L., J.~J. Dolado, and J.~Gonzalo (2021).
\newblock Quantile factor models.
\newblock {\em Econometrica\/}~{\em 89\/}(2), 875--910.

\bibitem[\protect\citeauthoryear{Corsi}{Corsi}{2009}]{corsi2009}
Corsi, F. (2009, 02).
\newblock {A Simple Approximate Long-Memory Model of Realized Volatility}.
\newblock {\em Journal of Financial Econometrics\/}~{\em 7\/}(2), 174--196.

\bibitem[\protect\citeauthoryear{Daniel and Moskowitz}{Daniel and
  Moskowitz}{2016}]{DANIEL2016221}
Daniel, K. and T.~J. Moskowitz (2016).
\newblock Momentum crashes.
\newblock {\em Journal of Financial Economics\/}~{\em 122\/}(2), 221--247.

\bibitem[\protect\citeauthoryear{de~Castro and Galvao}{de~Castro and
  Galvao}{2019}]{decastro2019}
de~Castro, L. and A.~F. Galvao (2019).
\newblock Dynamic quantile models of rational behavior.
\newblock {\em Econometrica\/}~{\em 87\/}(6), 1893--1939.

\bibitem[\protect\citeauthoryear{Delikouras and Kostakis}{Delikouras and
  Kostakis}{2019}]{Delikouras_Kostakis_2019}
Delikouras, S. and A.~Kostakis (2019).
\newblock A single-factor consumption-based asset pricing model.
\newblock {\em Journal of Financial and Quantitative Analysis\/}~{\em 54\/}(2),
  789--827.

\bibitem[\protect\citeauthoryear{Ding, Engle, Li, and Zheng}{Ding
  et~al.}{2025}]{ding2022factor}
Ding, Y., R.~Engle, Y.~Li, and X.~Zheng (2025).
\newblock Multiplicative factor model for volatility.
\newblock {\em Journal of Econometrics\/}~{\em 249}, 105959.

\bibitem[\protect\citeauthoryear{Dittmar}{Dittmar}{2002}]{dittmar2002}
Dittmar, R.~F. (2002).
\newblock Nonlinear pricing kernels, kurtosis preference, and evidence from the
  cross section of equity returns.
\newblock {\em The Journal of Finance\/}~{\em 57\/}(1), 369--403.

\bibitem[\protect\citeauthoryear{Fama and French}{Fama and
  French}{1993}]{FAMA19933}
Fama, E.~F. and K.~R. French (1993).
\newblock Common risk factors in the returns on stocks and bonds.
\newblock {\em Journal of Financial Economics\/}~{\em 33\/}(1), 3--56.

\bibitem[\protect\citeauthoryear{Fama and French}{Fama and
  French}{2015}]{FAMA20151}
Fama, E.~F. and K.~R. French (2015).
\newblock A five-factor asset pricing model.
\newblock {\em Journal of Financial Economics\/}~{\em 116\/}(1), 1--22.

\bibitem[\protect\citeauthoryear{Fama and French}{Fama and
  French}{2018}]{FAMA2018234}
Fama, E.~F. and K.~R. French (2018).
\newblock Choosing factors.
\newblock {\em Journal of Financial Economics\/}~{\em 128\/}(2), 234--252.

\bibitem[\protect\citeauthoryear{Fama and MacBeth}{Fama and
  MacBeth}{1973}]{famamacbeth}
Fama, E.~F. and J.~D. MacBeth (1973).
\newblock Risk, return, and equilibrium: Empirical tests.
\newblock {\em Journal of Political Economy\/}~{\em 81\/}(3), 607--636.

\bibitem[\protect\citeauthoryear{Farago and T{\'e}dongap}{Farago and
  T{\'e}dongap}{2018}]{FARAGO201869}
Farago, A. and R.~T{\'e}dongap (2018).
\newblock Downside risks and the cross-section of asset returns.
\newblock {\em Journal of Financial Economics\/}~{\em 129\/}(1), 69--86.

\bibitem[\protect\citeauthoryear{Frazzini and Pedersen}{Frazzini and
  Pedersen}{2014}]{FRAZZINI20141}
Frazzini, A. and L.~H. Pedersen (2014).
\newblock Betting against beta.
\newblock {\em Journal of Financial Economics\/}~{\em 111\/}(1), 1--25.

\bibitem[\protect\citeauthoryear{Freyberger, Neuhierl, and Weber}{Freyberger
  et~al.}{2020}]{10.1093/rfs/hhz123}
Freyberger, J., A.~Neuhierl, and M.~Weber (2020, 04).
\newblock Dissecting characteristics nonparametrically.
\newblock {\em The Review of Financial Studies\/}~{\em 33\/}(5), 2326--2377.

\bibitem[\protect\citeauthoryear{Giglio, Kelly, and Pruitt}{Giglio
  et~al.}{2016}]{GiglioKellyPruitt2016}
Giglio, S., B.~Kelly, and S.~Pruitt (2016).
\newblock Systemic risk and the macroeconomy: An empirical evaluation.
\newblock {\em Journal of Financial Economics\/}~{\em 119\/}(3), 457--471.

\bibitem[\protect\citeauthoryear{Gorodnichenko and Ng}{Gorodnichenko and
  Ng}{2017}]{gorodnichenko2017level}
Gorodnichenko, Y. and S.~Ng (2017).
\newblock Level and volatility factors in macroeconomic data.
\newblock {\em Journal of Monetary Economics\/}~{\em 91}, 52--68.

\bibitem[\protect\citeauthoryear{Han and Li}{Han and Li}{2025}]{Han_Li_2025}
Han, B. and G.~Li (2025).
\newblock Idiosyncratic volatility and the icapm covariance risk.
\newblock {\em Journal of Financial and Quantitative Analysis\/}, 1--48.

\bibitem[\protect\citeauthoryear{Harvey and Siddique}{Harvey and
  Siddique}{2000}]{harvey2000conditional}
Harvey, C.~R. and A.~Siddique (2000).
\newblock Conditional skewness in asset pricing tests.
\newblock {\em The Journal of Finance\/}~{\em 55\/}(3), 1263--1295.

\bibitem[\protect\citeauthoryear{He, Kelly, and Manela}{He
  et~al.}{2017}]{HeKellyManela2017}
He, Z., B.~Kelly, and A.~Manela (2017).
\newblock Intermediary asset pricing: New evidence from many asset classes.
\newblock {\em Journal of Financial Economics\/}~{\em 126\/}(1), 1--35.

\bibitem[\protect\citeauthoryear{He and Krishnamurthy}{He and
  Krishnamurthy}{2013}]{HeKrishnamurthy2013}
He, Z. and A.~Krishnamurthy (2013).
\newblock Intermediary asset pricing.
\newblock {\em American Economic Review\/}~{\em 103\/}(2), 732--770.

\bibitem[\protect\citeauthoryear{Herskovic, Kelly, Lustig, and {Van
  Nieuwerburgh}}{Herskovic et~al.}{2016}]{HERSKOVIC2016}
Herskovic, B., B.~Kelly, H.~Lustig, and S.~{Van Nieuwerburgh} (2016).
\newblock The common factor in idiosyncratic volatility: Quantitative asset
  pricing implications.
\newblock {\em Journal of Financial Economics\/}~{\em 119\/}(2), 249--283.

\bibitem[\protect\citeauthoryear{Hou, Mo, Xue, and Zhang}{Hou
  et~al.}{2020}]{q5factor}
Hou, K., H.~Mo, C.~Xue, and L.~Zhang (2020, 02).
\newblock An augmented q-factor model with expected growth*.
\newblock {\em Review of Finance\/}~{\em 25\/}(1), 1--41.

\bibitem[\protect\citeauthoryear{Kelly and Jiang}{Kelly and
  Jiang}{2014}]{kelly2014}
Kelly, B. and H.~Jiang (2014, 06).
\newblock {Tail Risk and Asset Prices}.
\newblock {\em The Review of Financial Studies\/}~{\em 27\/}(10), 2841--2871.

\bibitem[\protect\citeauthoryear{Kim, Korajczyk, and Neuhierl}{Kim
  et~al.}{2020}]{10.1093/rfs/hhaa102}
Kim, S., R.~A. Korajczyk, and A.~Neuhierl (2020, 09).
\newblock Arbitrage portfolios.
\newblock {\em The Review of Financial Studies\/}~{\em 34\/}(6), 2813--2856.

\bibitem[\protect\citeauthoryear{Langlois}{Langlois}{2020}]{LANGLOIS2020399}
Langlois, H. (2020).
\newblock Measuring skewness premia.
\newblock {\em Journal of Financial Economics\/}~{\em 135\/}(2), 399--424.

\bibitem[\protect\citeauthoryear{Manela and Moreira}{Manela and
  Moreira}{2017}]{MANELA2017137}
Manela, A. and A.~Moreira (2017).
\newblock News implied volatility and disaster concerns.
\newblock {\em Journal of Financial Economics\/}~{\em 123\/}(1), 137--162.

\bibitem[\protect\citeauthoryear{Massacci, Sarno, and Trapani}{Massacci
  et~al.}{2025}]{massacci2025factor}
Massacci, D., L.~Sarno, and L.~Trapani (2025).
\newblock Factor models of asset returns and bear market risk.
\newblock {\em Management Science\/}.

\bibitem[\protect\citeauthoryear{Merton}{Merton}{1973}]{Merton1973}
Merton, R.~C. (1973).
\newblock An intertemporal capital asset pricing model.
\newblock {\em Econometrica\/}~{\em 41\/}(5), 867--887.

\bibitem[\protect\citeauthoryear{Newey and West}{Newey and
  West}{1987}]{newey1987}
Newey, W.~K. and K.~D. West (1987).
\newblock A simple, positive semi-definite, heteroskedasticity and
  autocorrelation consistent covariance matrix.
\newblock {\em Econometrica\/}~{\em 55\/}(3), 703--708.

\bibitem[\protect\citeauthoryear{Newey and West}{Newey and
  West}{1994}]{newey1994}
Newey, W.~K. and K.~D. West (1994).
\newblock Automatic lag selection in covariance matrix estimation.
\newblock {\em The Review of Economic Studies\/}~{\em 61\/}(4), 631--653.

\bibitem[\protect\citeauthoryear{Oh and Patton}{Oh and
  Patton}{2017}]{oh2017modeling}
Oh, D.~H. and A.~J. Patton (2017).
\newblock Modeling dependence in high dimensions with factor copulas.
\newblock {\em Journal of Business \& Economic Statistics\/}~{\em 35\/}(1),
  139--154.

\bibitem[\protect\citeauthoryear{Pastor and Stambaugh}{Pastor and
  Stambaugh}{2003}]{pastor2003}
Pastor, L. and R.~F. Stambaugh (2003).
\newblock Liquidity risk and expected stock returns.
\newblock {\em Journal of Political Economy\/}~{\em 111\/}(3), 642--685.

\bibitem[\protect\citeauthoryear{Stambaugh and Yuan}{Stambaugh and
  Yuan}{2016}]{mispricing}
Stambaugh, R.~F. and Y.~Yuan (2016, 12).
\newblock Mispricing factors.
\newblock {\em The Review of Financial Studies\/}~{\em 30\/}(4), 1270--1315.

\bibitem[\protect\citeauthoryear{Welch and Goyal}{Welch and
  Goyal}{2007}]{goyal2007}
Welch, I. and A.~Goyal (2007, 03).
\newblock {A Comprehensive Look at The Empirical Performance of Equity Premium
  Prediction}.
\newblock {\em The Review of Financial Studies\/}~{\em 21\/}(4), 1455--1508.

\end{thebibliography}

%------------------------------------------------------------------------------------------------------------------------------------------------------%

%------------------------------------------------------------------------------------------------------------------------------------------------------%

\newpage
\clearpage
\appendix

\section{Quantile Factor Model}
\label{sec:qfm}

We assume a panel of returns of length $T$ and width $N$ after eliminating the common mean factors from the time series regression
\begin{align}
r_{i,t} = \alpha_{i} +  \beta_{i}^{\top} f_t +\epsilon_{i,t},
\end{align}
to have a $\tau$-dependent structure-$g_t (\tau)$-in idiosyncratic errors that we coin the common idiosyncratic quantile--CIQ($\tau$)-factors, which satisfies
\begin{align}
Q_{\epsilon_{i,t}}\Big[\tau \Big| g_t (\tau)\Big] = \gamma_i^{\top}(\tau) g_t (\tau),
\end{align}
which implies
\begin{align}
\epsilon_{i,t} = \gamma_i^{\top}(\tau) g_t (\tau) + u_{i, t} (\tau),
\end{align}
where $g_t(\tau)$ is an $r(\tau) \times 1$ vector of random common factors, $\gamma_i(\tau)$ is an $r(\tau) \times 1$ vector of nonrandom factor loadings with $r(\tau) \ll N$, $Q_{\epsilon_{i,t}}\Big[\tau \Big| g_t (\tau)\Big]$ is a conditional quantile function of $\epsilon_{i,t}$ at $\tau$, and the quantile-dependent idiosyncratic error $u_{i, t} (\tau)$ almost certainly satisfies the quantile restriction $P[u_{i, t} (\tau) < 0 | g_t(\tau)] = \tau$ for all $\tau \in (0,1)$.

To estimate the common factors that capture the comovement of quantile-specific features of distributions of the idiosyncratic parts of the stock returns, we use quantile factor analysis (QFA) introduced by \cite{chen2021}. In contrast to principal component analysis (PCA), QFA facilitates the capture of hidden factors that may shift more general characteristics, such as moments or quantiles of the distribution of returns other than the mean. The methodology is also suitable for large panels and requires less strict assumptions about the data generation process.

The quantile-dependent factors and their loadings can be estimated as follows:
\begin{equation}
	\underset{\left(\gamma_1, \ldots, \gamma_N, g_1, \ldots, g_T\right)}{\mathrm{argmin}} \, \frac{1}{NT} \sum_{i=1}^N \sum_{t=1}^T \rho_{\tau} \left(\epsilon_{it} - \gamma^{\top}_i(\tau) g_t(\tau)\right)
\end{equation}
where $\rho_{\tau}(u) = (\tau - \mathbf{1}\{ u \leq 0 \}) u$ is the check function while imposing the following normalizations: $\frac{1}{T} \sum_{t=1}^T g_t(\tau) g_t(\tau)^{\top} = \mathbb{I}_r$, and
$\frac{1}{N} \sum_{i=1}^N \gamma_i(\tau) \gamma_i(\tau)^{\top} $ is diagonal with nonincreasing diagonal elements. 

As discussed in \cite{chen2021}, this estimator is related to the PCA estimator studied in \cite{bai2002determining} and \cite{bai2003}, similar to quantile regression, which is related to classical least squares regression. Unlike the PCA estimator of \cite{bai2003}, this estimator does not yield an analytical closed-form solution. To solve for the stationary points of the objective function, \cite{chen2021} proposed a computational algorithm called iterative quantile regression. These authors also showed that the estimator possesses the same convergence rate as the PCA estimators for the approximate factor model. We follow their approach when estimating the quantile factors.\footnote{We employ the authors' MATLAB codes, which are provided on the Econometrica webpage.}

\section{Beyond Volatility Factors}
\label{sec:beyond_vol}

To illustrate the discussion and provide the link between volatility and quantiles, let's consider the data generating process to be a typical location-scale model with two unrelated factors in the first and second moments. Idiosyncratic returns $\epsilon_{i,t}$ of such model will be zero mean i.i.d. process independent of both factors with cumulative distribution function $F_{\epsilon_{i,t}}$. Further let $Q_{\epsilon_{i,t}}(\tau)=F^{-1}_{\epsilon_{i,t}}(\tau)=\inf\{s:F_{\epsilon_{i,t}}(s) \le \tau\}$ be a quantile function of $\epsilon_{i,t}$ and assume the median is zero. Then the following model that is typical for finance
\begin{align}
r_{i,t} = \beta_{i}f_{1,t} + (\sigma_{i,t}^{\top} f_{2,t})\epsilon_{i,t},
\end{align}
where $\sigma_{i,t}$ is time-varying volatility of an $i$th stock and $\sigma_{i,t} f_{2,t} >0$ can be assumed to generate returns. When $f_{1,t}$ and $f_{2,t}$ do not share common elements, then
\begin{align}
Q_{r_{i,t}}\Big[\tau \Big| f_t (\tau)\Big]  = \beta_i f_{1,t} +\sigma_{i,t}^{\top} f_{2,t} Q_{\epsilon_{i,t}}(\tau)
\end{align}
for $\tau \ne 0.5$ and $Q_{r_{i,t}}\Big[\tau \Big| f_t (\tau)\Big] = \beta_i f_{1,t}$ for $\tau=0.5$. Note that here loadings on the factor are the only quantile-dependent objects and structure in the mean and volatility describes well the structure in quantiles. While this is already restrictive example that operates with the assumption on first two moments, even in such case standard PCA will not provide consistent estimates if the distribution of $\epsilon_{i,t}$ is heavy-tailed \citep{chen2021}.

But what if the data follows more complicated models than the one implied by location-shift models? Consider adding asymmetric dependence such as  
\begin{align}
r_{i,t} = \beta_{i} f_{1,t} + f_{2,t}\epsilon_{i,t} + f_{3,t}\epsilon_{i,t}^3,
\end{align}
where $\epsilon_{i,t}$ is standard normal random variable with cumulative distribution function $\Phi(.)$. The quantiles of the returns will then follow
\begin{align}
Q_{r_{i,t}}\Big[\tau \Big| f_t (\tau)\Big]  = \beta_i f_{1,t} + \Phi^{-1}(\tau)\big[f_{2,t}+f_{3,t} \Phi^{-1}(\tau)^2],
\end{align}
for $\tau \ne 0.5$ and we can clearly see that second factor in $f(\tau)=[f_{1,t},f_{2,t}+f_{3,t} \Phi^{-1}(\tau)^2]^{\top}$ is quantile dependent.

%------------------------------------------------------------------------------------------------------------------------------------------------------%
\section{A Simple Intermediary-Fragility Model}
\label{app:model}

This appendix provides a simple model that formalizes the mechanism in
Section~\ref{sub:intermediary_constraints_and_the_pricing_kernel}. The goal is not to offer a fully calibrated structural model, but
to show that a continuum of intermediary balance-sheet constraints can aggregate into a
one-dimensional fragility state $s_t$ that (i) drives common shifts in firms' idiosyncratic
left-tail quantiles and (ii) appears in the stochastic discount factor through innovations
$\Delta s_{t+1}$.

\subsection{Setup}

There is a continuum of competitive intermediaries $j \in [0,1]$ with equity $E_{j,t}$. Intermediary
$j$ chooses a portfolio over a market factor $F_{t+1}$ and a continuum of firm claims with returns
$r_{i,t+1}$, subject to an expected shortfall (ES) constraint on portfolio losses. Let $\lambda_{j,t}$
denote the Lagrange multiplier on this constraint.

We define \emph{systemic fragility} as the equity-weighted average multiplier
\begin{equation}
  s_t := \int_0^1 \omega_{j,t}\,\lambda_{j,t}\,dj,
  \label{eq:def_s}
\end{equation}
where $\omega_{j,t}$ are equity weights. Intuitively, $s_t$ is the aggregate shadow cost of risk-bearing
in the intermediary sector.

Firm $i$'s idiosyncratic return $\varepsilon_{i,t+1}$ has a conditional quantile representation
\begin{equation}
  %q_{i,t}(\tau) = a_i(\tau) + b_i(\tau)\,s_t + u_{i,t}(\tau),
  \epsilon_{i,t} = b_i (\tau) s_t + \nu_{i,t}(\tau),
  \label{eq:appendix_quantile}
\end{equation}
with $b_i(\tau)<0$ for $\tau \le \tau_L$, $b_i(\tau)\approx 0$ for $\tau \ge 0.5$, and $\nu_{i,t}(\tau)$ satisfies the appropriate quantile restriction. This captures the
idea that tighter constraints shift firms' left-tail quantiles downward through funding, covenant, and
liquidity channels, but leave the center and right tail largely unaffected.

\subsection{Quantile factor representation}

\begin{proposition}[Quantile factor representation]
\label{prop:quantile_factor}
Under standard large-$N$ conditions for approximate factor models, the first quantile principal
component of idiosyncratic returns at $\tau_L$ consistently estimates an affine transformation of
$s_t$:
\[
  CIQ_t^L = \alpha + \kappa\,s_t + \eta_t,
\]
with $\kappa \neq 0$ and an error term $\eta_t$.
\end{proposition}

\noindent
\emph{Sketch of proof.} Equation~(\eqref{eq:appendix_quantile}) defines a one-factor structure in
the conditional $\tau_L$-quantiles of idiosyncratic returns. Under the assumptions of
\cite{chen2021} on cross-sectional pervasiveness and bounded idiosyncratic components, the
quantile factor analysis estimator recovers the space spanned by $s_t$ up to an affine
transformation. Normalization conditions pin down the scale and sign up to a constant.

\subsection{Pricing kernel}

Each intermediary's optimality condition with an ES constraint yields an SDF containing a term
proportional to its marginal shortfall loss $\ell_{j,t+1}$, in addition to a standard component
$m^{\text{std}}_{t+1}$. In equilibrium, all intermediaries share the same pricing kernel, which can be
written as
\begin{equation}
  m_{t+1} = m^{\text{std}}_{t+1} + \phi_t\,\ell_{t+1},
  \label{eq:appendix_sdf}
\end{equation}
where $\ell_{t+1}$ is the aggregate marginal shortfall and $\phi_t$ is proportional to the cross-sectional
average of $\lambda_{j,t}$, hence to $s_t$ by \eqref{eq:def_s}.

Assuming that portfolio shortfalls materialize precisely when the fragility state worsens, we can
approximate $\ell_{t+1}$ as a linear function of $\Delta s_{t+1}$:
\[
  \ell_{t+1} = a_0 + a_1\,\Delta s_{t+1} + \epsilon_{t+1},
\]
with $a_1>0$. Substituting into~\eqref{eq:appendix_sdf} yields
\begin{equation}
  m_{t+1} = m^{\text{std}}_{t+1} + \psi_t\,\Delta s_{t+1},
  \label{eq:appendix_sdf_state}
\end{equation}
with $\psi_t = \phi_t a_1 > 0$ in constrained states.

\begin{proposition}[CIQ beta as fragility exposure]
\label{prop:beta_fragility}
With $CIQ_t^L = \alpha + \kappa s_t + \eta_t$, the innovation $\Delta CIQ_t^L$ is proportional to
$\Delta s_t$ up to noise. The CIQ beta
\[
  \beta^{CIQ}_i =
  \frac{\mathrm{Cov}(r_{i,t+1},\Delta CIQ^L_{t+1})}{\mathrm{Var}(\Delta CIQ^L_{t+1})}
\]
is therefore proportional to the covariance of $r_{i,t+1}$ with $\Delta s_{t+1}$. Expected excess
returns satisfy
\[
  \mathbb{E}_t[r_{i,t+1}] = \lambda_M \beta^M_i + \lambda_{CIQ,t}\,\beta^{CIQ}_i,
\]
with $\lambda_{CIQ,t} > 0$ when constraints are tight.
\end{proposition}

\noindent
\emph{Sketch of proof.} Combine $\Delta CIQ_t^L = \kappa \Delta s_t + \Delta \eta_t$ with
\eqref{eq:appendix_sdf_state} and apply the standard linear beta pricing argument. Under mild
assumptions on the noise term $\Delta \eta_t$, the projection of $r_{i,t+1}$ on $\Delta CIQ^L_{t+1}$
coincides with its projection on $\Delta s_{t+1}$ up to a constant.

\begin{proposition}[Left-tail asymmetry]
\label{prop:left_tail_asymmetry}
Because ES constraints depend on losses, the marginal shortfall $\ell_{t+1}$ is insensitive to
upside co-movement and depends only on left-tail realizations. Under the quantile representation
\eqref{eq:appendix_quantile}, the resulting price of quantile-factor risk is monotone in $\tau$ for
$\tau < 0.5$ and negligible for $\tau \ge 0.5$. In particular, only the lower-tail CIQ factor is
priced, while median and upper-tail CIQ factors do not command premia.
\end{proposition}

\noindent
\emph{Sketch of proof.} The ES constraint binds only in states where portfolio returns fall below
a left-tail threshold. By construction, $\ell_{t+1}$ is orthogonal to purely central or right-tail
shifts. Under \eqref{eq:appendix_quantile}, $s_t$ affects $q_{i,t}(\tau)$ only for $\tau \le \tau_L$.
Thus, only quantile factors estimated from the left tail proxy for $\Delta s_{t+1}$ and receive a
non-zero price of risk in \eqref{eq:appendix_sdf_state}.

%------------------------------------------------------------------------------------------------------------------------------------------------------%

\newpage
\clearpage

\section{Simulation Study}
\label{sec:simul}

We present a simulation exercise to illustrate how the $\Delta$CIQ premia would look like if the driving force behind them were simply common volatility. We simulate stock returns from the following model
\begin{align}
  r_{i,t} = \alpha_i + \beta_i r_{m,t} + \gamma_{i} (V_{t} - \bar{V}) + \gamma_{i} \lambda^V  + e_{i,t}
\end{align}
where $V_t$ is the common variance factor, and the variance of the idiosyncratic error follows the factor structure proposed by \cite{ding2022factor}
\begin{align}
\begin{split}
  &e_{i,t} = \sqrt{V_{i,t}} z_{i,t}, \\
  &V_{i,t} = V_t \, \textrm{exp}(\mu_i + \sigma_i u_{i, t}) = V_t \tilde{V}_{i,t}, \\
  &z_{i,t}, \, u_{i,t} \sim i.i.d. \, N(0,1).
\end{split}
\end{align}
Time-series variation of the returns drive two common factors -- market factor, $r_{m,t}$, and variance factor $V_t$ with unconditional mean $\bar{V}$. The expected return of a stock is then equal to
\begin{align}
\mathbb{E}[r_i] = \alpha_i + \beta_i \mathbb{E}[r_m] + \gamma_i \lambda^V .
\end{align}

We assume that the market factor follows a simple GARCH(1,1) process of \cite{BOLLERSLEV1986307}, which we fit on the market return from the empirical analysis. We assume that the log of the variance factor follows a modified HAR model of \cite{corsi2009}
\begin{align}
\begin{split}
\label{eq:har_var}
  \log V_{t+1} &= \theta_0 + \theta_m x_t^m + \theta_y x_t^y + v_{t+1} \\
  v_{t+1} &\sim i.i.d. \, N(0,\sigma_v^2)
\end{split}
\end{align}
where $x_t^m$ and $x_t^y$ are the previous month's log-variance and average log-variance over the last 12-month period, respectively. The common variance process is approximated by the cross-sectional average of the squared residuals from the time series regression of stock returns on the market factor. We fit the model from Equation (\ref{eq:har_var}) on this time series. When simulating this time series, we initialise the process by randomly selecting 12 consequent observations of the common variance process estimated from the data and using those observations for iterating forward.

We calibrate the simulation setting to match the CRSP data sample we employ in the empirical investigation. We estimate stock-level market beta, $\beta_i$, using time-series regression of stock return on the market return. Exposure to the common variance, $\gamma_i$, is estimated by regressing the stock return on the estimate of the common variance process. Price of risk associated with the variance exposure, $\lambda^V$ is chosen to be equal to $3\times 10^{-3}$.\footnote{This value corresponds to approximately 6\% annual high minus low premium obtained from ten portfolios portfolios sorted on the exposure to the common variance. The choice of this value is not essential for the results that we present here.} We estimate stock-level parameters of the idiosyncratic error variance--$\mu_i$, $\sigma_i$--as the sample mean and standard deviation of $\log \tilde{V}_{i,t}$. To approximate the $\tilde{V}_{i,t}$, we use squared residuals from the time-series regression of the stock return on the market return. Then, to simulate these parameters, we approximate their distribution by normal distribution, with the mean equal to the estimates' cross-sectional average and the variance equal to the cross-sectional variance of the estimates.

We simulate a panel of 2,500 stocks with 120 observations. We repeat this simulation 1,000 times. Each time, we simulate stock returns by randomly choosing parameters for the stock-level process from the normal distribution with mean and variance corresponding to their sample counterparts. We remove the common time variation in stock returns by first forming the common linear factor 
\begin{align}
  f_t = \frac{1}{N} \sum_{i = 1}^N r_{i,t}, \quad t = 1,\ldots,T
\end{align}
and then regressing the returns on this factor
\begin{align}
  r_{i,t} = \alpha_i + \hat{\beta}_i f_t + \hat{e}_{i,t},
\end{align}
which yields the residuals $\hat{e}_{i,t}$. Those residuals are then used to form the common volatility and quantile factors. We construct the volatility factor as the first principal component of those squared residuals. $\Delta$CIQ$(\tau)$ factors are estimated as discussed in Section \ref{sec:ciq_factors}. Exposures to those factors are then estimated using univariate time-series regressions of stock returns on the increments of the volatility or quantile factors, respectively.

Similarly, as in the empirical investigation, we sort stocks into decile portfolios based on their estimated exposure to the factors to infer the associated risk premia. We proxy the premia by computing high-minus-low returns of the portfolios. Table \ref{tab:premium_sim} reports the average premia for the three $\Delta$CIQ factors that we investigate in the empirical analysis. We observe that the premium is positive for the lower-tail quantile factor, negative for the upper-tail factor and close to zero for the central factor. The magnitude of the premia are comparable across lower and upper tail factors and in absolute value approximately equal to 9.4\%. The premium associated with the exposure to the $\Delta$PCA-SQ factor is -6.09\%. We also compute associated $t$-statistics as a ratio between average premium and its standard deviation across all the simulations. premia for the lower and upper tail factors are significant, unlike the value for the central factor, with $t$-statistics of around 2.6 in absolute value. The $t$-value associated with the $\Delta$PCA-SQ factor is -2.33, so the premium estimated using this approach is also significant. Next, we present the proportion of rejections of non-significance of the $\Delta$CIQ betas at a 5\% significance level from multivariate cross-sectional regressions of average returns on those betas and market betas. We can see that the proportions are virtually identical for both lower- and upper-tail betas of around 96\%. The ratio for the $\Delta$PCA-SQ betas is 90\%.

\begin{table}[t!]
\caption{Simulated Risk premia}
\centering
\scriptsize
\begin{minipage}{\textwidth} 
The table contains risk average premia computed from high-minus-low returns of decile portfolios sorted on exposure to the CIQ($\tau$) risks. We simulate the returns using common variance factor model proposed by \cite{ding2022factor}. We simulate panel of 2,500 stocks with 120 monthly observations. We perform the simulation 1,000 times. $t$-statistics are obtained by dividing the average premium by its standard deviation. We also report proportion of rejections of non-significance of $\Delta$CIQ betas from multivariate cross-sectional regressions of average returns on those betas and market betas.
\end{minipage}
\vspace{1em}
\begin{tabular}{lccc}
  \toprule
  \midrule
$\tau$ & Premium & $t$-stat & Rejections \\ 
  \midrule
  $\Delta CIQ_{LT}$ & 9.37 & 2.44 & 0.96 \\ 
  $\Delta CIQ_{C}$ & 0.26 & 0.03 & 0.96 \\ 
  $\Delta CIQ_{UT}$ & -9.47 & -2.56 & 0.96 \\  
   \midrule
   \bottomrule
\end{tabular}
\label{tab:premium_sim}
\end{table}

As we can see from the results, if there was a simple common volatility element present in the returns, which is compensated in the cross-section, the $\Delta$CIQ risk premium would be symmetrical for lower and upper tail quantile factors. Moreover, the exposure to the $\Delta$PCA-SQ factor would be priced in this case, too. Overall, the evidence from the simulation exercise suggests that the $\Delta$CIQ risk premia we observe in the data are not attributable to the common volatility compensation.

%------------------------------------------------------------------------------------------------------------------------------------------------------%

\clearpage
\newpage

\section{Cross-Sectional Quantiles}
\label{sec:cs_quantiles}

\begin{table}[h!]
\caption{Quantiles of the Cross-Sectional Stock Returns} 
\centering
\scriptsize
\begin{minipage}{\textwidth} 
The table provides a summary of cross-sectional dispersion of excess stock returns. Each month, we compute cross-sectional quantile of raw or idiosyncratic excess stock returns with respect to market factor (CAPM), three factors of \cite{FAMA19933}, five factors of \cite{FAMA20151} (FF5) or six factors of \cite{FAMA2018234} (FF6). We report results for 20\% (lower-tail) and 80\% (upper-tail) quantiles. In Panel A, we report variance ratios between time-series of these quantile series. Rows correspond to the quantile series in the denominator, while columns corresponds to the quantile series in the nominator. In Panel B, we report regression results from regressing FF3 idiosyncratic quantiles on intercept, idiosyncratic volatility and CIQ (either lower- or upper-tail) factor. The data come from the CRSP database and cover the period from January 1968 to December 2024.
\end{minipage}
\vspace{1em}
%\resizebox{\textwidth}{!}{
%\begin{tabular}{lcccccccc}
\begin{tabularx}{0.8\linewidth}{l *{8}{>{\centering\arraybackslash}X}}
  \toprule
  \midrule
 {\textit{\textbf{Panel A:}} Variance ratios} & \multicolumn{4}{c}{Lower-Tail} & \multicolumn{4}{c}{Upper-Tail} \\
  \cmidrule(lr){2-5} \cmidrule(lr){6-9}
 & CAPM & FF3 & FF5 & FF6 & CAPM & FF3 & FF5 & FF6 \\ 
  \midrule
  Raw & 0.31 &  &  &  & 0.32 &  &  &  \\ 
  CAPM &  & 0.46 &  &  &  & 0.37 &  &  \\ 
  FF3 &  &  & 0.90 &  &  &  & 0.96 &  \\ 
  FF5 &  &  &  & 0.94 &  &  &  & 0.81 \\
   \midrule
\end{tabularx}
%}
\begin{tabularx}{0.8\linewidth}{l *{6}{>{\centering\arraybackslash}X}}
 %{\textit{\textbf{Panel B:}} Regressions} &&&&&& \\
 {\textit{\textbf{Panel B:}} Regressions} & \multicolumn{3}{c}{Lower-Tail} & \multicolumn{3}{c}{Upper-Tail} \\
  \cmidrule(lr){2-4} \cmidrule(lr){5-7}
  & (1) & (2) & (3) & (4) & (5) & (6) \\ 
  \midrule
  Intercept & -0.09 & -0.05 & -0.02 & 0.08 & 0.03 & 0.02 \\ 
   & (-106.88) & (-23.72) & (-4.35) & (88.70) & (15.96) & (5.37) \\ 
  Volatility &  & -0.23 &  &  & 0.26 &  \\ 
   &  & (-17.83) &  &  & (20.48) &  \\ 
  CIQ factor &  &  & 0.07 &  &  & 0.06 \\ 
   &  &  & (21.40) &  &  & (20.75) \\ 
  adj. $R^2$ & 0.00 & 0.32 & 0.40 & 0.00 & 0.38 & 0.39 \\
   \midrule
   \bottomrule
  \end{tabularx}
\label{tab:quantiles_summary}
\end{table}

\begin{figure}[h!]
  \caption{Cross-Sectional Dispersion of Stock Returns}
  \centering
  \scriptsize
  \begin{minipage}{\textwidth} 
  The figure shows various measures of cross-sectional stock returns dispersion through time. For each month, we plot 20\% and 80\% cross-sectional quantiles of either excess or idiosyncratic excess returns. The idiosyncratic returns are computed with respect to the three-factor model of \cite{FAMA19933}. The data come from the CRSP database and cover the period from January 1968 to December 2024. The shaded areas represent NBER recessions.
  \end{minipage}
  \vspace{1em}
  \includegraphics[scale=0.45]{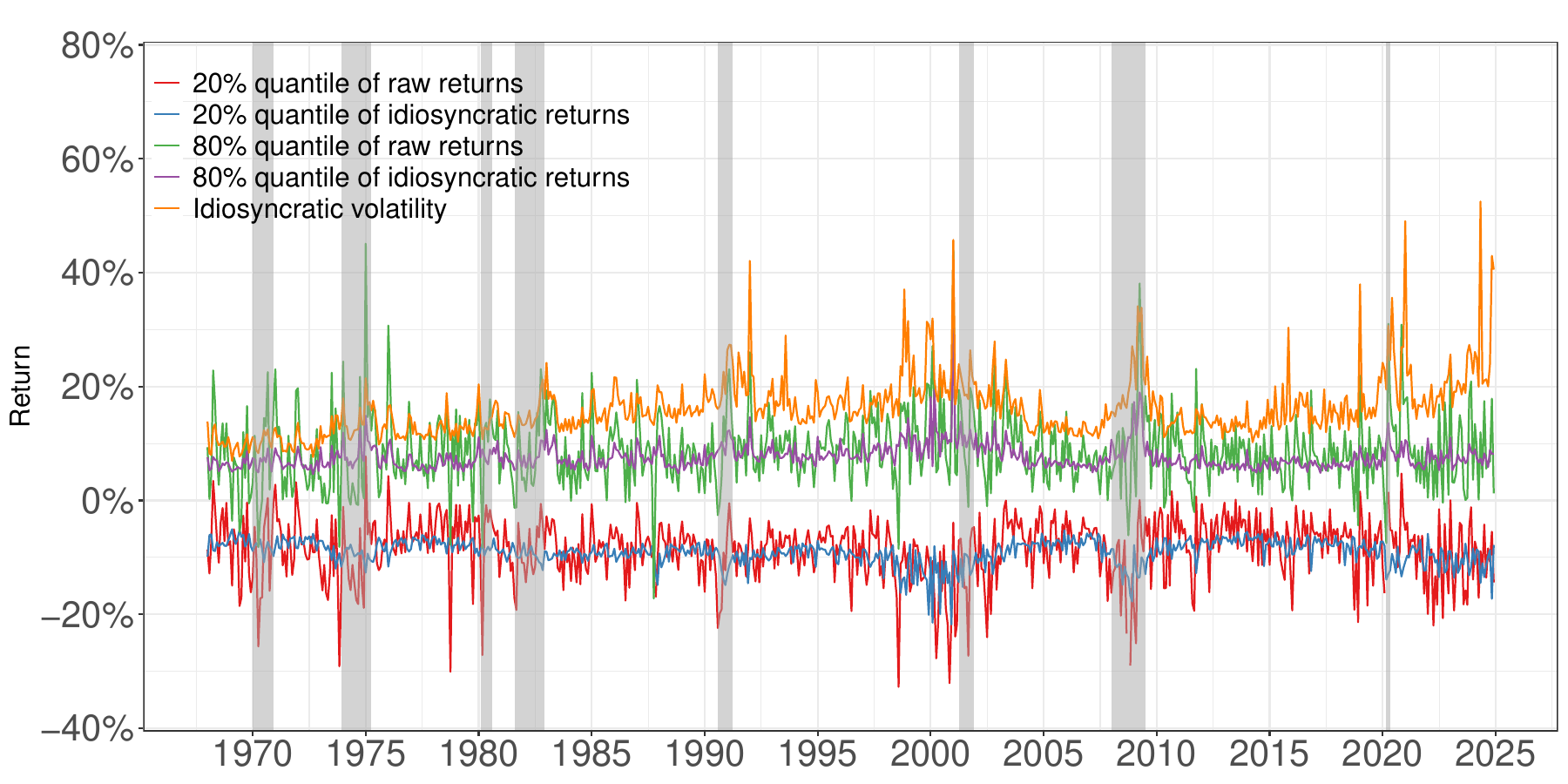}
  \label{fig:quantiles_monthly}
\end{figure}

%------------------------------------------------------------------------------------------------------------------------------------------------------%

\clearpage
\newpage

\section{Risk Measures Definitions}
\label{sec:definitions}

This Appendix provides a brief exposition of the estimation process of each of the control risk measures employed in the main text. For further details regarding the nuances of the related computations, consult the original papers.

We use two sources of data to compute these measures. First, we use either daily or monthly data of stock returns from the CRSP database. Second, we utilise the value-weighted return of the CRSP stocks from Kenneth French's online library to approximate the overall market return.

Variables are estimated using moving windows of various lengths following the procedures proposed in their original papers. In the case of measures estimated from the daily stock returns, we use mostly a moving window of one year. We require at least 200 daily observations during the window to be included. If we estimate a measure based on monthly return data we use a window of at least 60 months and demand at least 36 monthly observations.

The measures are estimated following the definition proposed in the literature. In some cases, we slightly change the requirements regarding the minimal history of stocks to be included in the analysis. This modification aims at the precision of the estimates as well as the broadest possible dataset.

Throughout this section, we use $r_{i,t}$ ($r_{i,t}^e$) to denote a raw (excess) return of an asset $i$ at time $t$. The raw (excess) market return is denoted by $f_t$ ($f_t^e$). Corresponding variables with a bar denote their time-series averages computed in a given window.

\subsection{Market Beta}
Market beta is estimated using daily data over the previous year for stocks that possess at least 200 observations as
\begin{align}
\beta_i^{CAPM} = \frac{\sum_{t} (r_{i,t}^e - \bar{r}_{i}^e) (f_{t}^e - \bar{f}^e)}{\sum_{t}(f_{t}^e - \bar{f}^e)^2}.
\end{align}

\subsection{Idiosyncratic Volatility}
Following \cite{ang2006vol}, idiosyncratic volatility is estimated using daily data over the previous month relative to the model of \cite{FAMA19933}
\begin{align}
r_{i,t}^e = \alpha_{i} + \beta^{MKT}_i f_t^e + \beta^{SMB}_i SMB_t + \beta^{HML}_i HML_t + e_{i, t}
\end{align}
and taking standard deviation of the estimated residuals, $IVOL_i = \sqrt{var(e_i)}$.

\subsection{Total and Idiosyncratic Skewness}
Following \cite{LANGLOIS2020399}, we estimate total skewness as mean of the cubed standardised daily returns $r_{i,t}^e$, and idiosyncratic skewness from the model
\begin{align}
r_{i,t}^e = \alpha_i + \beta_{1, i} f_t^e + \beta_{2, i}(f_t^e)^2 + e_{i, t}
\end{align}
and taking mean of the cubed standardised residuals $e_{i,t}$. We estimate these quantities using one year of data and requiring at least 200 observations.

\subsection{Co-skewness}
Co-skewness of \cite{harvey2000conditional} is estimated using daily excess returns and is defined as
\begin{align}
CSK_{i} &= \frac{\frac{1}{T} \sum_{t=1}^{T} (r_{i,t}^e - \bar{r}_{i}^e) (f_{t}^e - \bar{f}^e)^2} {\sqrt{\frac{1}{T}\sum_{t=1}^{T} (r_{i,t}^e - \bar{r}_{i}^e)^2} \frac{1}{T} \sum_{t=1}^{T}(f_{t}^e - \bar{f}^e)^2} .
\end{align}
Estimation window is set to one year, at least 200 daily observations are required.

\subsection{Co-kurtosis}
Co-kurtosis of \cite{dittmar2002} is estimated using daily data and is defined as
\begin{align}
CKT_{i} &= \frac{\frac{1}{T} \sum_{t=1}^{T} (r_{i,t}^e - \bar{r}_{i}^e) (f_{t}^e - \bar{f}^e)^3} {\sqrt{\frac{1}{T} \sum_{t=1}^{T} (r_{i,t}^e - \bar{r}_{i}^e)^2} \big(\frac{1}{T} \sum_{t=1}^{T}(f_{t}^e - \bar{f}^e)^2 \big)^{3/2}}.
\end{align}
Estimation window is set to 1 year, at least 200 daily observations are required.

\subsection{PCA-SQ Betas}
Both the factor and the exposures are estimated using 60-month moving window of monthly data similarly as in the case of the $\Delta$CIQ factors. Exposure is estimated from time series regression of regressing excess stock returns on a constant and $\Delta$PCA-SQ factor. We require at least 48 observations during the estimation period.

\subsection{CIV Beta}
Following \cite{HERSKOVIC2016}, CIV beta is estimated by regressing monthly excess stock returns on a constant, increments of the CIV factor and increments of the monthly market variance
\begin{equation}
r_{i,t}^e = \alpha_{i} + \beta^{CIV}_i \Delta CIV_t + \beta^{MV}_i \Delta MV_t + e_{i, t}.
\end{equation}
 We use 60-month rolling window and require at least 48 observations.

\subsection{VIX Beta}
VIX beta is estimated following \cite{ang2006down} using daily data over the previous month by regressing stock returns on a constant, market factor and increments of the CBOE volatility index as
\begin{align}
r_{i, t}^e = \alpha_i + \beta_{i}^{MKT} f_t^e + \beta_{i}^{VIX} \Delta VIX_t + e_{i, t}.
\end{align}
We require at least 17 observations during the estimation month.

\subsection{Downside Beta}
Downside beta of \cite{ang2006down} is estimated using daily data and is defined as
\begin{align}
\beta_i^{down} = \frac{\displaystyle\sum_{f_t^e < \bar{f}^e} (r_{i,t}^e - \bar{r}_{i}^e) (f_{t}^e - \bar{f}^e)}{\displaystyle\sum_{f_t^e < \bar{f}^e}(f_{t}^e - \bar{f}^e)^2}.
\end{align}
Estimation window is set to one year, at least 200 daily observations are required.

%\subsection{Semibeta}
%Negative semibeta of \cite{BOLLERSLEV2021} is estimated using daily data within the previous month as
%\begin{align}
%\beta_{i}^N = \frac{\displaystyle\sum_{r_{i,t} < 0, f_t < 0 } r_{i,t} f_{i,t}}{\displaystyle\sum_{t} f_t^2}.
%\end{align}

\subsection{Tail Risk Beta}
Tail risk beta of \cite{kelly2014} is estimated using monthly return data using 120-month window with requirement of at least 36 monthly observations. Following the original setting, we require stocks to have price higher than \$5. Beta is computed by means of least-square estimator from the predictive regression of the form
\begin{align}
r_{i,t+1} = \mu_i + \beta_i^{tail} \lambda_t + \epsilon_{t+1,i}
\end{align}
where the tail risk factor is obtained as
\begin{align}
\lambda_t = \frac{1}{K_t} \sum_{k=1}^{K_t} \textrm{ln} \frac{e_{k,t}}{u_t}
\end{align}
where $e_{k,t}$ is the $k$th daily idiosyncratic return that falls below an extreme value threshold $u_t$ during month $t$, and $K_t$ is the total number of such exceedences within month $t$. Idiosyncratic return is computed relative to the three-factor model of \cite{FAMA19933}, and the threshold value is taken to be 5\% quantile from the monthly cross-section of daily returns.

%\subsection{Downside Correlation}
%Downside correlation (\texttt{down\_corr}) based on \cite{hong2006} and \cite{jiang_wu_zhou_2018} is estimated using daily data and is defined as
%\begin{align}
%\mathbb{C}or_i^{-} = \mathbb{C}or(r_i, f | r_i < 0, f < 0) - \mathbb{C}or(r_i, f | r_i > 0, f > 0)
%\end{align}
%using empirical counterpart of the correlation. Minimum of 200 observations in the 1-year window is demanded.

\subsection{Hybrid Tail Covariance Risk}
Hybrid tail covariance risk of \cite{htcr2014} is estimated using daily data using 6-month window with at least 80 daily observations as
\begin{align}
HTCR_i = \sum_{r_{i,t} < h_i} (r_{i,t} - h_i) (f_t - h_f)
\end{align}
where $h_i$ and $h_f$ are the 10\% empirical quantiles of stock and market return, respectively.

%\sum\limits_{r_{i,t} < 0, f_t < 0 }

%\subsection{Exceedance Coentropy}
%Exceedance coentropy (\texttt{coentropy}) measure based on \cite{BACKUS20181} and \cite{jiang_wu_zhou_2018} using daily data and 1-year estimation window with at least 200 observations is based on
%\begin{align}
%C^+(0, r_{i}, f) = \frac{L(r_if) - L(r_i) - L(f)}{L(r_i) + L(f)} \Big| (r_i > 0, y > 0) \\
%C^-(0, r_{i}, f) = \frac{L(r_if) - L(r_i) - L(f)}{L(r_i) + L(f)} \Big| (r_i < 0, y < 0)
%\end{align}
%where $L(x) = \textrm{ln} \mathbb{E}(x) - \mathbb{E} (\textrm{ln} x)$. The measure is then defined as
%\begin{align}
%Coentropy = C^-(0, r_{i}, f) - C^+(0, r_{i}, f).
%\end{align}

%\subsection{Downside Asymmetry}
%Please, refer to \cite{jiang_wu_zhou_2018} for the full specification of the downside asymmetry estimation.

\subsection{Multivariate Crash Risk}
Multivariate crash risk of \cite{chabi2022multivariate} is estimated using daily data with 1-year window and minimum of 200 nonzero observations in the following steps. First, for each stock separately, using stock and $N$ factor returns, we estimate $N+1$ GARCH(1,1) models of \cite{BOLLERSLEV1986307} to obtain a series of conditional distribution functions $F_{i,t} (h) = \mathbb{P}_{t-1}[r_{i,t} \leq h] $ and use it to compute probability integral transforms as $\hat{u}_{i,t} = F_{i,t} (r_{i,t})$. Second, we estimate MCRASH as
\begin{align}
\textrm{MCRASH}_{i,t} = \frac{\sum_t \mathbb{I} (\{ \hat{u}_{1,t} \leq p \})\cdot \mathbb{I} ( \cup_{j=2}^{N+1} \{ \hat{u}_{j,t} \leq p \} )}{\sum_t \mathbb{I} ( \cup_{j=2}^{N+1} \{ \hat{u}_{j,t} \leq p \} )}
\end{align}
where $\mathbb{I}$ denotes the indicator function and $p$ is set to 0.05. We follow the baseline specification of \cite{chabi2022multivariate} and use the five factors of \cite{FAMA20151}, momentum factor of \cite{carhart1997} and betting-against-beta factor of \cite{FRAZZINI20141}.

\subsection{Predicted Systematic Co-skewness}
Predicted systematic co-skewness of \cite{LANGLOIS2020399} is based on
\begin{align}
\label{eq:cos}
Cos_{i,t} = \mathbb{C}ov_{t-1}\big(r_{i,t}, f_t^2\big),
\end{align}
then, each month we run the panel regression using all available stocks and history of data
\begin{align}
F\big(Cos_{i, k-12 \rightarrow k-1} \big) = \kappa + F\big(Y_{i, k-24 \rightarrow k-13} \big) \theta + F\big(X_{i, k-13} \big) \phi + \epsilon_{i, k-12 \rightarrow k-1}
\end{align}
where $Cos_{i, k-12 \rightarrow k-1}$ is the co-skewness from Equation (\ref{eq:cos}) computed using daily returns from month $k-12$ to month $k-1$, $Y_{i, k-24 \rightarrow k-13}$ are risk measures (volatility, market beta, etc.) estimated using daily data from month $k-24$ to month $k-13$, and $X_{i, k-13}$ are characteristics (size, book-to-price, etc.) observed at the end of month $k-13$. The function $F(x_{i,t}) = \frac{Rank(x_{i,t})}{N_t + 1}$ transforms the original variable into its normalised rank in the cross-section of variable $x_t$, which posses $N_t$ observations.

The predicted systematic co-skewness for each stock is then obtained using the estimated coefficients of $\hat{\kappa}, \hat{\theta}, \hat{\phi}$ as
\begin{align}
F\big(\widehat{Cos_{i, t \rightarrow t+11}} \big) = \hat{\kappa} + F\big(Y_{i, t-12 \rightarrow t-1} \big) \hat{\theta} + F\big(X_{i, t-1} \big) \hat{\phi}.
\end{align}
The choice of risk measures and characteristics employed in the prediction of systematic skewness follows closely \cite{LANGLOIS2020399}.

%------------------------------------------------------------------------------------------------------------------------------------------------------%

\clearpage
\newpage
\section{Summary of the CIQ Factors}
\label{sec:ciq_summary}

\begin{table}[h!]
\caption{Summary of the $\Delta$CIQ$(\tau)$ Factors} 
\centering
\scriptsize
\begin{minipage}{\textwidth} 
The table provides summary of the estimated $\Delta$CIQ$(\tau)$ factors. In Panel A, we report descriptive statistics of the $\Delta$CIQ$(\tau)$ factors including their means, standard deviations, skewness, kurtosis and autocorrelation coefficients of order between one and three. In Panel B, we report correlations between $\Delta$CIQ$(\tau)$ factors. The data cover the period from January 1968 to December 2024.
\end{minipage}
\vspace{1em}
\resizebox{\textwidth}{!}{
\begin{tabular}{lccccccccccccccccccc}
  \toprule
  \midrule
 & 0.05 & 0.1 & 0.15 & 0.2 & 0.25 & 0.3 & 0.35 & 0.4 & 0.45 & 0.5 & 0.55 & 0.6 & 0.65 & 0.7 & 0.75 & 0.8 & 0.85 & 0.9 & 0.95 \\
 \midrule
 \multicolumn{20}{l}{\textit{\textbf{Panel A:}} Descriptive statistics} \\ % align: l,c,r
 \cmidrule(lr){2-20}
  Mean $\times 10^3$ & -5.83 & -5.17 & -3.45 & -3.21 & -4.02 & -4.07 & -6.18 & -9.23 & -35.29 & -41.48 & -78.70 & -52.38 & -16.06 & -9.55 & -7.92 & -8.34 & -5.93 & -6.25 & 0.57 \\ 
  St. Dev. & 0.15 & 0.16 & 0.17 & 0.19 & 0.22 & 0.26 & 0.33 & 0.49 & 1.27 & 1.46 & 1.46 & 1.32 & 0.57 & 0.36 & 0.28 & 0.23 & 0.20 & 0.18 & 0.17 \\ 
  Skewness & -0.10 & -0.06 & 0.06 & 0.06 & 0.22 & 0.15 & 0.25 & 0.27 & -0.27 & -0.35 & -0.30 & -0.06 & 0.19 & 0.19 & 0.03 & -0.05 & -0.19 & -0.01 & 0.26 \\ 
  Kurtosis & 4.42 & 4.21 & 4.57 & 5.09 & 5.80 & 5.69 & 6.06 & 6.09 & 8.09 & 6.41 & 6.55 & 5.46 & 6.90 & 7.20 & 6.63 & 6.86 & 6.78 & 6.18 & 5.57 \\ 
  AR(1) & -0.36 & -0.40 & -0.41 & -0.44 & -0.43 & -0.46 & -0.46 & -0.46 & -0.29 & -0.27 & -0.27 & -0.33 & -0.47 & -0.46 & -0.44 & -0.42 & -0.40 & -0.37 & -0.33 \\ 
  AR(2) & -0.14 & -0.08 & -0.05 & -0.02 & -0.00 & 0.03 & 0.03 & 0.05 & 0.07 & 0.07 & 0.02 & -0.01 & 0.01 & -0.00 & -0.05 & -0.07 & -0.10 & -0.13 & -0.13 \\ 
  AR(3) & 0.17 & 0.12 & 0.08 & 0.03 & 0.00 & -0.02 & -0.02 & -0.04 & -0.07 & -0.02 & -0.00 & -0.01 & -0.00 & 0.02 & 0.06 & 0.07 & 0.09 & 0.09 & 0.08 \\
  \midrule
  \multicolumn{20}{l}{\textit{\textbf{Panel B:}} Correlations} \\ % align: l,c,r
   \cmidrule(lr){2-20}
  0.05 & 1.00 & 0.93 & 0.89 & 0.85 & 0.79 & 0.72 & 0.65 & 0.54 & 0.19 & 0.07 & 0.05 & 0.11 & 0.15 & 0.05 & -0.06 & -0.15 & -0.24 & -0.33 & -0.39 \\ 
  0.10 &  & 1.00 & 0.97 & 0.93 & 0.89 & 0.84 & 0.78 & 0.67 & 0.28 & 0.14 & 0.11 & 0.18 & 0.25 & 0.16 & 0.03 & -0.06 & -0.17 & -0.26 & -0.34 \\ 
  0.15 &  &  & 1.00 & 0.98 & 0.95 & 0.91 & 0.85 & 0.76 & 0.35 & 0.19 & 0.16 & 0.26 & 0.35 & 0.26 & 0.13 & 0.03 & -0.08 & -0.19 & -0.27 \\ 
  0.20 &  &  &  & 1.00 & 0.98 & 0.95 & 0.91 & 0.82 & 0.40 & 0.22 & 0.20 & 0.32 & 0.44 & 0.36 & 0.23 & 0.13 & 0.01 & -0.10 & -0.20 \\ 
  0.25 &  &  &  &  & 1.00 & 0.98 & 0.95 & 0.87 & 0.45 & 0.26 & 0.23 & 0.36 & 0.51 & 0.43 & 0.31 & 0.21 & 0.09 & -0.03 & -0.14 \\ 
  0.30 &  &  &  &  &  & 1.00 & 0.98 & 0.92 & 0.51 & 0.31 & 0.28 & 0.41 & 0.58 & 0.52 & 0.40 & 0.31 & 0.19 & 0.07 & -0.07 \\ 
  0.35 &  &  &  &  &  &  & 1.00 & 0.96 & 0.58 & 0.38 & 0.34 & 0.47 & 0.66 & 0.61 & 0.50 & 0.41 & 0.29 & 0.16 & 0.02 \\ 
  0.40 &  &  &  &  &  &  &  & 1.00 & 0.69 & 0.47 & 0.43 & 0.55 & 0.74 & 0.69 & 0.60 & 0.52 & 0.41 & 0.29 & 0.14 \\ 
  0.45 &  &  &  &  &  &  &  &  & 1.00 & 0.77 & 0.72 & 0.71 & 0.67 & 0.61 & 0.56 & 0.51 & 0.45 & 0.36 & 0.26 \\ 
  0.50 &  &  &  &  &  &  &  &  &  & 1.00 & 0.90 & 0.73 & 0.53 & 0.47 & 0.44 & 0.41 & 0.38 & 0.31 & 0.24 \\ 
  0.55 &  &  &  &  &  &  &  &  &  &  & 1.00 & 0.80 & 0.52 & 0.47 & 0.44 & 0.41 & 0.38 & 0.32 & 0.26 \\ 
  0.60 &  &  &  &  &  &  &  &  &  &  &  & 1.00 & 0.69 & 0.61 & 0.57 & 0.53 & 0.48 & 0.40 & 0.29 \\ 
  0.65 &  &  &  &  &  &  &  &  &  &  &  &  & 1.00 & 0.96 & 0.92 & 0.87 & 0.80 & 0.69 & 0.54 \\ 
  0.70 &  &  &  &  &  &  &  &  &  &  &  &  &  & 1.00 & 0.97 & 0.93 & 0.87 & 0.77 & 0.62 \\ 
  0.75 &  &  &  &  &  &  &  &  &  &  &  &  &  &  & 1.00 & 0.97 & 0.93 & 0.86 & 0.72 \\ 
  0.80 &  &  &  &  &  &  &  &  &  &  &  &  &  &  &  & 1.00 & 0.97 & 0.91 & 0.78 \\ 
  0.85 &  &  &  &  &  &  &  &  &  &  &  &  &  &  &  &  & 1.00 & 0.96 & 0.85 \\ 
  0.90 &  &  &  &  &  &  &  &  &  &  &  &  &  &  &  &  &  & 1.00 & 0.91 \\ 
  0.95 &  &  &  &  &  &  &  &  &  &  &  &  &  &  &  &  &  &  & 1.00 \\
   \midrule
   \bottomrule
\end{tabular}
}
\label{tab:ciq_summary}
\end{table}

%------------------------------------------------------------------------------------------------------------------------------------------------------%
\clearpage
\newpage
\section{Predicting the CIQ Premium}
\label{sec:pred_ciq_premium}

\begin{table}[h!]
\caption{Predicting the $\Delta$CIQ Premium}
\centering
\scriptsize
\begin{minipage}{\textwidth} 
The table reports predictive regressions of the high-minus-low equal-weighted portfolio returns from decile sorts on the exposures to the $\Delta CIQ^{LT}$ factor. We regress the portfolio's one-month-ahead returns on the lagged market return ($Mkt$), on the CBOE variance factor ($VIX^2$) and on its two components: the conditional variance of stock returns (CV) and the equity variance premium (VP) based on the procedure of \cite{BEKAERT2014181}. Coefficients are scaled to capture the effect of one-standard-deviation increase in the factor on the annualized portfolio return in percentage points. $t$-statistics (in parentheses) are based on standard errors of \cite{newey1994}. The  data cover the period between February 1990 and January 2022.
\end{minipage}
\vspace{1em}
\begin{tabular}{lcccccc}
  \toprule
  \midrule
  & (1) & (2) & (3) & (4) & (5) & (6) \\
  \midrule
  $VIX^2$ & 6.10 &  &  &  &  &  \\ 
   & (1.90) &  &  &  &  &  \\ 
  $VP$ &  & 7.28 &  &  & 8.61 & 7.81 \\ 
   &  & (2.27) &  &  & (2.71) & (2.34) \\ 
  $CV$ &  &  & 4.30 &  & -1.86 & 1.23 \\ 
   &  &  & (1.37) &  & (-0.84) & (0.43) \\ 
  $Mkt$ &  &  &  & 2.74 &  & 5.36 \\ 
   &  &  &  & (1.14) &  & (2.23) \\
   %Const & 2.42 & 1.50 & 4.77 & 8.47 & 1.94 & -1.12 \\ 
   %& (0.72) & (0.47) & (1.58) & (3.62) & (0.58) & (-0.32) \\
  $R^2$ IS & 2.22 & 3.16 & 1.10 & 0.45 & 3.26 & 4.58 \\ 
  %$R^2$ IS CT & 2.22 & 3.16 & 1.10 & 0.41 & 3.28 & 4.58 \\ 
  %oosConst & 4.29 & 4.10 & 4.66 & 4.92 & 5.18 & 4.33 \\ 
   %& (1.50) & (1.45) & (1.60) & (1.60) & (1.77) & (1.45) \\ 
  $R^2$ OOS & 0.90 & 2.44 & -0.77 & -2.45 & -0.10 & -0.83 \\ 
  $R^2$ OOS CT & 1.49 & 2.44 & 0.42 & -0.97 & 2.34 & 2.00 \\
   \midrule
   \bottomrule
\end{tabular}
\label{tab:ciq_port_regressions}
\end{table}

%------------------------------------------------------------------------------------------------------------------------------------------------------%

\clearpage
\newpage
\section{CIQ premia}
\label{sec:ciq_all}

\begin{table}[h!]
\caption{Decile Portfolios Sorted on the Exposure to the $\Delta$CIQ$(\tau)$} 
\centering
\scriptsize
\begin{minipage}{\textwidth} 
The table contains annualized out-of-sample excess returns of ten portfolios sorted on the exposure to the $\Delta$CIQ($\tau$) factors. We also report returns of the high minus low (H - L) portfolios, their $t$-statistics, and annualized alphas with respect to the six-factor model of \cite{FAMA2018234}. All $t$-statistics are based on the correction of \cite{newey1994}. The data cover the period from January 1968 to December 2024. Each month, we use all the CRSP stocks with at least 48 monthly observations over the last 60 months, and exclude penny stocks with prices below \$1.
\end{minipage}
\vspace{1em}
%\adjustbox{max height=\dimexpr\textheight-5.5cm\relax,
%           max width=\textwidth}{
\resizebox{\textwidth}{!}{%
\begin{tabular}{lcccccccccccccc}
  \toprule
  \midrule
$\tau$ & Low & 2 & 3 & 4 & 5 & 6 & 7 & 8 & 9 & High & H - L & $t$-stat & $\alpha$ & $t$-stat \\ 
  \midrule
%  \multicolumn{15}{l}{Panel A: One-month returns} \\ % align: l,c,r
%  \midrule
  \multicolumn{15}{c}{\textit{Equal-weighted}} \\ % align: l,c,r
	0.05 & 5.68 & 8.17 & 9.25 & 9.51 & 9.93 & 9.55 & 10.08 & 10.95 & 11.55 & 10.87 & 5.19 & 2.97 & 5.35 & 3.05 \\ 
  0.10 & 5.14 & 7.91 & 8.76 & 9.59 & 9.50 & 10.28 & 10.68 & 10.72 & 11.98 & 10.96 & 5.82 & 3.29 & 6.07 & 3.68 \\ 
  0.15 & 4.78 & 8.02 & 9.13 & 9.79 & 10.01 & 9.69 & 10.60 & 10.72 & 11.22 & 11.58 & 6.80 & 3.76 & 7.13 & 4.25 \\ 
  0.20 & 4.36 & 7.94 & 9.47 & 9.63 & 9.46 & 10.36 & 10.41 & 10.74 & 11.39 & 11.77 & 7.41 & 4.30 & 8.31 & 4.97 \\ 
  0.25 & 4.78 & 7.76 & 9.26 & 9.93 & 10.20 & 10.15 & 10.41 & 10.45 & 11.06 & 11.54 & 6.76 & 4.09 & 7.86 & 5.10 \\ 
  0.30 & 4.86 & 7.98 & 9.14 & 9.89 & 10.03 & 10.51 & 10.35 & 10.87 & 10.64 & 11.27 & 6.41 & 3.90 & 7.86 & 5.05 \\ 
  0.35 & 4.93 & 8.37 & 9.63 & 9.65 & 9.74 & 10.47 & 10.46 & 10.40 & 11.00 & 10.89 & 5.96 & 3.83 & 7.46 & 4.81 \\ 
  0.40 & 5.06 & 8.57 & 9.49 & 9.56 & 9.73 & 10.28 & 10.77 & 10.84 & 10.72 & 10.53 & 5.46 & 3.72 & 6.79 & 4.60 \\ 
  0.45 & 5.80 & 7.95 & 10.31 & 9.34 & 10.00 & 10.28 & 10.49 & 10.46 & 10.71 & 10.19 & 4.38 & 2.52 & 4.80 & 2.60 \\ 
  0.50 & 6.90 & 9.05 & 10.35 & 9.05 & 10.23 & 9.85 & 10.31 & 10.24 & 10.30 & 9.26 & 2.36 & 1.36 & 1.68 & 0.85 \\ 
  0.55 & 6.36 & 8.85 & 10.03 & 9.27 & 10.18 & 9.99 & 10.43 & 10.28 & 10.53 & 9.61 & 3.25 & 1.86 & 2.27 & 1.09 \\ 
  0.60 & 7.16 & 9.26 & 10.92 & 9.61 & 10.32 & 10.00 & 9.93 & 9.58 & 9.95 & 8.82 & 1.66 & 1.01 & 0.95 & 0.51 \\ 
  0.65 & 6.84 & 9.89 & 9.69 & 10.31 & 9.77 & 9.62 & 9.97 & 10.08 & 10.24 & 9.13 & 2.29 & 1.51 & 1.73 & 1.08 \\ 
  0.70 & 7.80 & 9.37 & 10.39 & 10.26 & 9.49 & 9.72 & 10.05 & 9.45 & 10.21 & 8.79 & 0.99 & 0.63 & -0.01 & -0.01 \\ 
  0.75 & 8.26 & 9.79 & 10.35 & 9.78 & 9.81 & 9.88 & 9.81 & 9.78 & 9.04 & 9.03 & 0.77 & 0.50 & -0.49 & -0.30 \\ 
  0.80 & 8.86 & 10.09 & 10.46 & 9.62 & 9.73 & 9.70 & 9.90 & 9.00 & 9.74 & 8.43 & -0.42 & -0.26 & -1.60 & -0.92 \\ 
  0.85 & 9.07 & 10.28 & 9.79 & 10.03 & 9.67 & 9.59 & 9.77 & 9.69 & 9.02 & 8.63 & -0.44 & -0.26 & -1.65 & -0.90 \\ 
  0.90 & 9.31 & 10.25 & 9.88 & 9.97 & 9.33 & 10.22 & 10.04 & 9.34 & 8.76 & 8.43 & -0.89 & -0.50 & -2.16 & -1.18 \\ 
  0.95 & 9.65 & 10.44 & 10.28 & 9.67 & 10.03 & 9.36 & 10.41 & 8.78 & 9.14 & 7.77 & -1.88 & -0.97 & -3.20 & -1.75 \\
   \midrule
     \multicolumn{15}{c}{\textit{Value-weighted}} \\ % align: l,c,r
   0.05 & 6.39 & 5.95 & 6.45 & 7.62 & 6.64 & 7.58 & 7.52 & 9.51 & 9.96 & 8.75 & 2.36 & 0.96 & 0.70 & 0.28 \\ 
  0.10 & 5.16 & 6.12 & 5.98 & 7.81 & 6.80 & 7.95 & 7.83 & 8.66 & 10.35 & 8.96 & 3.80 & 1.51 & 3.02 & 1.22 \\ 
  0.15 & 4.41 & 6.59 & 7.11 & 6.79 & 7.47 & 7.75 & 8.84 & 7.49 & 10.26 & 9.76 & 5.35 & 2.05 & 5.21 & 2.03 \\ 
  0.20 & 3.35 & 7.29 & 6.65 & 6.56 & 7.17 & 8.11 & 7.98 & 8.24 & 9.81 & 10.73 & 7.37 & 2.73 & 7.86 & 2.84 \\ 
  0.25 & 4.40 & 6.68 & 6.80 & 6.88 & 8.15 & 7.84 & 7.90 & 7.30 & 9.47 & 10.35 & 5.95 & 2.27 & 6.57 & 2.43 \\ 
  0.30 & 4.05 & 7.65 & 6.98 & 7.36 & 7.71 & 8.07 & 7.51 & 7.06 & 8.73 & 10.23 & 6.18 & 2.27 & 6.69 & 2.37 \\ 
  0.35 & 5.18 & 7.16 & 7.45 & 7.36 & 7.22 & 8.19 & 7.37 & 6.77 & 8.42 & 9.86 & 4.68 & 1.77 & 5.83 & 2.18 \\ 
  0.40 & 4.81 & 7.57 & 6.50 & 8.16 & 7.56 & 7.95 & 6.73 & 7.37 & 7.47 & 9.99 & 5.18 & 2.15 & 6.96 & 2.83 \\ 
  0.45 & 4.43 & 5.41 & 7.10 & 6.92 & 7.76 & 8.16 & 8.26 & 7.29 & 7.58 & 10.13 & 5.70 & 2.44 & 6.55 & 2.71 \\ 
  0.50 & 5.02 & 5.18 & 6.63 & 6.58 & 7.51 & 8.43 & 8.15 & 7.73 & 6.49 & 9.38 & 4.36 & 1.81 & 3.99 & 1.56 \\ 
  0.55 & 5.03 & 4.72 & 6.90 & 6.21 & 7.85 & 7.60 & 8.31 & 8.53 & 6.86 & 9.06 & 4.03 & 1.60 & 3.46 & 1.22 \\ 
  0.60 & 4.61 & 6.73 & 7.90 & 6.50 & 7.43 & 7.47 & 8.07 & 7.50 & 6.94 & 7.90 & 3.29 & 1.40 & 2.45 & 1.02 \\ 
  0.65 & 4.62 & 7.98 & 7.98 & 8.94 & 7.32 & 7.17 & 6.40 & 7.22 & 6.92 & 7.32 & 2.70 & 1.22 & 2.30 & 0.95 \\ 
  0.70 & 5.64 & 7.48 & 8.84 & 8.98 & 7.55 & 7.34 & 6.93 & 6.89 & 6.49 & 7.08 & 1.44 & 0.59 & 0.87 & 0.36 \\ 
  0.75 & 4.86 & 8.91 & 7.78 & 8.74 & 7.98 & 7.58 & 6.48 & 6.63 & 5.38 & 6.69 & 1.83 & 0.76 & 0.90 & 0.37 \\ 
  0.80 & 7.23 & 9.11 & 8.15 & 7.07 & 8.11 & 7.84 & 6.81 & 6.37 & 5.87 & 6.33 & -0.90 & -0.36 & 0.03 & 0.01 \\ 
  0.85 & 6.91 & 8.92 & 7.71 & 6.76 & 8.25 & 7.22 & 7.36 & 7.01 & 6.01 & 5.63 & -1.27 & -0.51 & 0.08 & 0.03 \\ 
  0.90 & 6.74 & 8.84 & 7.44 & 6.75 & 7.98 & 7.80 & 6.82 & 6.69 & 6.79 & 5.49 & -1.24 & -0.50 & 0.03 & 0.01 \\ 
  0.95 & 5.99 & 8.62 & 7.72 & 8.08 & 7.15 & 7.39 & 7.68 & 6.16 & 6.03 & 5.14 & -0.85 & -0.31 & 0.09 & 0.04 \\
   \midrule
   \bottomrule
\end{tabular}%
}
\label{tab:port_univ_10}
\end{table}

\begin{table}[h!]
\caption{Quintile Portfolios Sorted on the Exposure to the $\Delta$CIQ$(\tau)$} 
\centering
\scriptsize
\begin{minipage}{\textwidth} 
The table contains annualized out-of-sample excess returns of five portfolios sorted on the exposure to the $\Delta$CIQ($\tau$) factors. We also report returns of the high minus low (H - L) portfolios, their $t$-statistics, and annualized alphas with respect to the six-factor model of \cite{FAMA2018234}. All $t$-statistics are based on the correction of \cite{newey1994}. The data cover the period from January 1968 to December 2024. Each month, we use all the CRSP stocks with at least 48 monthly observations over the last 60 months, and exclude penny stocks with prices below \$1.
\end{minipage}
\vspace{1em}
%\adjustbox{max height=\dimexpr\textheight-5.5cm\relax,
%           max width=\textwidth}{
%\resizebox{\textwidth}{!}{%
\begin{tabular}{lccccccccc}
  \toprule
  \midrule
$\tau$ & Low & 2 & 3 & 4 & High & H - L & $t$-stat & $\alpha$ & $t$-stat \\ 
  \midrule
%  \multicolumn{15}{l}{Panel A: One-month returns} \\ % align: l,c,r
%  \midrule
  \multicolumn{10}{c}{\textit{Equal-weighted}} \\ % align: l,c,r
	0.05 & 6.93 & 9.38 & 9.74 & 10.51 & 11.21 & 4.28 & 2.93 & 4.91 & 3.36 \\ 
  0.10 & 6.53 & 9.17 & 9.89 & 10.70 & 11.47 & 4.95 & 3.28 & 5.66 & 4.09 \\ 
  0.15 & 6.40 & 9.46 & 9.85 & 10.66 & 11.40 & 5.00 & 3.33 & 5.89 & 4.33 \\ 
  0.20 & 6.15 & 9.55 & 9.91 & 10.57 & 11.58 & 5.43 & 3.76 & 6.83 & 5.11 \\ 
  0.25 & 6.27 & 9.59 & 10.18 & 10.43 & 11.30 & 5.03 & 3.67 & 6.65 & 5.50 \\ 
  0.30 & 6.42 & 9.51 & 10.27 & 10.61 & 10.96 & 4.54 & 3.43 & 6.23 & 5.25 \\ 
  0.35 & 6.65 & 9.64 & 10.11 & 10.43 & 10.94 & 4.29 & 3.49 & 6.07 & 5.43 \\ 
  0.40 & 6.81 & 9.52 & 10.01 & 10.80 & 10.62 & 3.81 & 3.33 & 5.42 & 5.07 \\ 
  0.45 & 6.88 & 9.82 & 10.14 & 10.48 & 10.45 & 3.57 & 2.72 & 4.18 & 3.06 \\ 
  0.50 & 7.97 & 9.70 & 10.04 & 10.27 & 9.78 & 1.81 & 1.36 & 1.56 & 1.04 \\ 
  0.55 & 7.60 & 9.65 & 10.09 & 10.36 & 10.07 & 2.47 & 1.87 & 1.94 & 1.25 \\ 
  0.60 & 8.21 & 10.26 & 10.16 & 9.76 & 9.38 & 1.17 & 0.98 & 0.87 & 0.64 \\ 
  0.65 & 8.36 & 10.00 & 9.70 & 10.02 & 9.68 & 1.32 & 1.12 & 1.18 & 0.96 \\ 
  0.70 & 8.59 & 10.32 & 9.61 & 9.75 & 9.50 & 0.92 & 0.75 & 0.26 & 0.20 \\ 
  0.75 & 9.03 & 10.07 & 9.85 & 9.79 & 9.03 & 0.01 & 0.01 & -0.85 & -0.64 \\ 
  0.80 & 9.48 & 10.03 & 9.72 & 9.45 & 9.09 & -0.39 & -0.29 & -1.36 & -0.99 \\ 
  0.85 & 9.67 & 9.91 & 9.63 & 9.73 & 8.82 & -0.85 & -0.60 & -1.91 & -1.33 \\ 
  0.90 & 9.78 & 9.93 & 9.77 & 9.69 & 8.59 & -1.19 & -0.79 & -2.31 & -1.59 \\ 
  0.95 & 10.05 & 9.98 & 9.69 & 9.60 & 8.45 & -1.59 & -0.98 & -2.90 & -2.03 \\
   \midrule
     \multicolumn{10}{c}{\textit{Value-weighted}} \\ % align: l,c,r
   0.05 & 6.03 & 7.01 & 6.93 & 8.38 & 9.46 & 3.42 & 1.65 & 2.43 & 1.28 \\ 
  0.10 & 5.77 & 6.96 & 7.40 & 8.11 & 9.94 & 4.17 & 2.04 & 3.75 & 1.96 \\ 
  0.15 & 5.91 & 6.91 & 7.62 & 8.21 & 10.02 & 4.12 & 2.00 & 4.28 & 2.31 \\ 
  0.20 & 5.90 & 6.61 & 7.69 & 7.98 & 9.96 & 4.06 & 2.02 & 4.82 & 2.71 \\ 
  0.25 & 6.01 & 6.92 & 8.02 & 7.32 & 9.61 & 3.60 & 1.84 & 4.49 & 2.51 \\ 
  0.30 & 6.37 & 6.99 & 8.02 & 7.11 & 9.13 & 2.77 & 1.45 & 3.72 & 2.08 \\ 
  0.35 & 6.48 & 7.28 & 7.63 & 6.95 & 8.87 & 2.38 & 1.27 & 3.84 & 2.19 \\ 
  0.40 & 6.52 & 7.29 & 7.55 & 6.98 & 8.24 & 1.72 & 0.95 & 3.25 & 1.83 \\ 
  0.45 & 4.92 & 6.88 & 7.86 & 7.64 & 8.82 & 3.90 & 2.31 & 4.96 & 2.72 \\ 
  0.50 & 5.06 & 6.55 & 7.89 & 8.04 & 7.57 & 2.51 & 1.46 & 2.25 & 1.13 \\ 
  0.55 & 4.73 & 6.46 & 7.65 & 8.40 & 7.62 & 2.89 & 1.70 & 2.75 & 1.40 \\ 
  0.60 & 6.07 & 6.99 & 7.43 & 7.66 & 7.23 & 1.16 & 0.70 & 0.74 & 0.40 \\ 
  0.65 & 6.80 & 8.40 & 7.13 & 6.61 & 7.21 & 0.42 & 0.25 & 0.66 & 0.36 \\ 
  0.70 & 6.97 & 8.95 & 7.31 & 6.85 & 6.70 & -0.27 & -0.16 & -0.64 & -0.36 \\ 
  0.75 & 7.78 & 8.44 & 7.61 & 6.53 & 6.04 & -1.74 & -1.04 & -2.24 & -1.24 \\ 
  0.80 & 8.34 & 7.75 & 7.74 & 6.61 & 6.34 & -2.00 & -1.12 & -2.15 & -1.13 \\ 
  0.85 & 8.53 & 7.31 & 7.63 & 7.21 & 6.07 & -2.47 & -1.31 & -2.29 & -1.20 \\ 
  0.90 & 8.24 & 6.88 & 8.02 & 6.76 & 6.51 & -1.73 & -0.88 & -1.11 & -0.59 \\ 
  0.95 & 7.83 & 8.15 & 7.30 & 6.98 & 5.82 & -2.01 & -0.94 & -1.46 & -0.76 \\
   \midrule
   \bottomrule
\end{tabular}%
%}
\label{tab:port_univ_5}
\end{table}

\begin{table}[h!]
\caption{Correlations between CIQ Premia} 
\centering
\scriptsize
\begin{minipage}{\textwidth} 
The table provides time-series correlations between the CIQ premia. We use the high-minus-low equal-weighted portfolios from decile sorts based on the exposures to the $\Delta$CIQ factors and report the correlations for range of values of $\tau$. The data cover the period from January 1968 to December 2024.
\end{minipage}
\vspace{1em}
\resizebox{\textwidth}{!}{
\begin{tabular}{lccccccccccccccccccc}
  \toprule
  \midrule
 & 0.05 & 0.1 & 0.15 & 0.2 & 0.25 & 0.3 & 0.35 & 0.4 & 0.45 & 0.5 & 0.55 & 0.6 & 0.65 & 0.7 & 0.75 & 0.8 & 0.85 & 0.9 & 0.95 \\
 \midrule
  0.05 & 1.00 & 0.93 & 0.90 & 0.87 & 0.82 & 0.77 & 0.69 & 0.56 & 0.25 & 0.15 & 0.16 & 0.14 & 0.04 & -0.16 & -0.32 & -0.46 & -0.55 & -0.60 & -0.66 \\ 
  0.10 &  & 1.00 & 0.96 & 0.93 & 0.90 & 0.85 & 0.78 & 0.67 & 0.35 & 0.22 & 0.24 & 0.23 & 0.14 & -0.06 & -0.23 & -0.37 & -0.47 & -0.54 & -0.62 \\ 
  0.15 &  &  & 1.00 & 0.96 & 0.93 & 0.89 & 0.84 & 0.72 & 0.41 & 0.28 & 0.30 & 0.26 & 0.18 & -0.02 & -0.19 & -0.33 & -0.43 & -0.51 & -0.61 \\ 
  0.20 &  &  &  & 1.00 & 0.97 & 0.94 & 0.88 & 0.79 & 0.44 & 0.30 & 0.32 & 0.32 & 0.26 & 0.06 & -0.12 & -0.27 & -0.38 & -0.47 & -0.58 \\ 
  0.25 &  &  &  &  & 1.00 & 0.97 & 0.93 & 0.84 & 0.49 & 0.33 & 0.36 & 0.36 & 0.34 & 0.13 & -0.04 & -0.20 & -0.31 & -0.41 & -0.53 \\ 
  0.30 &  &  &  &  &  & 1.00 & 0.96 & 0.89 & 0.53 & 0.36 & 0.39 & 0.43 & 0.42 & 0.23 & 0.05 & -0.11 & -0.23 & -0.34 & -0.47 \\ 
  0.35 &  &  &  &  &  &  & 1.00 & 0.94 & 0.59 & 0.41 & 0.44 & 0.47 & 0.50 & 0.33 & 0.17 & 0.02 & -0.10 & -0.21 & -0.35 \\ 
  0.40 &  &  &  &  &  &  &  & 1.00 & 0.71 & 0.51 & 0.53 & 0.57 & 0.60 & 0.46 & 0.32 & 0.18 & 0.07 & -0.04 & -0.19 \\ 
  0.45 &  &  &  &  &  &  &  &  & 1.00 & 0.77 & 0.77 & 0.69 & 0.58 & 0.47 & 0.42 & 0.34 & 0.28 & 0.20 & 0.09 \\ 
  0.50 &  &  &  &  &  &  &  &  &  & 1.00 & 0.91 & 0.72 & 0.55 & 0.48 & 0.43 & 0.36 & 0.33 & 0.26 & 0.16 \\ 
  0.55 &  &  &  &  &  &  &  &  &  &  & 1.00 & 0.75 & 0.59 & 0.51 & 0.45 & 0.39 & 0.35 & 0.28 & 0.16 \\ 
  0.60 &  &  &  &  &  &  &  &  &  &  &  & 1.00 & 0.75 & 0.65 & 0.56 & 0.49 & 0.42 & 0.33 & 0.17 \\ 
  0.65 &  &  &  &  &  &  &  &  &  &  &  &  & 1.00 & 0.92 & 0.82 & 0.73 & 0.62 & 0.50 & 0.32 \\ 
  0.70 &  &  &  &  &  &  &  &  &  &  &  &  &  & 1.00 & 0.94 & 0.87 & 0.78 & 0.68 & 0.51 \\ 
  0.75 &  &  &  &  &  &  &  &  &  &  &  &  &  &  & 1.00 & 0.95 & 0.89 & 0.81 & 0.68 \\ 
  0.80 &  &  &  &  &  &  &  &  &  &  &  &  &  &  &  & 1.00 & 0.96 & 0.90 & 0.79 \\ 
  0.85 &  &  &  &  &  &  &  &  &  &  &  &  &  &  &  &  & 1.00 & 0.95 & 0.86 \\ 
  0.90 &  &  &  &  &  &  &  &  &  &  &  &  &  &  &  &  &  & 1.00 & 0.92 \\ 
  0.95 &  &  &  &  &  &  &  &  &  &  &  &  &  &  &  &  &  &  & 1.00 \\
   \midrule
   \bottomrule
\end{tabular}
}
\label{tab:ciq_corrs_all}
\end{table}

%------------------------------------------------------------------------------------------------------------------------------------------------------%

\clearpage
\newpage
\section{$\Delta$CIQ Betas and Firm Characteristics}
\label{sec:chars}

\begin{table}[h!]
\caption{Cross-Sectional Determinants of the Upper-Tail CIQ Exposure}
\centering
\scriptsize
\begin{minipage}{\textwidth}
This table reports Fama–MacBeth cross-sectional regressions explaining firm-level exposure to innovations in the upper-tail CIQ factor. The dependent variable is each stock’s rolling 60-month beta with respect to $\Delta CIQ^{UT}$. Firm characteristics are ranked cross-sectionally each month and linearly scaled to lie between $-1$ and $1$. Regressions are estimated monthly, and reported coefficients are time-series averages. $t$-statistics based on \cite{newey1994} are reported in parentheses. Firm-level characteristics are from \cite{10.1093/rfs/hhz123} and \cite{10.1093/rfs/hhaa102}. The sample covers the period from January 1968 to December 2018.
\end{minipage}
\vspace{0.8em}
\begin{tabular}{lccccc}
  \toprule
  \midrule
 & (1) & (2) & (3) & (4) & (5) \\
 & Liquidity & Funding & Risk & Vulnerability & Full \\ 
  \midrule
  \multicolumn{6}{l}{\textit{Liquidity fragility}}  \\
  $\texttt{dto}$ & 0.01 &  &  &  & 0.00 \\ 
   & (2.35) &  &  &  & (1.75) \\ 
  $\texttt{lturnover}$ & 0.10 &  &  &  & 0.08 \\ 
   & (5.40) &  &  &  & (4.53) \\ 
  $\texttt{std\_turn}$ & -0.02 &  &  &  & -0.01 \\ 
   & (-1.06) &  &  &  & (-1.10) \\ 
  $\texttt{suv}$ & 0.00 &  &  &  & 0.00 \\ 
   & (-1.19) &  &  &  & (-0.13) \\
      \multicolumn{1}{l}{\textit{Funding and financial slack}} &&&&& \\ 
  $\texttt{c}$ &  & 0.01 &  &  & 0.01 \\ 
   &  & (0.56) &  &  & (0.62) \\ 
  $\texttt{nop}$ &  & -0.06 &  &  & -0.03 \\ 
   &  & (-3.24) &  &  & (-3.37) \\
      \multicolumn{6}{l}{\textit{Risk controls}} \\ 
  $\texttt{beta}$ &  &  & 0.05 &  & 0.03 \\ 
   &  &  & (5.93) &  & (5.77) \\ 
  $\texttt{rel\_to\_high\_price}$ &  &  & -0.03 &  & -0.01 \\ 
   &  &  & (-2.20) &  & (-1.57) \\
   \multicolumn{6}{l}{\textit{Vulnerability control}} \\ 
  $\texttt{pm\_adj}$ &  &  &  & -0.01 & -0.02 \\ 
   &  &  &  & (-1.11) & (-2.61) \\
   Intercept & 0.00 & -0.00 & 0.00 & 0.00 & -0.00\\
 & (0.09) & (-0.03) & (0.52) & (0.04) & (-0.11)\\
 %\midrule
$R_{adj}^2$ & 3.54 & 2.52 & 2.64 & 1.44 & 5.87\\
$\bar{n}$ & 2578 & 2578 & 2578 & 2578 & 2578\\
$T$ & 612 & 612 & 612 & 612 & 612\\
   \midrule
   \bottomrule
\end{tabular}
\label{tab:ciq_exp_ut_det}
\end{table}

\begin{table}[h!]
\caption{Cross-Sectional Determinants of Lower-Tail CIQ Exposure -- Full Results}
\centering
\tiny
\begin{minipage}{\textwidth}
This table reports Fama–MacBeth cross-sectional regressions explaining firm-level exposure to innovations in the lower-tail CIQ factor. The dependent variable is each stock’s rolling 60-month beta with respect to $\Delta CIQ^{LT}$. Firm characteristics are ranked cross-sectionally each month and linearly scaled to lie between $-1$ and $1$. Regressions are estimated monthly, and reported coefficients are time-series averages. $t$-statistics based on \cite{newey1994} are reported in parentheses. Firm-level characteristics are from \cite{10.1093/rfs/hhz123} and \cite{10.1093/rfs/hhaa102}. The sample covers the period from January 1968 to December 2018.
\end{minipage}
\vspace{0.8em}
\begin{tabular}{lccccc}
  \toprule
  \midrule
 & (1) & (2) & (3) & (4) & (5) \\
 & Liquidity & Funding & Risk & Fundamental & Full \\ 
  \midrule
  \multicolumn{6}{l}{\textit{Liquidity fragility}}  \\
  $\texttt{spread\_mean}$ & -0.03 &  &  &  &  \\ 
   & (-1.17) &  &  &  &  \\ 
  $\texttt{dto}$ & -0.02 &  &  &  & -0.01 \\ 
   & (-7.50) &  &  &  & (-5.66) \\ 
  $\texttt{lturnover}$ & -0.22 &  &  &  & -0.21 \\ 
   & (-11.08) &  &  &  & (-10.64) \\ 
  $\texttt{std\_turn}$ & 0.06 &  &  &  & 0.09 \\ 
   & (4.57) &  &  &  & (7.00) \\ 
  $\texttt{std\_volume}$ & 0.02 &  &  &  &  \\ 
   & (0.58) &  &  &  &  \\ 
  $\texttt{suv}$ & 0.02 &  &  &  & 0.01 \\ 
   & (7.36) &  &  &  & (3.77) \\
   \multicolumn{1}{l}{\textit{Funding and financial slack}} &&&&& \\ 
  $\texttt{debt2p}$ &  & 0.00 &  &  &  \\ 
   &  & (-0.31) &  &  &  \\ 
  $\texttt{roc}$ &  & -0.07 &  &  & -0.03 \\ 
   &  & (-5.49) &  &  & (-4.81) \\ 
  $\texttt{c}$ &  & -0.03 &  &  & -0.03 \\ 
   &  & (-2.92) &  &  & (-3.17) \\ 
  $\texttt{c2d}$ &  & 0.00 &  &  &  \\ 
   &  & (0.13) &  &  &  \\ 
  $\texttt{free\_cf}$ &  & 0.01 &  &  &  \\ 
   &  & (0.67) &  &  &  \\ 
  $\texttt{nop}$ &  & 0.09 &  &  & 0.06 \\ 
   &  & (2.95) &  &  & (3.65) \\ 
  $\texttt{d\_so}$ &  & -0.01 &  &  &  \\ 
   &  & (-0.28) &  &  &  \\
   \multicolumn{6}{l}{\textit{Risk controls}} \\ 
  $\texttt{beta}$ &  &  & -0.06 &  & -0.03 \\ 
   &  &  & (-11.23) &  & (-7.53) \\ 
  $\texttt{total\_vol}$ &  &  & -0.22 &  & -0.08 \\ 
   &  &  & (-8.72) &  & (-4.64) \\ 
  $\texttt{idio\_vol}$ &  &  & 0.15 &  & 0.05 \\ 
   &  &  & (4.97) &  & (2.27) \\ 
  $\texttt{ret\_max}$ &  &  & 0.01 &  &  \\ 
   &  &  & (1.38) &  &  \\ 
  $\texttt{rel\_to\_high\_price}$ &  &  & 0.05 &  & 0.03 \\ 
   &  &  & (6.28) &  & (4.80) \\ 
  $\texttt{cum\_return\_1\_0}$ &  &  & 0.00 &  &  \\ 
   &  &  & (-1.34) &  &  \\
   \multicolumn{6}{l}{\textit{Fundamental controls}} \\ 
  $\texttt{lme}$ &  &  &  & -0.02 &  \\ 
   &  &  &  & (-0.43) &  \\ 
  $\texttt{beme}$ &  &  &  & 0.07 &  \\ 
   &  &  &  & (1.22) &  \\ 
  $\texttt{e2p}$ &  &  &  & 0.05 & 0.01 \\ 
   &  &  &  & (3.94) & (0.74) \\ 
  $\texttt{q}$ &  &  &  & -0.02 &  \\ 
   &  &  &  & (-0.58) &  \\ 
  $\texttt{prof}$ &  &  &  & 0.00 &  \\ 
   &  &  &  & (0.09) &  \\ 
  $\texttt{roa}$ &  &  &  & 0.03 &  \\ 
   &  &  &  & (1.88) &  \\ 
  $\texttt{pm\_adj}$ &  &  &  & -0.06 & -0.04 \\ 
   &  &  &  & (-9.06) & (-7.77) \\ 
  $\texttt{investment}$ &  &  &  & -0.06 & -0.02 \\ 
   &  &  &  & (-5.28) & (-2.76) \\ 
  $\texttt{noa}$ &  &  &  & 0.01 &  \\ 
   &  &  &  & (1.79) &  \\ 
   Intercept & -0.00 & -0.00 & -0.00 & -0.00 & -0.00\\
 & (-1.37) & (-0.56) & (-0.72) & (-0.59) & (-1.10)\\
 %\midrule
$R_{adj}^2$ & 9.47 & 6.88 & 8.90 & 9.17 & 12.86\\
$\bar{n}$ & 2578 & 2578 & 2578 & 2578 & 2578\\
$T$ & 612 & 612 & 612 & 612 & 612\\
   \midrule
   \bottomrule
\end{tabular}
\label{tab:ciq_exp_det_full}
\end{table}

\begin{table}[h!]
\caption{Firm Characteristics} 
\centering
\scriptsize
\begin{minipage}{\textwidth} 
The table provides a list of firm characteristics of \cite{10.1093/rfs/hhz123} and \cite{10.1093/rfs/hhaa102}. We employ them to explain the firm-level exposures to the $\Delta$CIQ factors.
\end{minipage}
\vspace{1em}
\centering
\begin{tabular}{@{}l l p{0.6\textwidth}@{}}
\toprule
\midrule
% Category: Trading Frictions
\\[-0.8em] % Add vertical space before category
\multicolumn{3}{l}{\uline{\textbf{Liquidity Fragility and Trading}}} \\[0.3em]
(1) & \texttt{spread\_mean}          & Average daily bid-ask spread \\
(2)  & \texttt{dto}                   & De-trended Turnover - market Turnover \\  
(2)  & \texttt{lturnover}             & Last month’s volume to shares outstanding \\  
(4) & \texttt{std\_turn}             & Standard deviation of daily turnover \\  
(5) & \texttt{std\_volume}           & Standard deviation of daily volume \\  
(6) & \texttt{suv}                   & Standard unexplained volume \\  

\\[-0.8em] 
\multicolumn{3}{l}{\uline{\textbf{Funding and Financial Slack}}} \\[0.3em] 
(7) & \texttt{debt2p}                & Total debt to Size \\
(8) & \texttt{roc}                   & Size + long-term debt - total assets to cash \\  
(9) & \texttt{c}                     & Cash to AT \\ 
(10) & \texttt{c2d}                   & Cash flow to total liabilities \\  
(11) & \texttt{free\_cf}              & Free cash flow to BE \\
(12) & \texttt{nop}                   & Net payouts to Size \\  
(13) & \texttt{d\_so}                 & Log change in split-adjusted shares outstanding \\

\\[-0.8em] 
\multicolumn{3}{l}{\uline{\textbf{Risk Controls}}} \\[0.3em]
(14)  & \texttt{beta}           & CAPM beta using daily returns \\
(15) & \texttt{total\_vol}            & Standard deviation of daily returns \\ 
(16)  & \texttt{idio\_vol}             & Idio vol of Fama-French 3 factor model \\
(17) & \texttt{ret\_max}              & Maximum daily return \\
(18)  & \texttt{rel\_to\_high\_price}  & Price to 52 week high price \\
(19) & \texttt{cum\_return\_1\_0}     & Return 1 month before prediction \\

\\[-0.8em] 
\multicolumn{3}{l}{\uline{\textbf{Fundamental Controls}}} \\[0.3em]
(20)  & \texttt{lme}                   & Price times shares outstanding \\
(21) & \texttt{beme}                  & Book to market ratio \\
(22) & \texttt{e2p}                   & Income before extraordinary items to Size \\  
(23) & \texttt{q}                     & Tobin’s Q \\
(24) & \texttt{prof}                  & Gross profitability over BE \\  
(25) & \texttt{roa}                   & Return on assets \\
(26) & \texttt{pm\_adj}               & Profit margin - mean PM in Fama-French 48 industry \\
(27) & \texttt{investment}            & \% change in AT \\
(28) & \texttt{noa}                   & Net-operating assets over lagged AT \\  
\midrule 
\bottomrule
\end{tabular}
\label{tab:chars_list}
\end{table}

%------------------------------------------------------------------------------------------------------------------------------------------------------%

\end{document}